\documentclass[prl,a4paper,11pt]{article}
\pdfoutput=1

\usepackage{bm}        
\usepackage{amsmath}						
\usepackage{amssymb}						
\usepackage{amsthm}						
\usepackage{array}							
\usepackage[english]{babel}					
\usepackage{cancel}						
\usepackage{eufrak}							
\usepackage{fullpage}						
\usepackage{graphicx}						
\usepackage{multirow}						
\usepackage{nccmath}						
\usepackage{nicefrac}						
\usepackage[section]{placeins}                         
\usepackage{booktabs}						
\usepackage[format=hang]{caption}				
\usepackage[hang,flushmargin]{footmisc}		
\usepackage{hyperref}						
\usepackage{mathtools}						
\usepackage[usenames]{color}
\usepackage[noadjust]{cite}
\numberwithin{equation}{section}				
\def\d{\mathrm d}							

\newcommand{\be}{\begin{equation}}
\newcommand{\ee}{\end{equation}}
\newcommand{\bea}{\begin{eqnarray}}
\newcommand{\eea}{\end{eqnarray}}

\begin{document}

\thispagestyle{empty}

\vspace*{2cm}

\begin{center}
{\bf \LARGE  Lumpy AdS$\bf{_5\times}$ S$\bf{^5}$ Black Holes and Black Belts}\\
\vspace*{2.5cm}

{ {\bf\'Oscar J. C.~Dias$^{1}$, } {\bf Jorge E.~Santos$^{2}$}, {\bf Benson Way$^{2}$}

\vspace*{1cm}

{\it $1$ STAG research centre and Mathematical Sciences, University of Southampton, UK}\\ \vspace{0.3cm}
{\it $2$ Department of Applied Mathematics and Theoretical Physics, University of Cambridge, Wilberforce Road, Cambridge CB3 0WA, UK}\\ \vspace{0.3cm}}

\vspace*{0.5cm} {\tt ojcd1r13@soton.ac.uk, jss55@cam.ac.uk, bw356@cam.ac.uk}

\end{center}

\vspace*{1cm}

\begin{abstract}
Sufficiently small Schwarzschild black holes in global AdS$_5\times$S$^5$ are Gregory-Laflamme unstable. We construct new families of black hole solutions that bifurcate from the onset of this instability and break the full SO$(6)$ symmetry group of the S$^5$ down to SO$(5)$. These new ``lumpy" solutions are labelled by the harmonics $\ell$. We find evidence that the $\ell = 1$ branch never dominates the microcanonical/canonical ensembles and connects through a topology-changing merger to a localised black hole solution with S$^8$ topology. We argue that these S$^8$ black holes should become the dominant phase in the microcanonical ensemble for small enough energies, and that the transition to Schwarzschild black holes is first order. Furthermore, we find two branches of solutions with $\ell = 2$.  We expect one of these branches to connect to a solution containing two localised black holes, while the other branch connects to a black hole solution with horizon topology $\mathrm S^4\times\mathrm S^4$ which we call a ``black belt".
\end{abstract}
\noindent

\newpage
\thispagestyle{empty}
\tableofcontents

\setcounter{page}{1} \setcounter{footnote}{0}

\section{Introduction \label{sec:Intro}}
Gauge-gravity duality \cite{Maldacena:1997re} provides a unique arena to study quantum gravity in its most extreme regimes. Its best understood formulation stipulates an equivalence between ten dimensional IIB String Theory on AdS$_5\times \mathrm{S}^5$ and four-dimensional $\mathcal{N}=4$ Super-Yang-Mills (SYM) with gauge group $SU(N)$. This AdS$_5$/CFT$_4$ duality is the most concrete example of the holographic principle and provides a non-perturbative definition of string theory.  

States of the field theory at large $N$ and large t'Hooft coupling correspond to solutions of classical gravity in the bulk.  In particular, bulk black holes describe thermal states on the field theory, with the field theory temperature $T$ identified with the Hawking temperature of the AdS black hole. We thus expect black holes in AdS$_5\times$S$^5$ to play an important role in understanding the phase diagram of $\mathcal{N} = 4$ SYM.

According to the correspondence, the background spacetime for the field theory is specified by the four-dimensional boundary of AdS$_5$.  Since $\mathcal{N}=4$ SYM is a conformal field theory, it does not exhibit phase transitions on a scale-invariant background like Minkowski space $\mathbb{M}^{1,3}$.  Instead, a different background spacetime can be chosen which allows for a more interesting phase structure. The phase structure of such solutions is a well-studied topic (see, e.g. the review \cite{Marolf:2013ioa} and references therein).  However, much of this study neglects effects set by the curvature scale of the S$^5$.  

In this manuscript, we construct new thermal phases where the S$^5$ plays an important role.  One of the most well-studied backgrounds for the field theory is the Einstein static universe $\mathbb{R}_t\times \mathrm{S}^3$.  We will therefore be concerned with solutions that are asymptotically \emph{global} AdS$_5\times \mathrm{S}^5$. These solutions must satisfy the type IIB SUGRA equations of motion.  With only the metric $g$ and Ramond-Ramond 5-form $F_{(5)}=\mathrm{d} C_{(4)}$ turned on, these equations are given by:
\begin{equation} \label{EOM:IIB}
G_{MN}\equiv R_{MN} - \frac{1}{48} F_{MPQRS} F_M{}^{PQRS}=0, \qquad  \nabla_M F^{MPQRS}=0\,, \qquad 
F_{(5)}=\star F_{(5)}\,,
\end{equation}
where we have imposed self-duality on the 5-form.  Perhaps the most well-known solution to these equations is 
AdS$_5 \times$S$^5$ which in global coordinates is described by\footnote{Capital indices $M,N,...$ are $d=10$ indices, Greek indices $\mu,\nu,...$ are AdS$_5$ indices, and small case Latin indices $a,b,...$ are indices on the S$^5$.}
\be\label{AdS5xS5}
 \mathrm{d}s^2 = -\left(1+\frac{r^2}{L^2}\right) \mathrm{d}t^2+\frac{\mathrm{d}r^2}{1+\frac{r^2}{L^2}} + r^2 \mathrm{d}\Omega_3^2 +L^2 \mathrm{d}\Omega_5^2\;,\qquad F_{\mu\nu\rho\sigma\tau}=\epsilon_{\mu\nu\rho\sigma\tau},\qquad F_{abcde}=\epsilon_{abcde}\;,
\ee
where $\epsilon_{\mu \nu \rho\sigma\tau}\mathrm{d}y^\mu\wedge \mathrm{d}y^\nu\wedge \mathrm{d}y^\rho\wedge \mathrm{d}y^\sigma \wedge \mathrm{d}y^\tau$ and $\epsilon_{abcde} \mathrm{d}x^a \wedge \mathrm{d}x^b \wedge \mathrm{d}x^c \wedge \mathrm{d}x^d \wedge \mathrm{d}x^e$ are the volume forms of the AdS$_5$ and S$^5$ base spaces, respectively.  $L$ is the AdS$_5$ length scale and the S$^5$ radius.  These are required to be the same by the equations of motion. 

We are interested in black hole solutions that are asymptotically global AdS$_5 \times$S$^5$.  There is, of course, the Schwarzschild BH family (hereafter abbreviated as AdS$_5$-Schw$\times$S$^5$ BH or simply the AdS$_5$-Schw BH) with horizon topology S$^3\times$S$^5$.  This solution can be written as
\begin{eqnarray}\label{Schw}
&& \mathrm{d}s^2 = -f \mathrm{d}t^2+\frac{\mathrm{d}r^2}{f} + r^2 \mathrm{d}\Omega_3^2 +L^2 \mathrm{d}\Omega_5^2\;,\qquad F_{\mu \nu \rho\sigma\tau} = \epsilon_{\mu \nu \rho\sigma\tau}, \qquad
F_{abcde} = \epsilon_{abcde}\,,\\
&& \hbox{where}
 \quad f=1+\frac{r^2}{L^2}-\frac{r_+^2}{r^2}\left(\frac{r_+^2}{L^2}+1\right)\,. \nonumber
\end{eqnarray}
When the horizon radius $r_+$ vanishes, this solution reduces to global AdS$_5 \times$S$^5$.  

Let us review the phase diagram of this family of solutions in a given thermodynamic ensemble.  There are two ensembles that differ significantly: the canonical ensemble (at fixed temperature) and the microcanonical ensemble (at fixed energy).

In the canonical ensemble, we must consider Euclidean solutions at fixed temperature $T$, that asymptote to S$^1\times$S$^3\times$S$^5$, with the radius of the S$^1$ identified as the inverse temperature $T^{-1}$.  At a given temperature, there are up to three solutions in this family that compete: small black holes, large black holes, and thermal AdS (which corresponds to a thermal gas of gravitons).  Computing the free energy from the Euclidean action determines which solution is preferred.  At high temperatures, the large black hole phase dominates, while at low temperatures the gas of AdS gravitons is the dominant phase. Small black holes are never preferred.  This indicates the existence of a critical temperature where a first order phase transition -- the Hawking-Page transition -- occurs \cite{Hawking:1982dh,Witten:1998qj}.  In the dual CFT$_4$, this is interpreted as a confinement/deconfinement transition \cite{Witten:1998qj}. 

In the microcanonical ensemble one fixes the energy and computes the entropy to find the dominant phase. In the AdS$_5$-Schw$\times$S$^5$ family, there is only one solution at a given energy, so its phase diagram is trivial.  

However, the AdS$_5$-Schw$\times$S$^5$ family assumes a large amount of symmetry.  There may be other solutions with AdS$_5 \times$S$^5$ asymptotics which do not globally respect the $SO(6)$ symmetry of the S$^5$. 

Why would one expect such solutions to exist?  As first observed  by \cite{Banks:1998dd,Peet:1998cr,Hubeny:2002xn}, small AdS$_5$-Schw$\times$S$^5$ BHs are unstable to a Gregory-Laflamme (GL) instability \cite{Gregory:1993vy,Gregory:2000gf} when the ratio between the horizon radius and the S$^5$ radius, $r_+/L$, is smaller than a critical value. Indeed, a topologically S$^3\times$S$^5$ black hole with a large S$^5$ and small S$^3$ resembles a black brane, and systems with such a large hierarchy of scales are generically unstable to GL-type instabilities. 

In analogous systems, such instabilities typically contain a zero mode which indicates the existence of a new family of black holes.  This new family often connects to other black holes with different topologies through conical transitions.  For example, in the phase diagram of Kaluza Klein black holes that are asymptotically $\mathbb{M}^{1,3}\times$S$^1$ \cite{Horowitz:2001cz,Gubser:2001ac,Kol:2002xz,Wiseman:2002zc,Kol:2003ja,Harmark:2002tr,Harmark:2003yz,Gorbonos:2004uc,Kudoh:2004hs,Sorkin:2004qq,Asnin:2006ip,Harmark:2007md,Horowitz:2011cq,Headrick:2009pv,Figueras:2012xj}, there are S$^2\times$S$^1$ black strings that are GL unstable when the S$^2$ is small relative to the S$^1$.  The associated zero mode connects these solutions to non-uniform strings which themselves connect to topologically spherical S$^3$ black holes.

As another example, asymptotically flat Myers-Perry BHs in $d\geq 6$ dimensions with a single non-trivial (but large) angular momentum also suffer from a GL-type instability \cite{Emparan:2003sy,Emparan:2007wm,Dias:2009iu,Dias:2010eu,Dias:2010maa,Dias:2010gk,Dias:2011jg,Emparan:2011ve,Dias:2014cia} (known in this setup as the ultraspinning instability). The zero mode in this case connects the Myers-Perry BHs to lumpy (a.k.a. bumpy or rippled) rotating BHs with S$^{d-2}$ horizons.  Some of these lumpy solutions then connect to black rings with horizon topology $S^1\times S^{d-3}$ \cite{Emparan:2007wm,Emparan:2011ve,Dias:2014cia}.  There are also zero modes corresponding to higher harmonics on the S$^{d-2}$, which yield other lumpy solutions that connect to multi-horizon solutions like black Saturns or di-rings \cite{Emparan:2014pra}. 

In the present case, zero modes in AdS$_5$-Schw$\times$S$^5$ BHs would lead to new black holes with the same S$^3\times$S$^5$ topology, but break the $SO(6)$ rotation symmetry of the S$^5$. We call these new solutions ``lumpy" black holes.  These lumpy black holes are expected to connect to other (possibly multi-horizon) solutions with horizon topology S$^8$, or other products of spheres like S$^4\times$S$^4$.
From a dimensional reduction, the $S^5$ is interpreted in the dual field theory as a number of scalar operators that respect an $SO(6)$ symmetry.   The GL instability is therefore associated with spontaneous symmetry breaking, where these operators develop non-trivial VEVs. In this case, we will argue that this phase transition is first order.  

Thus, there are many different solutions that compete in a given thermodynamic ensemble.  Since these new solutions appear only for black holes that are sufficiently small compared to the S$^5$, they are not expected to dominate the canonical ensemble, which favours large black holes at similar temperatures.  However, in the microcanonical ensemble at small energies, we expect the solution with the most entropy to be a black hole with spherical topology S$^8$.  More specifically, the most entropic spherical black hole is expected to be the most symmetric, preserving the full $SO(4)$ symmetry of global AdS$_5$ as well as the largest subgroup of $SO(6)$, which is $SO(5)$.  We therefore focus on solutions which preserve the S$^3$ of AdS$_5$ and a round S$^4$ within the S$^5$.  

Let us write the metric of S$^5$ suggestively as
\be\label{S5}
\mathrm d\Omega_5^2=\frac{\mathrm d \tilde x^2}{1-\tilde x^2}+(1-\tilde x^2)\d\Omega_4^2 \;.
\ee
These are just the usual spherical coordinates with a redefined polar angle $\tilde x=\cos\theta$.  
Now we must isolate the perturbative modes that preserve the round S$^4$.  Any smooth perturbation about AdS$_5$-Schw$\times$S$^5$ can be decomposed into a sum of perturbations of scalar, vector and tensor types.  These types are defined according to how they transform under diffeomorphisms of the S$^5$.  The modes we are searching for appear in the scalar sector.  In the coordinates of \eqref{Schw} and \eqref{S5}, these take the simple form $\delta g_{MN}=\sum_\ell h^{(\ell)}_{MN}(r)Y_\ell(\tilde{x})$, where $Y_\ell(\tilde{x})$ is a scalar harmonic on the S$^5$ and $\ell$ is its quantum number. If we focus on harmonics that only depend on the polar angle $\tilde{x}$ (and hence preserve an S$^4$), $\ell$ labels the number of nodes along $\tilde{x}$.  

\begin{figure}[ht]
\centering
\includegraphics[width=.48\textwidth]{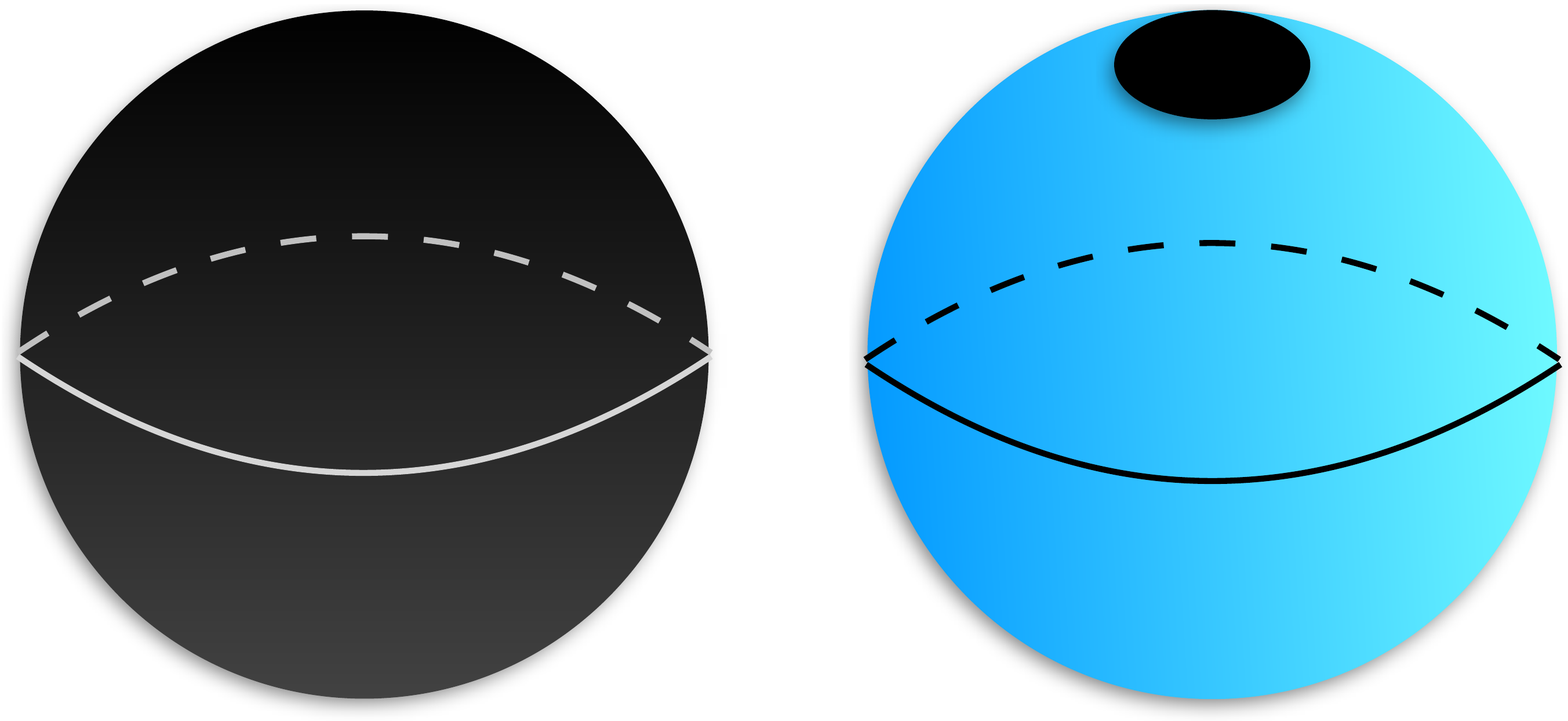}
\includegraphics[width=.48\textwidth]{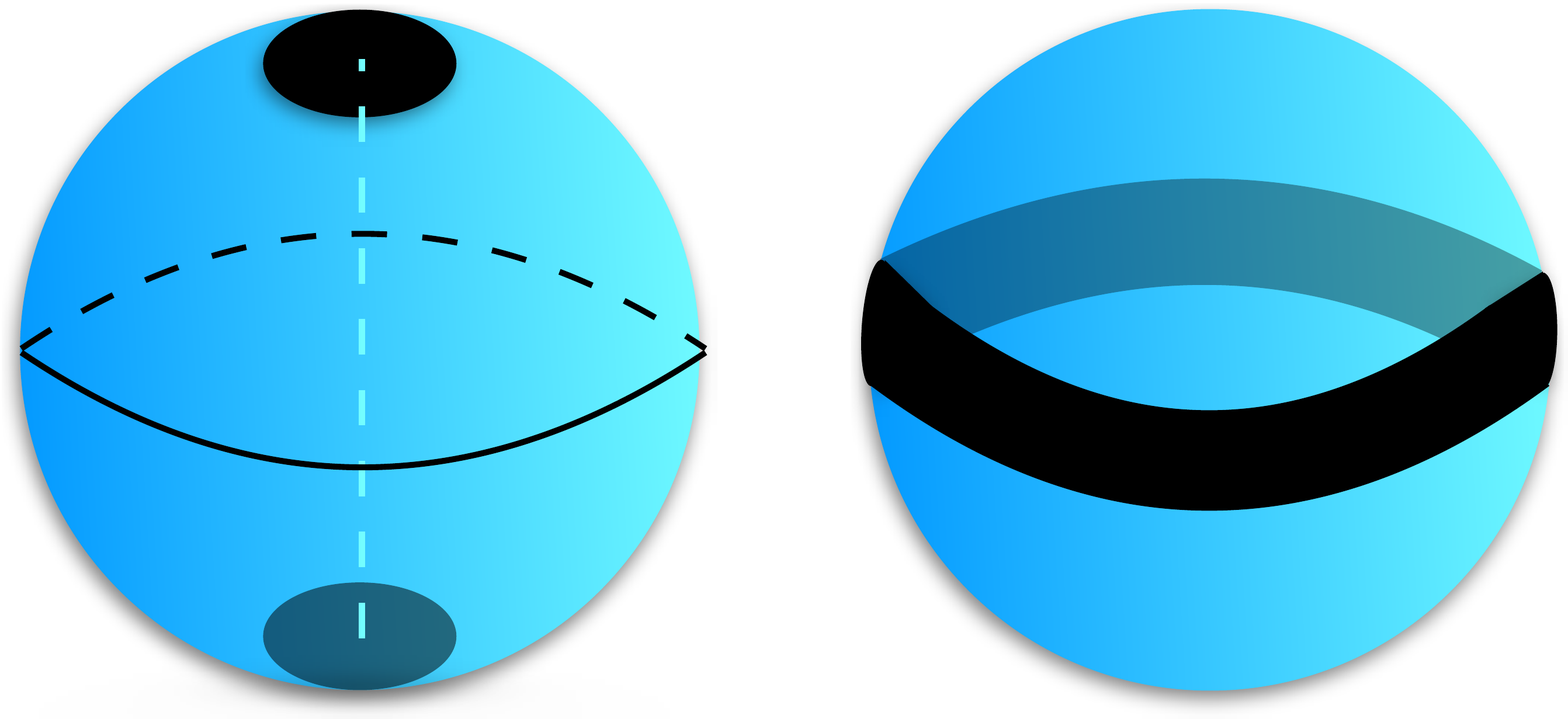}
\caption{A pictorial representation of the $\mathrm S^5$ for some black hole solutions that asymptote to global AdS$_5 \times$S$^5$ (the last three are conjectured solutions). The first one is the familiar  AdS$_5$-Schw$\times$S$^5$ that is smeared over the S$^5$. The second is a $\ell=1$ BH localized on the north pole of the S$^5$. The third is a $\ell=2$ localised BH solution and the fourth is a $\ell=2$ black belt. In this manuscript, we construct the $\ell=1,2$ ``lumpy" BH solutions, with horizon topology S$^3\times$S$^5$ that we conjecture to be connect to these localised BH solutions in a phase diagram.}\label{Fig:phases}
\end{figure}

The zero mode of AdS$_5$-Schw$\times$S$^5$ that appears with the largest horizon radius is  $\ell=1$. The linearised field equations reduce to a single ODE in the radial coordinate $r$ and can be solved for any positive integer value of $\ell$. For $\ell=1$, we find that the zero mode appears at the critical horizon radius
\begin{equation}\label{GLradius}
r_+{\bigl |}_{\ell=1}\simeq 0.4402373 \,L\,,
\end{equation}
as first found by Hubeny and Rangamani in \cite{Hubeny:2002xn}; AdS$_5$-Schw$\times$S$^5$ with $r_+\leq r_+{\bigl |}_{\ell=1}$ are unstable. At this zero mode, we expect a family of lumpy black holes to emerge.  By drawing an analogy with similar situations, we conjecture that this family leads to a conical merger with a family of black holes with S$^8$ topology, which can be thought of as being localised on the S$^5$; see Fig. \ref{Fig:phases}.b.  

Of course, the $\ell=1$ zero mode is only the first mode that appears.  There are an infinite number of such modes, with higher modes appearing at ever smaller horizon radii.  For example, the $\ell=2$ mode appears at
\be
r_+{\bigl |}_{\ell=2}\simeq 0.3238898 \,L\,.
\ee
We note, however, that there are important differences between the even $\ell$ modes and odd $\ell$ modes.  If $\delta g$ is a linear perturbation, then $-\delta g$ is also a linear perturbation.  In the odd $\ell$ modes, these can be mapped to each other via a $\mathbb Z_2$ symmetry of the S$^5$, and so are equivalent.  For example, among the $\ell=1$ solutions, the choice of sign merely selects whether the localised S$^8$ black hole will develop on the north or south pole of the S$^5$.  In the even $\ell$ modes, however, these perturbations map to themselves under this symmetry.  The $\delta g$ and $-\delta g$ perturbations are not equivalent, which means we have \emph{two} branches of solutions emanating from the even $\ell$ zero modes. In the $\ell=2$ modes, we expect one branch to lead to two disconnected S$^8$ black holes localised on the poles of the S$^5$;  see Fig. \ref{Fig:phases}.c.  We expect the other branch to lead to an s$^4\times$S$^4$ black hole (the s$^4$ being a smaller sphere than the S$^4$);  see Fig. \ref{Fig:phases}.d. The larger S$^4$ wraps around (coincides with) the S$^4$ equator of the S$^5$,  so we call these conjectured solutions ``black belts" (the s$^4$ gives the transverse directions of the belt).  Higher $\ell$ modes lead to various other multi-horizon solutions with some combination of S$^8$ holes and s$^4\times$S$^4$ belts.

In this paper, we construct these lumpy black holes connected to the $\ell=1$ and $\ell=2$ zero modes and study their thermodynamic properties.  We detail our numerical construction in section 2, and compute the phase diagram in section 3.  In appendix A, we give the technical details of Kaluza-Klein holography necessary to interpret our results on the CFT$_4$ \cite{Skenderis:2006uy} (see also \cite{Kim:1985ez,Gunaydin:1984fk,Lee:1998bxa,Lee:1999pj,Arutyunov:1999en,Skenderis:2006di,Skenderis:2007yb}).  Numerical checks are in appendix B.  
\section{Numerical Construction \label{sec:construction}}

\subsection{Lumpy solutions with $\ell=1$ \label{sec:constructionL1}}
Here, we give the numerical details of our construction of the $\ell=1$ lumpy black holes (BHs).  We use the DeTurck method, which proceeds as follows.  First, we choose a reference metric $\bar g$ that satisfies our desired boundary conditions (i.e. contains a regular horizon, has the correct asymptotics, and has the desired symmetry axes).  Then, rather than solve the equations \eqref{EOM:IIB}, we instead solve the similar equations
\begin{equation} \label{EOM:deTurck}
G^{(H)}_{MN} \equiv R_{MN} - \frac{1}{48} F_{MPQRS} F_M{}^{PQRS}-\nabla_{(M}\xi_{N)}=0, \qquad \nabla_M F^{MPQRS}=0\,, \qquad 
F_{(5)}=\star F_{(5)}\;,
\end{equation}
where  $\xi^M = g^{PQ}[\Gamma^M_{PQ}(g)-\bar{\Gamma}^M_{PQ}(\bar{g})]$ and $\bar{\Gamma}(\bar{g})$ is the Levi-Civita connection associated with the reference metric $\bar{g}$.

The benefit of this method is that unlike the equations \eqref{EOM:IIB}, the above equations \eqref{EOM:deTurck} form a set of elliptic PDEs \cite{Headrick:2009pv}.  For solutions of \eqref{EOM:deTurck} to also be solutions of \eqref{EOM:IIB}, we must have $\xi^M=0$.  In some cases (such as vacuum-Einstein), there is a proof that all solutions to \eqref{EOM:deTurck} also have $\xi^M=0$.  In our case, we do not have such a proof, so we must verify that $\xi^M=0$ after solving the equations.  The local uniqueness property of elliptic equations guarantees that solutions with $\xi^M\neq0$ cannot be arbitrarily close to those with $\xi^M=0$.

Now we must find a suitable reference metric $\bar g$.  The most obvious choice is to use AdS$_5$-Schw$\times$S$^5$, but rather than use the coordinates in \eqref{Schw} and \eqref{S5}, we instead redefine
\be
r=\frac{r_+ }{1-y^2},\qquad\tilde{x}=x\sqrt{2-x^2}\;,
\label{eq:coordinate}
\ee
and set $y_+=r_+/L$.  With these new coordinates, the solution \eqref{AdS5xS5} can be written
\begin{align}
\d s^2& = \frac{L^2}{\left(1-y^2\right)^2}\left[-y^2 \left(2-y^2\right)\,G(y)\,\d t^2  +\frac{4 y_+^2\,\d y^2}{\left(2-y^2\right) G(y)}+y_+^2\, \d\Omega_3^2 \right] \nonumber\\
& \qquad\qquad\qquad+L^2 \left(\frac{4 \d x^2}{2-x^2}+\left(1-x^2\right)^2 \d\Omega_4^2 \right) \,,\nonumber \\
C_{(4)}&=\frac{L^4 y_+^4}{\sqrt{2}} \frac{y^2 \left(2-y^2\right)}{\left(1-y^2\right)^4}\, H(y) \, \d t\wedge \d S_{(3)}-\frac{L^4}{\sqrt{2}}\,\d S_{(4)}\;,\label{schwarzschild2}
\end{align}
where 
\begin{equation}\label{lumpyansatz:aux}
G(y)=\left(1-y^2\right)^2+y_+^2 H(y)\,,\qquad H(y)=2-2 y^2+y^4\,.
\end{equation}
We choose this to be our reference metric $\bar g$.  Here, $x\in[-1,1]$, with $x=-1$ and $x=1$ corresponding to the north and south poles of the S$^5$, and $y\in[0,1]$ with $y=0$ corresponding to the horizon and $y=1$ to the boundary of the AdS$_5$.

Using this reference metric, we write down a general ansatz which is static and preserves the symmetries of the S$^3$ and S$^4$:
\begin{eqnarray}\label{lumpyansatz}
\d s^2 &=& \frac{L^2}{\left(1-y^2\right)^2}\nonumber\\
&& \times \left[-y^2 \left(2-y^2\right)\,G(y)\,Q_1 \,\d t^2  +\frac{4 y_+^2}{\left(2-y^2\right) G(y)}\,Q_2 \,\left[\d y + (1-y^2)^2 Q_3 \,\d x\right]^2+y_+^2 Q_5 \, \d\Omega_3^2 \right] \nonumber\\
&& \qquad+L^2 \left[Q_4\,\frac{4 \d x^2}{2-x^2}+ Q_6\, \left(1-x^2\right)^2 \d\Omega_4^2 \right] \,,\nonumber \\
C_{(4)}&=&\frac{L^4 y_+^4}{\sqrt{2}} \frac{y^2 \left(2-y^2\right)}{\left(1-y^2\right)^4}\, H(y) \,Q_7 \, \d t\wedge \d S_{(3)}-\frac{L^4}{\sqrt{2}}\, W\,\d S_{(4)}\;,
\end{eqnarray}
where $\left\{Q_I,W\right\}$ ($I=1,\ldots, 7$) are functions of ${x,y}$ which will be determined numerically. If we set $Q_1=Q_2=Q_4=Q_5=Q_6=Q_7=W=1$ and $Q_3=0$, we recover the AdS$_5$-Schw$\times$S$^5$ solution as written in \eqref{schwarzschild2}.  On a solution different from AdS$_5$-Schw$\times$S$^5$, our ansatz would preserve the full $SO(4)$ symmetry of the S$^3$ and an $SO(5)$ symmetry of the S$^5$, allowing deformations along the polar direction $x$.  Also, note that $L$ drops out of the equations of motion, so we are left with one parameter $y_+$ and 8 functions $\left\{Q_I,W\right\}$.  

At this point, let us count the number of equations we have.  The $tt$, $xx$, $xy$, $yy$, $\Omega_3\Omega_3$, and $\Omega_4\Omega_4$ components of the Einstein-DeTurck equation give 6 equations, the $t\Omega_3$ and $\Omega_4$ components of the five-form equation give 2 equations, and the  $x\Omega_4$ and $y\Omega_4$ components of the self-duality condition give us an additional 2.  With 10 equations and 8 functions, this seems like too many equations.  Furthermore, the self-duality equations do not yield second-order differential equations, so our equations are not manifestly elliptic.  

It turns out that these issues do not pose a problem.  One can use the two independent components of the self-duality constraint to algebraically solve for $\partial_xW$ and $\partial_yW$.  After substituting these derivatives into the five-form $F_{(5)}$, one finds that $F_{(5)}$ is now independent of $W$, i.e. just a function of $Q_{1,\ldots,7}$ and their first derivatives. This means that differentiating to obtain the second derivatives of $W$ and substituting those into the remaining equations of motion eliminates $W$ completely.  The five-form equations of motion reduce to a single equation and, together with the Einstein-DeTurck equations, we are left with 7 elliptic equations for the remaining 7 functions $Q_I$.  Since the metric and $F_{(5)}$ are now independent of $W$, there is also no need to compute $W$ to extract physical quantities.  

Now let us discuss boundary conditions (BCs).  The BCs at the horizon $y=0$ and the poles $x=\pm 1$ are determined by regularity.  The conformal boundary $y=1$ is determined by demanding that the solution is asymptotically AdS$_5\times$S$^5$.  More specifically, we would like the various operators in the dual field theory to be unsourced.  Determining the correct powers of $1-y$ that accomplishes this is a subtle issue which we defer to section \ref{sec:KKrenormalization} of the appendix.  Here, we simply give our boundary conditions in full:

\begin{equation}\label{BCs:poles}
\hspace{-0.2cm}  \hbox{BCs at the poles ($x=\pm1$)}: \left\{
\begin{array}{ll}
\partial_x Q_I(y,\pm 1)=0 \,, & \qquad I=1,2,5,6,7  \,; \\
Q_3(y,\pm 1)=0 \,, & \qquad I= 3 \,; \\
Q_4(y,\pm 1)= Q_6(y,\pm 1)\,, & \qquad I= 4 \,.
\end{array}
\right.
\end{equation}
\begin{equation}\label{BCs:H}
\hspace{-2.2cm} \hbox{BCs at horizon ($y=0$)}: \left\{
\begin{array}{ll}
Q_1(0,x)= Q_2(0,x) \,, & \qquad I= 1 \,; \\
\partial_y Q_I(0,x)=0 \,, & \qquad I=2,4,5,6,7  \,; \\
Q_3(0,x)= 0\,, & \qquad I= 3 \,.
\end{array}
\right.
\end{equation}
\begin{equation}\label{BCs:confBdry}
\hbox{BCs at conformal boundary ($y=1$)}: \left\{
\begin{array}{ll}
Q_1(1,x)=1 \,, & \qquad I=1,2,4,5,6,7 \,; \\
Q_3(1,x)= 0\,, & \qquad I= 3 \,.
\end{array}
\right.
\end{equation}
Now we are in a position to solve the PDE system \eqref{EOM:deTurck} subject to the boundary conditions \eqref{BCs:poles}-\eqref{BCs:confBdry}.  To solve the equations, we use a standard Newton-Raphson iteration algorithm based on pseudo-spectral collocation on a Chebyshev grid.  We obtain a seed by perturbing the AdS$_5$-Schw$\times$S$^5$ solution near the zero mode.  Our solutions are parametrised by $y_+$.  We will present our numerical results in section \ref{sec:results}, and discuss numerical checks in section \ref{sec:numerics} of the appendix.

\subsection{Lumpy solutions with $\ell=2$ \label{sec:constructionL2}}
It turns out that black hole solutions branching from modes with even values of $\ell$ are easier to construct than those with odd values of $\ell$. In order to see this we note the following two $\mathbb Z_2$ symmetries: (1) modes with even $\ell$ preserve the symmetry $\tilde{x}\to-\tilde{x}$ of the line element (\ref{S5}); (2) they also preserve the symmetry $r\to-r$ of the line element (\ref{Schw}). While the first symmetry is easy to understand\footnote{Recall that the parity of scalar spherical harmonics under the symmetry $\tilde{x}\to-\tilde{x}$ is simply $(-1)^{\ell}$.}, the second requires some explanation.

It turns out that generic infinitesimal perturbations about AdS$_5\times$S$^5$ can always be reduced to the study of a set of decoupled ODEs \cite{Kim:1985ez}. Solutions to these ODEs can then be used to reconstruct metric perturbations, which in turn allows us to read off the decay of the several metric functions as we approach the boundary. Indeed, these decays are related to the mass of each perturbation from the AdS$_5$ perspective. It turns out that for even values of $\ell$, all perturbations lead to masses that cause the symmetry $r\rightarrow-r$ to be preserved.

We can thus take advantage of these symmetries to design a better line element for a reference metric. We will still choose as a reference metric that AdS$_5$-Schw$\times$S$^5$, but rather than use the coordinates in \eqref{eq:coordinate}, we instead redefine
\be
r=\frac{L\,y_+ }{\sqrt{1-y^2}},\qquad\tilde{x}=x\sqrt{2-x^2}\;.
\ee
With these new coordinates, the solution \eqref{AdS5xS5} can be written
\begin{align}
\d s^2& = \frac{L^2}{1-y^2}\left[-y^2\widehat{G}(y)\,\d t^2  +\frac{y_+^2\,\d y^2}{\left(1-y^2\right) \widehat{G}(y)}+y_+^2\, \d\Omega_3^2 \right]+L^2 \left[\frac{4 \d x^2}{2-x^2}+\left(1-x^2\right)^2 \d\Omega_4^2 \right] \,,\nonumber \\
C_{(4)}&=\frac{L^4 y_+^4}{\sqrt{2}} \frac{y^2}{\left(1-y^2\right)^2}\, \widehat{H}(y) \, \d t\wedge \d S_{(3)}-\frac{L^4}{\sqrt{2}}\,\d S_{(4)}\;,
\end{align}
where 
\begin{equation}
\widehat{G}(y)=1-y^2+y_+^2 \widehat{H}(y)\,,\qquad \widehat{H}(y)=2-y^2\,.
\end{equation}
We choose this to be our reference metric $\bar g$.  Here, $x\in[0,1]$, with $x=0$ and $x=1$ corresponding respectively to the $\mathbb Z_2$ axis of symmetry and north pole of the S$^5$, and $y\in[0,1]$ with $y=0$ corresponding to the horizon and $y=1$ to the boundary of the AdS$_5$.

Using this metric, we write down a general ansatz which is static and preserves the above $\mathbb Z_2$ symmetries and also those of the S$^3$ and S$^4$:
\begin{eqnarray}\label{ansatzL2}
\d s^2 &=& \frac{L^2}{1-y^2} \left[-y^2\widehat{G}(y)\,Q_1 \,\d t^2  +\frac{y_+^2}{\left(1-y^2\right) \widehat{G}(y)}\,Q_2 \,\left[\d y + (1-y^2) Q_3 \,\d x\right]^2+y_+^2 Q_5 \, \d\Omega_3^2 \right] \nonumber\\
&& \qquad+L^2 \left[Q_4\,\frac{4 \d x^2}{2-x^2}+ Q_6\, \left(1-x^2\right)^2 \d\Omega_4^2 \right]\,,\nonumber \\
C_{(4)}&=&\frac{L^4 y_+^4}{\sqrt{2}} \frac{y^2}{\left(1-y^2\right)^2}\, \widehat{H}(y) \,Q_7 \, \d t\wedge \d S_{(3)}-\frac{L^4}{\sqrt{2}}\, W\,\d S_{(4)}\;,
\end{eqnarray}
where $\left\{Q_I,W\right\}$ ($I=1,\ldots, 7$) are functions of ${x,y}$ which will be determined numerically. The construction of these solutions and respective boundary conditions parallel those the previous section, and as such will not be presented here. The only change occurs at the new boundary $x=0$, since there we must demand reflection symmetry. This is reduces to
\begin{equation}
\hspace{-0.2cm}  \hbox{BCs at the reflection plane ($x=0$)}: \left\{
\begin{array}{ll}
\partial_x Q_I(y,0)=0 \,, & \qquad I=1,2,4,5,6,7  \,; \\
Q_3(y,0)=0 \,, & \qquad I= 3 \,.
\end{array}
\right.
\end{equation}
\section{Phase diagram of AdS$_5\times$S$^5$ BHs. \label{sec:results}}

In this section we wish to compute the phase diagram of the lumpy BH families and the AdS$_5$-$\mathrm{Schw}$ BHs.  We first need the thermodynamic quantities of the lumpy BHs.  The BH temperature is proportional to its surface gravity at the horizon, $2\pi T=\left[ -(\partial K^2)^2/(4 K^2)\right]^{1/2}$ with $K$ being the generator of the Killing horizon. It follows from \eqref{lumpyansatz}, with $K=\partial_t$, that the temperature of the lumpy BHs and AdS$_5$-Schw is then
\begin{equation}\label{temp}
T=\frac{1}{L}\frac{2 y_+^2+1}{2 \pi  y_+}\,.
\end{equation}

The entropy of a BH is proportional to its horizon area, $S=A_H/(4 G_{10})$. From the AdS/CFT dictionary, the 10-dimensional ($G_{10}$) and 5-dimensional ($G_5$) Newton's constants are related to $N$ by
\begin{equation}\label{newton}
G_{10}=\frac{\pi^4}{2}\frac{L^8}{N^2}\,,\qquad G_5=\frac{G_{10}}{\pi^3 L^5}\,.
\end{equation}
The entropy of the $\ell=1,2$ lumpy BHs can then be written as
\begin{equation}\label{entropy}
S=L^3\,N^2 \, \frac{16}{3}\,y_+^3\,\int_{-1}^1 dx \,\frac{\left(1-x^2\right)^4}{\sqrt{2-x^2}}\sqrt{Q_4(0,x)Q_5(0,x)^3 Q_6(0,x)^4}\,,
\end{equation}
while the AdS$_5$-$\mathrm{Schw}$ BH entropy is still given by this expression with $Q_4=Q_5=Q_6\equiv 1$ i.e., $S_{\mathrm{Schw}}=L^3\,N^2 \pi \,y_+^3$. 

Computing the energy of the lumpy BH is non-trivial. It can be read from the asymptotic expansion of the fields at the holographic boundary using the formalism of Kaluza-Klein (KK) holography and holographic renormalisation \cite{Skenderis:2006uy} (see also \cite{Kim:1985ez,Gunaydin:1984fk,Lee:1998bxa,Lee:1999pj,Arutyunov:1999en,Skenderis:2006di,Skenderis:2007yb}). A detailed discussion of this formalism applied to this system is given in Appendix \ref{KKholography}. It culminates with the expression for the energy \eqref{energy0}, that we reproduce here (this is valid for the $\ell=1,2$ lumpy solutions), 
\begin{equation} \label{energy}
E=\frac{N^2}{3072 L^2}{\biggl [}576+2304 \,y_+^2(1+y_+^2) -y_+^4 {\biggl (}5\, \beta_2+30 \,\beta_2^2+12\left(16 \,\delta_{0}+\delta_4\right){\biggr)}{\biggr ]}.
\end{equation}
In this expression, $\{\beta_2,\delta_{0},\delta_4 \}$ are undetermined coefficients that appear in a Taylor expansion of the fields about the holographic boundary (see \eqref{coef:H}-\eqref{coef:lumpy}) after imposing appropriate Dirichlet boundary conditions (these correspond to having no sources in the dual CFT$_4$). These coefficients can be obtained by differentiating our numerical results. The energy of the AdS$_5$-$\mathrm{Schw}$ BH is given by 
\be
E_{\mathrm{Schw}}=\left[(3/4) y_+^2 \left(y_+^2+1\right) + 3/16\right] N^2/L^2\;,
\ee
and is recovered when $\beta_2=\delta_{0}=\delta_4=0$.  The AdS$_5$-$\mathrm{Schw}$ BH energy contains a contribution from the AdS$_5$ background, $E_{AdS_5}=\frac{N^2}{L^2} \frac{3}{16}$, which is the well known Casimir energy of the dual $\mathcal N=4$ SYM on $\mathbb R\times S^3$.  

\begin{figure}[ht]
\centering
\includegraphics[width=.6\textwidth]{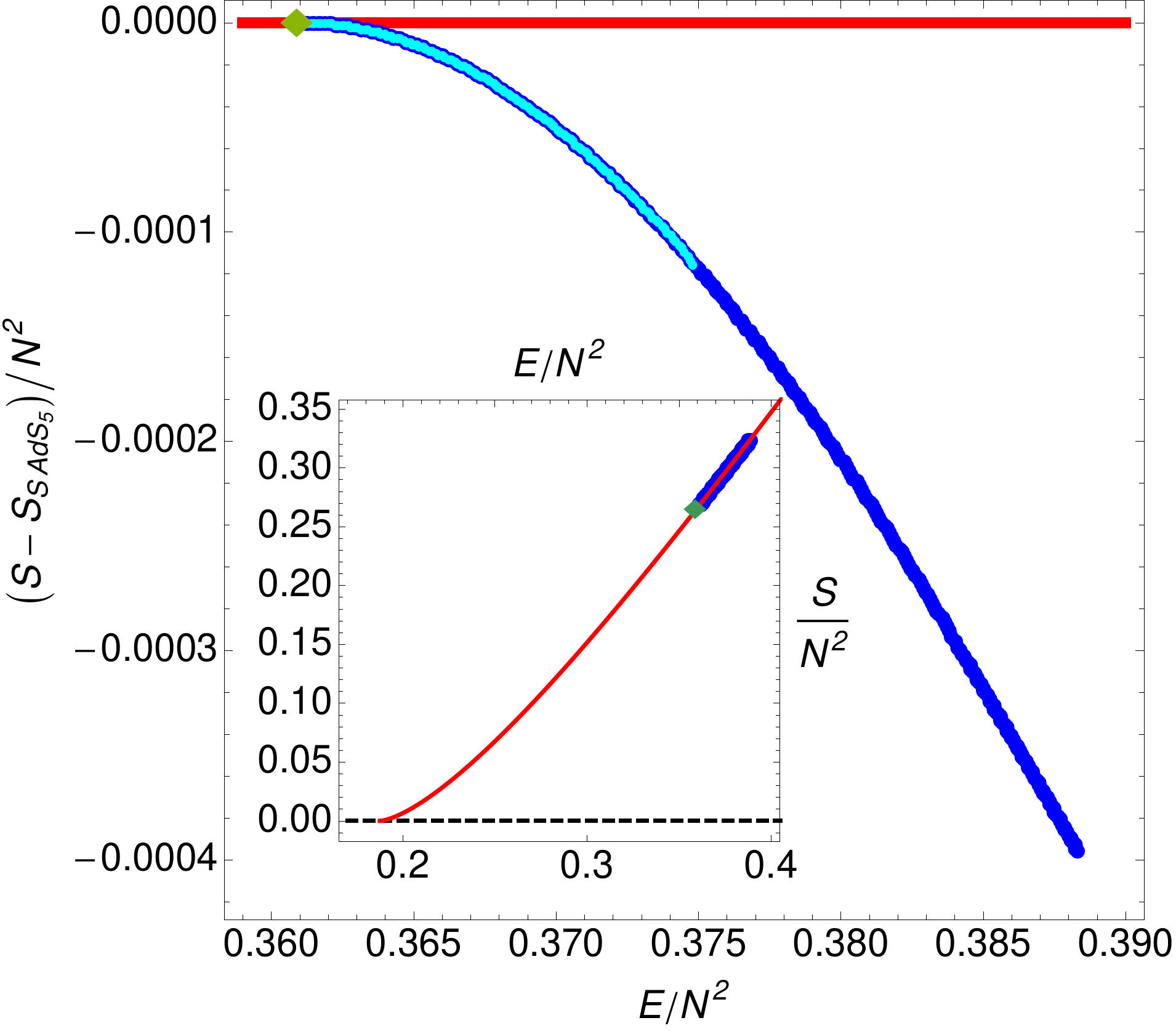}
\caption{Phase diagram in the microcanonical ensemble with the $\ell=1$ lumpy BH family. We plot the entropy difference between each solution and that of the AdS$_5$-$\mathrm{Schw}\times$S$^5$ BHs as a function of energy. The red line represents the AdS$_5$-$\mathrm{Schw}\times$S$^5$ BH, while the blue dots describe the $\ell=1$ lumpy BH family. The green diamond marks the GL zero mode where the lumpy and AdS$_5$-$\mathrm{Schw}\times$S$^5$ families merge. We have computed the thermodynamic quantities both by integrating the first law (dark blue curve) and by reading the asymptotic energy (light blue curve; see text).  The inset plot is a zoomed out version where we plot the entropy of the solutions as a function of their energy.  Here, we fix $L=1$.}\label{Fig:microcanonicalL1}
\end{figure}

\begin{figure}[ht]
\centering
\includegraphics[width=.65\textwidth]{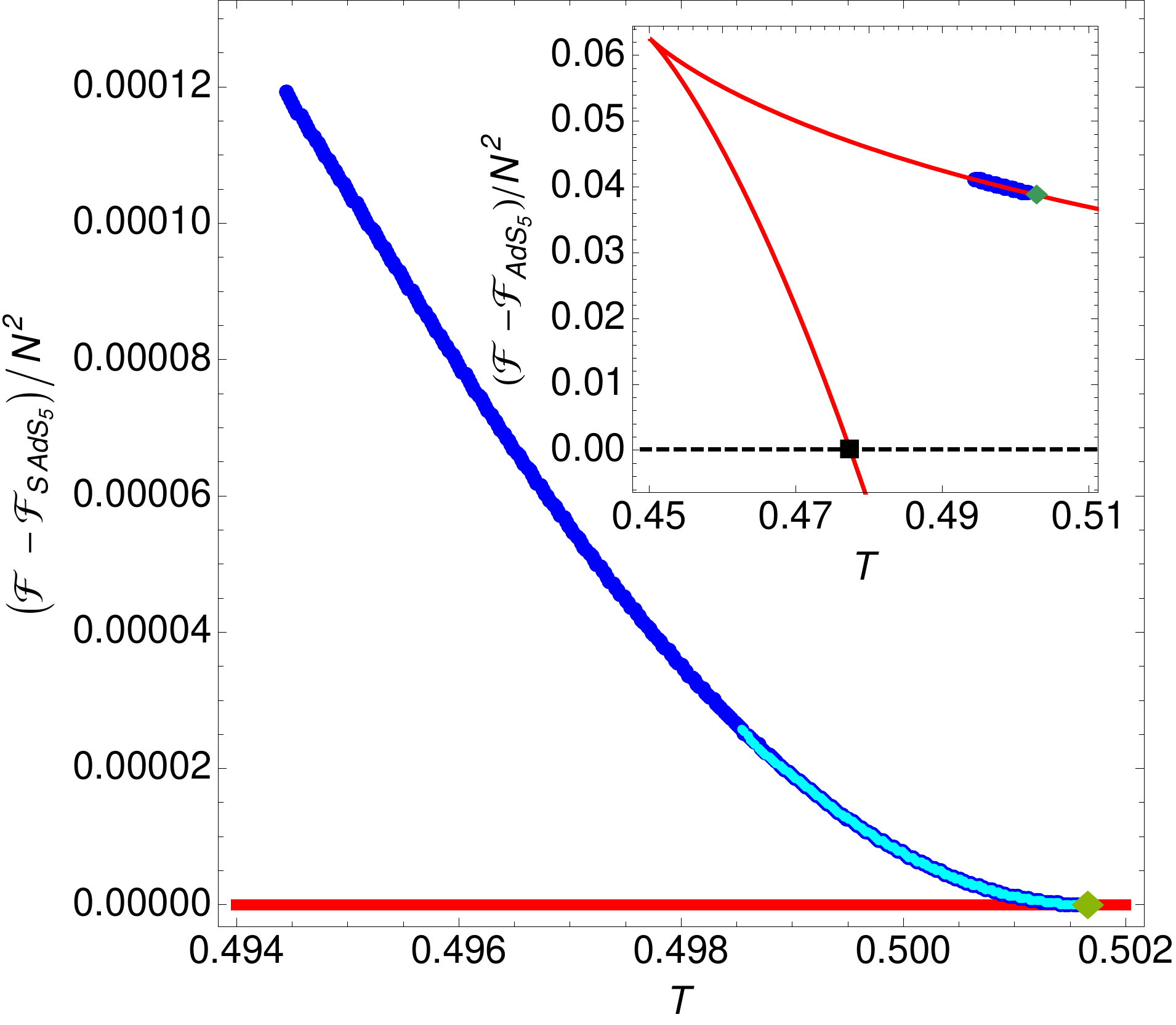}
\caption{Phase diagram in the canonical ensemble with the $\ell=1$ lumpy BH family. We plot the difference in free energy between each solution and that of the small AdS$_5$-Schw$\times$S$^5$ BHs as a function of temperature.  The red line represents the small  AdS$_5$-$\mathrm{Schw}\times$S$^5$ BHs.  Again, the green diamond marks the zero mode where the $\ell=1$ lumpy and AdS$_5$-$\mathrm{Schw}\times$S$^5$ families merge. We compute the thermodynamic quantities both by integrating the first law (dark blue curve) and also by reading the asymptotic energy (light blue curve; see text). The inset shows a wider range of the phase diagram and plots the difference in free energy between each solution and thermal global AdS$_5\times$S$^5$ as a function of temperature.  The black square marks the Hawking-Page transition, where large BHs dominate at higher temperatures and thermal AdS$_5\times$S$^5$ dominates at lower temperatures.  We again set $L=1$. 
}\label{Fig:canonicalL1}
\end{figure}

Extracting the constants $\{\beta_2,\delta_{0},\delta_4 \}$ requires accurately evaluating four or two derivatives, for odd or even $\ell$, respectively.  Rather than evaluate up to four derivatives, a numerically simpler way is to integrate the first law $\d E=T\d S$.  Some of our data has been extracted using this simpler method.  Where we can accurately do both, these energies agree (this comparison is in Figs. \ref{Fig:microcanonicalL1} and \ref{Fig:canonicalL1}).  

With these thermodynamic variables, we can compute the (Helmholtz) free energy using
\begin{equation} \label{freeenergy}
{\cal F}=E-T S\,.
\end{equation}
For the AdS$_5$-$\mathrm{Schw}$ BH this is given by
\begin{equation} \label{freeenergy:SAdS}
{\cal F}_{S AdS_5} = \frac{N^2}{L^2}\left( \frac{3}{16}+\frac{1}{4}y_+^2 \left(1-y_+^2 \right) \right).
\end{equation}
When we set the horizon radius $y_+=0$ we get the free energy of AdS$_5$, ${\cal F}_{AdS_5} = \frac{N^2}{L^3} \frac{3}{16}$.

Now let us discuss the various thermodynamic ensembles.  For our system, these are the microcanonical and canonical ensembles.  In the microcanonical ensemble, the energy is held fixed and solutions with higher entropy are preferred.  In the canonical ensemble, the temperature is fixed and the solutions with the lower free energy are preferred.  To obtain adimensional quantities, we scale by factors of $G_{10}\sim 1/N^2$ and $L$.  The relevant phase diagrams are therefore $S/(L^3N^2)$ vs $EL^2/N^2$ for the microcanonical ensemble and $\mathcal FL^2/N^2$ vs $TL$ for the canonical ensemble.

Before giving the phase diagram with the lumpy BHs, it is instructive to first discuss the AdS$_5$-$\mathrm{Schw}$ phases.  In the microcanonical ensemble, the AdS$_5$-Schw BHs have increasing entropy with increasing energy.  As mentioned earlier, the zero entropy solution corresponds to the Casimir energy $E_{AdS_5}=\frac{N^2}{L^2} \frac{3}{16}$.  At a given energy, there is only one solution in this family, so the phase diagram is trivial. 

AdS$_5$-$\mathrm{Schw}$ BHs are more complex in the canonical ensemble.  From \eqref{temp}, one can see that AdS$_5$-$\mathrm{Schw}$ BHs have a minimum temperature at
\begin{equation} \label{Tcv}
y_+{\bigl |}_{c_V}=\frac{1}{\sqrt{2}}\simeq 0.707107 \,,\qquad T_{c_V}L=\frac{\sqrt{2}}{\pi}\simeq 0.450158\,.
\end{equation}
There are thus two AdS$_5$-$\mathrm{Schw}$ BH solutions with any given temperature above $T_{c_V}$.  These can be distinguished by their size, so AdS$_5$-$\mathrm{Schw}$ BHs with horizon radius $y_+<y_+{\bigl |}_{c_V}$ are called \emph{small} BH's, while those with $y_+>y_+{\bigl |}_{c_V}$ are \emph{large}.  The free energies of large AdS$_5$-$\mathrm{Schw}$ BHs are always lower than that of the corresponding small AdS$_5$-$\mathrm{Schw}$ BH with the same temperature.  That is, large BHs are preferred over small BHs.  In the $\mathcal FL^2/N^2$ vs $TL$ phase diagram, the large and small BHs meet at a cusp at $T_{c_V}$.  There is, however, a third phase which is thermal AdS$_5\times$S$^5$. This is just the Euclidean solution of AdS$_5\times$S$^5$ with an arbitrary period chosen for the Euclidean time circle (i.e. at any temperature).  Below a temperature corresponding to
\begin{equation} \label{HP}
y_+{\bigl |}_{HP}=1 \,,\qquad T_{HP}L=\frac{3}{2\pi}\simeq 0.477465\,,
\end{equation}
thermal AdS$_5\times$S$^5$ has lower free energy than both large and small AdS$_5$-$\mathrm{Schw}$ BHs. At temperatures above $T_{HP}$, large AdS$_5$-$\mathrm{Schw}$ BHs are preferred.  This is a first-order phase transition known as the Hawking-Page (HP) phase transition \cite{Hawking:1982dh}.  In the dual CFT, this is interpreted as a confinement/deconfinement transition \cite{Witten:1998qj}.

This phase diagram will become more complex if we further allow for solutions that break the symmetries of the S$^5$.  As discussed earlier and in \cite{Hubeny:2002xn}, the largest wavelength GL zero mode  ($\ell=1$) that preserves an $SO(5)$ symmetry of the S$^5$ appears at
\begin{equation}\label{GLradius2}
y_+{\bigl |}_{\ell=1}\simeq 0.440237\,, \qquad T_{GL}L\simeq 0.501653\,,
\end{equation}
which is a horizon radius corresponding to a small BH.  A branch of lumpy BHs emerges from this zero mode.  

Now let us consider these lumpy BHs in the microcanonical ensemble whose phase diagram is in Fig. \ref{Fig:microcanonicalL1}.  Because the entropy between these lumpy solutions and AdS$_5$-$\mathrm{Schw}$ are close, we have instead plotted the entropy difference between these solutions at the same energy. The actual entropy is in the inset plot.  There, we can see that these lumpy BHs always have lower entropy than AdS$_5$-$\mathrm{Schw}\times$S$^5$ BHs and therefore do not dominate the microcanonical ensemble.

The phase diagram in the canonical ensemble with the $\ell=1$ lumpy BHs is shown in  Fig. \ref{Fig:canonicalL1}.  The zero mode lies in the small BH branch, so we plot the difference in free energy between the lumpy BH and that of the small AdS$_5$-$\mathrm{Schw}\times$S$^5$ BH as a function of temperature.  In the inset, we show a wider view of the actual free energy vs the temperature.  We can see that the lumpy BHs have higher free energy than both large and small AdS$_5$-Schw BHs as well as thermal AdS.  They therefore never dominate this ensemble either.
 
\begin{figure}[ht]
\centering
\includegraphics[width=.4805\textwidth]{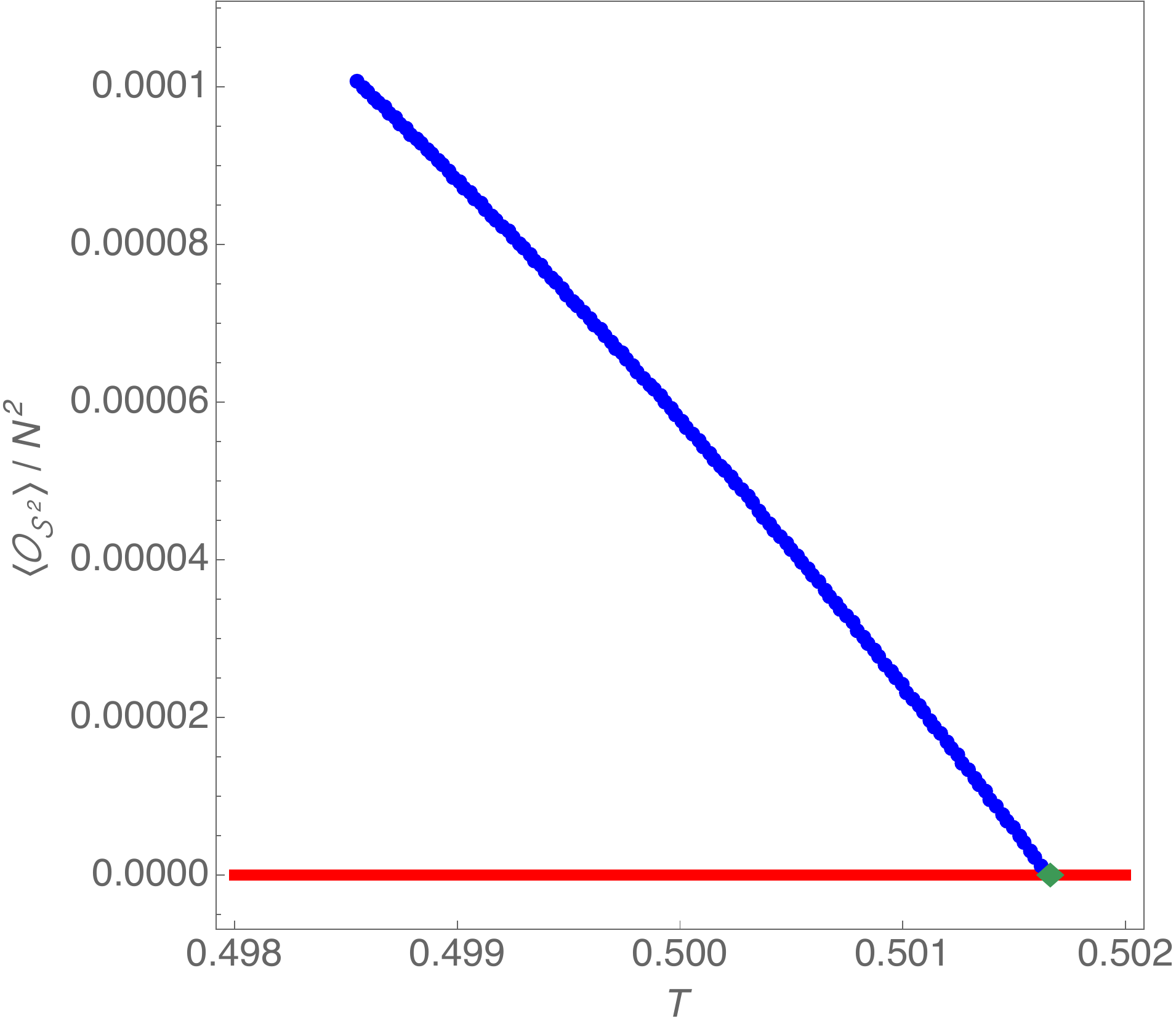}
\includegraphics[width=.502\textwidth]{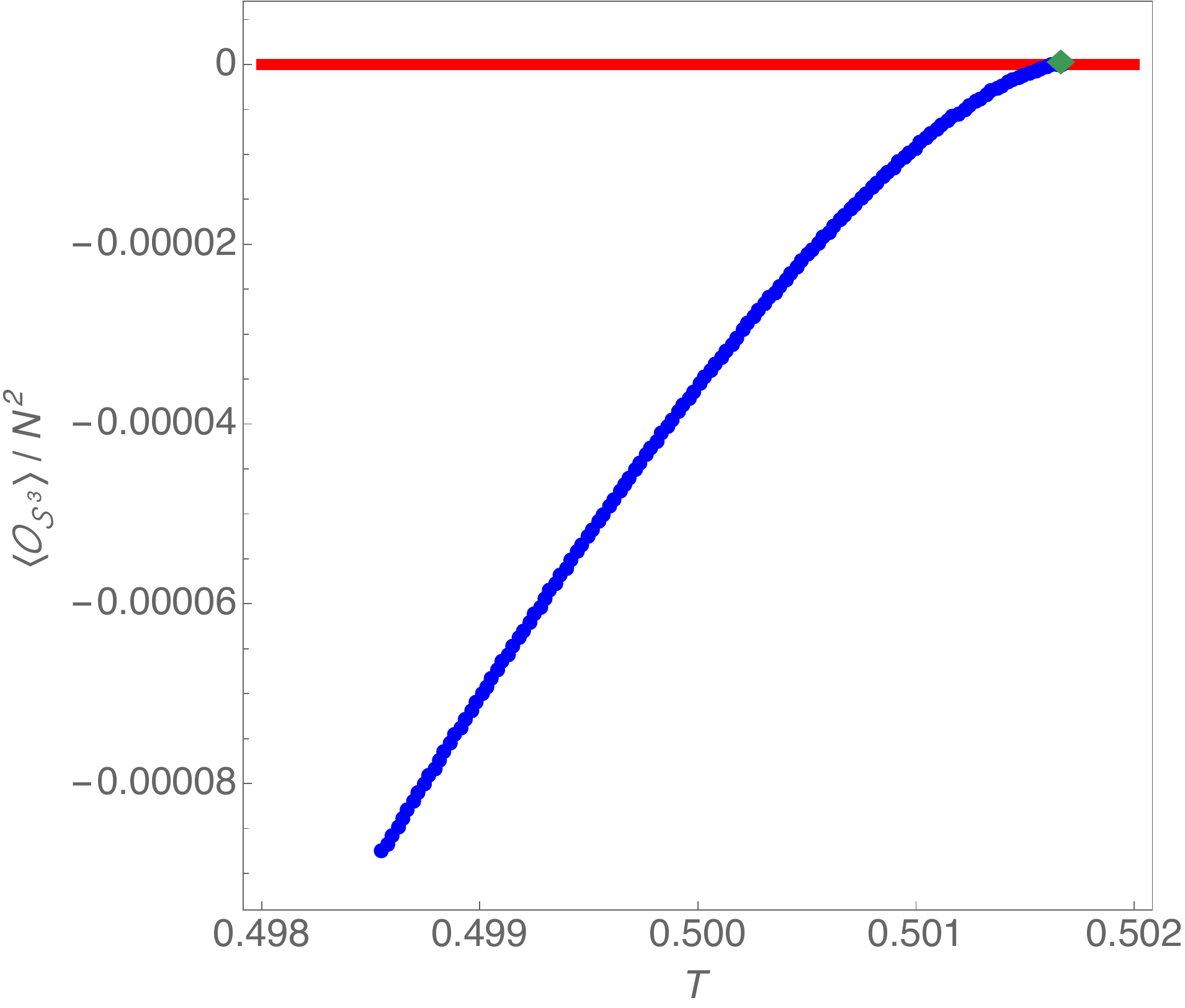}
\caption{Expectation values $\left \langle {\cal O}_{{\cal S}^2} \right\rangle$ \emph{(Left)} and  $\left \langle {\cal O}_{{\cal S}^3} \right\rangle$ (\emph{Right})  of the dual operators to the KK scalar fields ${\cal S}^2$ and ${\cal S}^3$ (here we set $L= 1$) for the $\ell=1$ lumpy BH. We find that near the merger we have the fitting:  $\left \langle {\cal O}_{{\cal S}^2} \right\rangle \simeq A \left( 1-T/T_c \right)^\alpha$  with $A\simeq 0.0147 \pm 0.0001 $ and $\alpha\simeq 0.966 \pm 0.001$; and   $\left \langle {\cal O}_{{\cal S}^3} \right\rangle \simeq B \left( 1-T/T_c \right)^\beta$  with $B\simeq -0.1643 \pm 0.0006 $ and $\beta\simeq 1.4739 \pm 0.0006$.}\label{Fig:vevO}
\end{figure}

If the full 10-dimensional theory is dimensionally reduced to a theory on AdS$_5$, the lumpy AdS$_5\times$S$^5$ BHs are reinterpreted as five dimensional BHs with non-trivial scalar fields.  From the perspective of the field theory dual to the AdS$_5$, there are nonzero expectation values for scalar operators.  This is spontaneous symmetry breaking.  These expectations values can be computed using the tools of Kaluza-Klein holography \cite{Skenderis:2006uy}, whose discussion we defer to Appendix \ref{KKholography}  (see also \cite{Kim:1985ez,Gunaydin:1984fk,Lee:1998bxa,Lee:1999pj,Arutyunov:1999en,Skenderis:2006uy,Skenderis:2006di,Skenderis:2007yb}).  We find that some scalar operators develop non-trivial VEVs.  The expectation values of two of these scalar operators are plotted in Fig. \ref{Fig:vevO} as a function of the  temperature. As expected, we find that these VEVs vanish at the zero mode (and for the AdS$_5$-$\mathrm{Schw}\times$S$^5$ BH family) and then their amplitude grows monotonically along the lumpy BH family as we move away from the zero mode. 
   
Now let us move on to the $\ell=2$ lumpy solutions.  As we have mentioned in the introduction, there should be two branches of black holes that emanate from the $\ell=2$ zero mode.  We conjecture that one of these branches connects to two S$^8$ BHs, so we call this the ``double black hole (BH)" branch.  We expect that the other branch connects to topologically s$^4\times$S$^4$ BHs with the S$^4$ wrapping around the equator of the S$^5$, so we call this the ``black belt" branch.

\begin{figure}[ht]
\centering
\includegraphics[width=.6\textwidth]{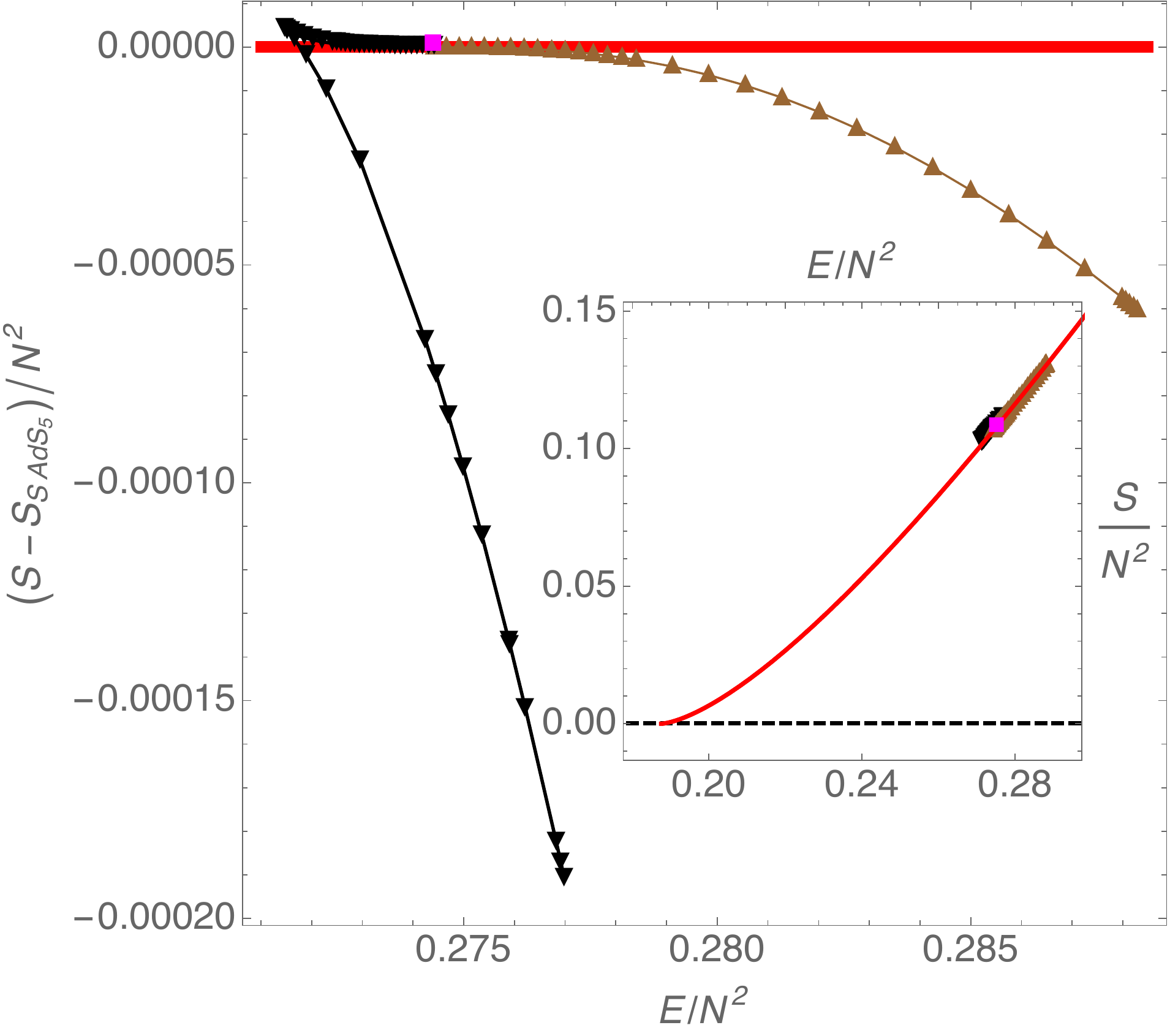}
\caption{Phase diagram in the microcanonical ensemble similar to Fig. \ref{Fig:microcanonicalL1} but now with the $\ell=2$ lumpy solutions. The red line with $\Delta S(E)=0$ represents the AdS$_5$-$\mathrm{Schw}\times$S$^5$ BH.  Black inverted triangles describe the $\ell=2$ double BH branch while the brown triangles represent the black belt branch. The magenta square marks the $\ell=2$ zero mode where these lumpy BHs and AdS$_5$-Schw$\times$S$^5$ families merge. The inset plot is, again, a zoomed out plot.}\label{Fig:microcanonicalL2}
\end{figure}

The phase diagram of the $\ell=2$ solutions in the microcanonical ensemble is displayed in Fig.\ref{Fig:microcanonicalL2}.  Near the zero mode, the black belt branch extends towards higher energy, but with \emph{lower} entropy than AdS$_5$-$\mathrm{Schw}\times$S$^5$.  On the other hand, the double BH branch extends towards lower energy, but with \emph{higher} entropy than AdS$_5$-$\mathrm{Schw}\times$S$^5$.  Some of these lumpy BHs are therefore favoured over AdS$_5$-$\mathrm{Schw}\times$S$^5$ (but might not be the dominant phase of the ensemble).  Furthermore, the double BH branch contains a turning point at a cusp.  Thus, there can be two lumpy BHs for a given energy.  

\begin{figure}[ht]
\centering
\includegraphics[width=.65\textwidth]{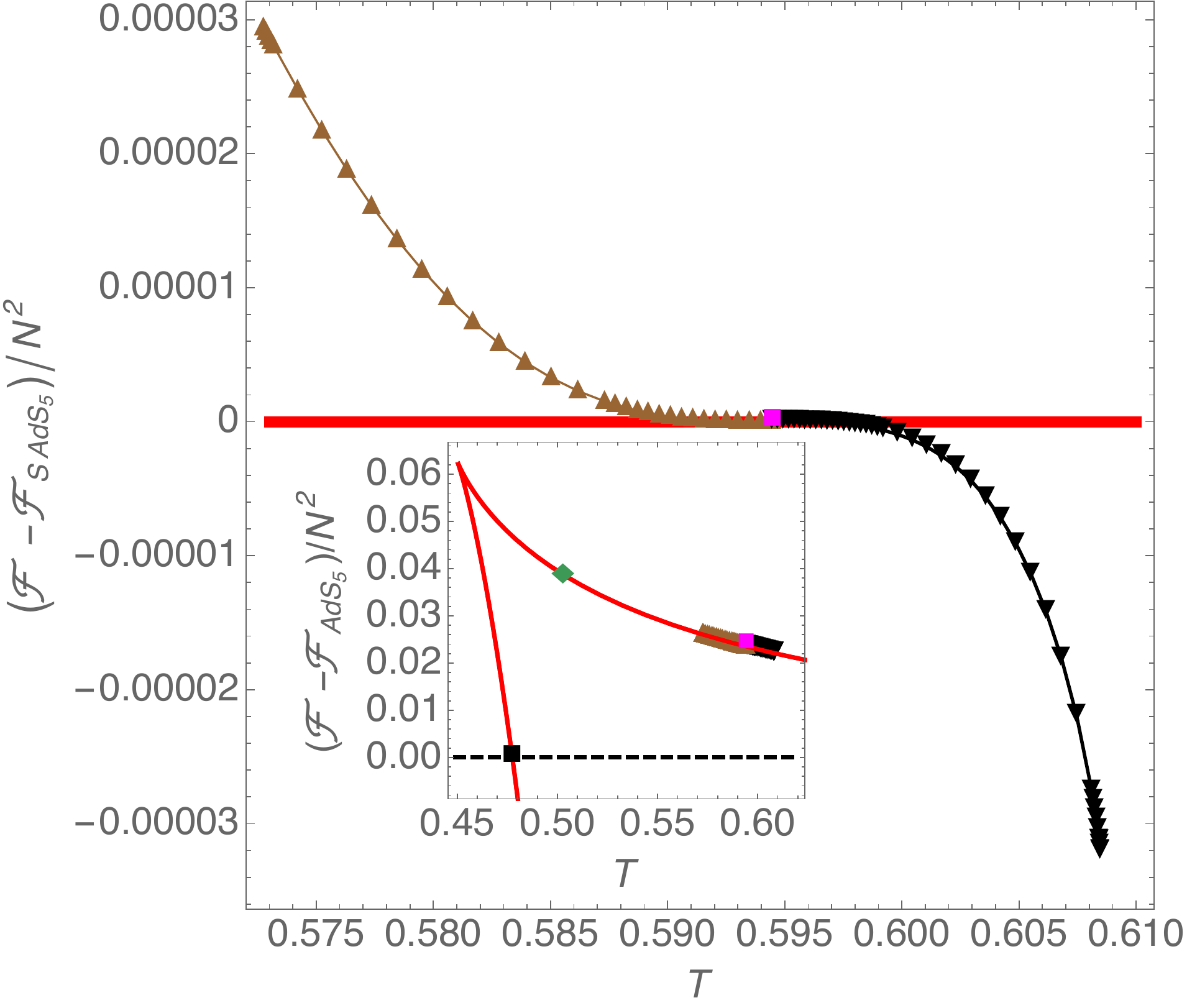}
\caption{Phase diagram in the canonical ensemble similar to Fig. \ref{Fig:canonicalL1} but this time with the $\ell=2$ lumpy solutions. The red line with $\Delta {\cal F}(T)=0$ represents the AdS$_5$-Schw$\times$S$^5$ BH.  Black inverted triangles describe the $\ell=2$ double BH branch while the brown triangles represent the black belt branch. The magenta square marks the $\ell=2$ zero mode. Like in Fig. \ref{Fig:canonicalL1} the inset plot gives a broader view of the phase diagram.}\label{Fig:canonicalL2}
\end{figure}     

Now let us continue with the $\ell=2$ solutions in the canonical ensemble shown in Fig. \ref{Fig:canonicalL2}.  Near the GL zero mode, the black belt branch extends towards lower temperature and \emph{higher} free energy than the corresponding small AdS$_5$-$\mathrm{Schw}\times$S$^5$ BH.  The double BH branch extends towards higher temperature and \emph{lower} free energy than the small AdS$_5$-Schw$\times$S$^5$ BHs.  So also in this ensemble, the double BH branch of the lumpy BHs is therefore favoured over small AdS$_5$-$\mathrm{Schw}\times$S$^5$ BHs, but still have higher free energy than the large BHs.  

The $\ell=2$ solutions also have non-vanishing scalar operators whose magnitude grows monotonically away from the merger.  This is much like the behaviour displayed in Fig. \ref{Fig:vevO} for the $\ell=1$ case, so we do not present a separate plot.

\begin{figure}[ht]
\centering
\includegraphics[width=.6\textwidth]{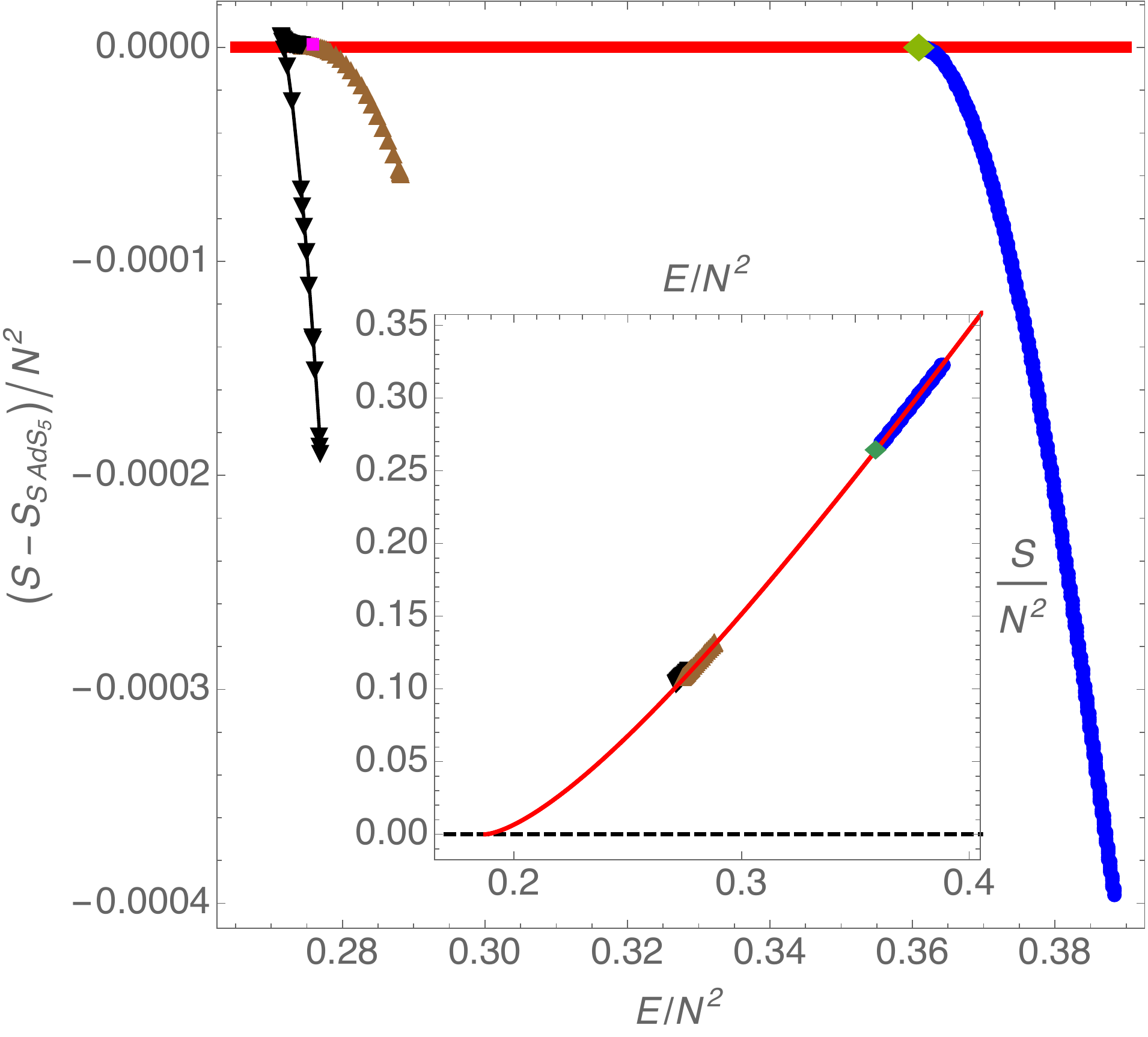}
\caption{Phase diagram in the microcanonical ensemble that collects the information displayed both in Fig. \ref{Fig:microcanonicalL1} and Fig.  \ref{Fig:microcanonicalL2}. The red line represents the AdS$_5$-Schw$\times$S$^5$ BH; blue dots represent the $\ell=1$ lumpy BH family;  black inverted triangles describe the $\ell=2$ lumpy double BH branch; and the brown triangles represent the black belt branch.}\label{Fig:microcanonicalTot}
\end{figure}

\begin{figure}[ht]
\centering
\includegraphics[width=.65\textwidth]{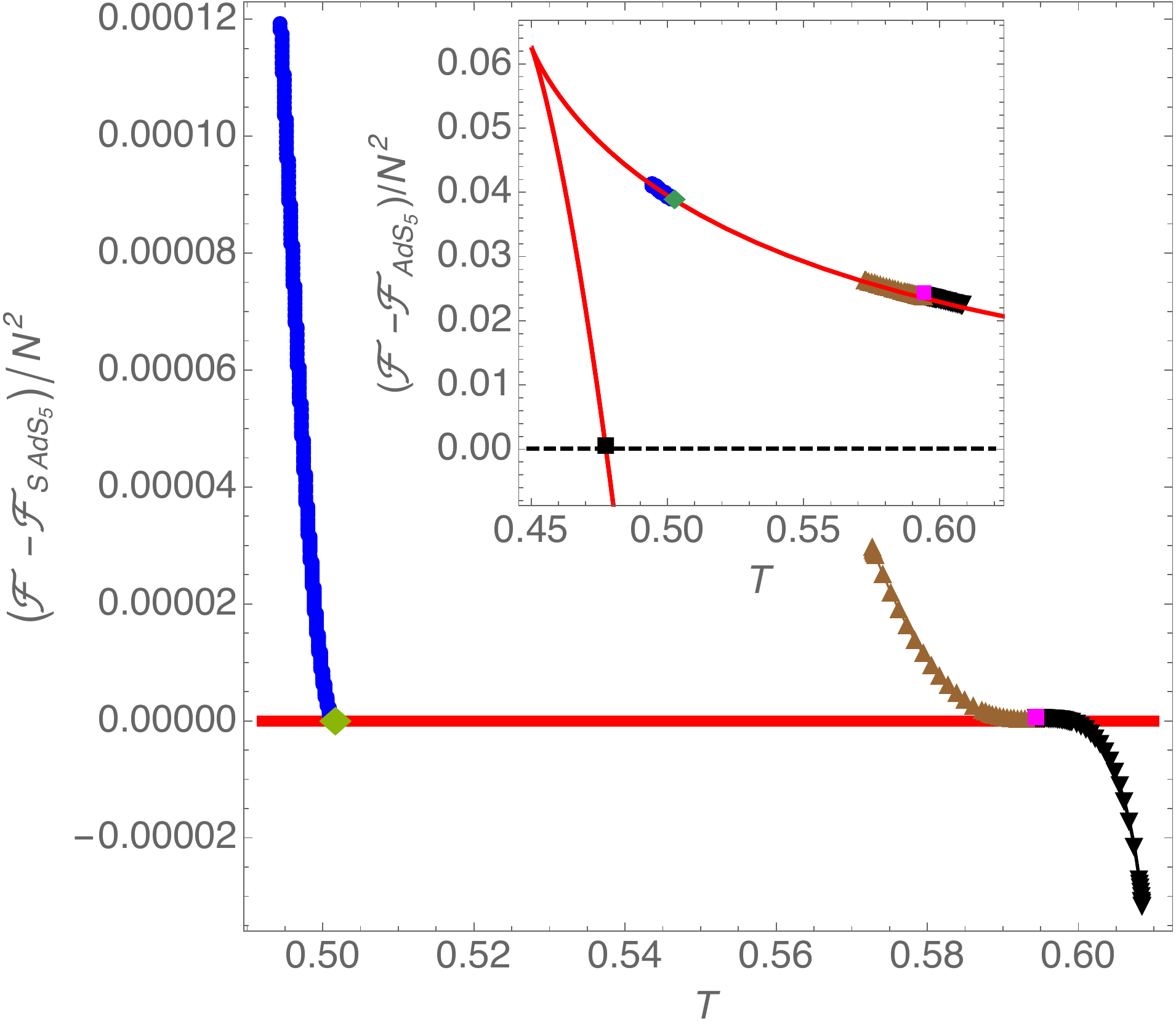}
\caption{Phase diagram in the canonical ensemble  that collects the information displayed both in Fig. \ref{Fig:canonicalL1} and Fig.  \ref{Fig:canonicalL2}. The red line represents the AdS$_5$-Schw$\times$S$^5$ BH; blue disks represent the $\ell=1$ lumpy BH family;  black inverted triangles describe the $\ell=2$ lumpy double BH branch; and the brown triangles represent the black belt branch.}\label{Fig:canonicalTot}
\end{figure}

Putting the $\ell=1$ and $\ell=2$ plots together, we get Fig. \ref{Fig:microcanonicalTot} for the microcanonical ensemble and Fig. \ref{Fig:canonicalTot} for the canonical ensemble.  We note that at the GL zero modes, the slope of the entropy and free energy of the lumpy solutions match that of the AdS$_5$-Schw$\times$S$^5$ BH.  This indicates a second-order phase transition and is consistent with the fact that these phases arise perturbatively.  

\begin{figure}[ht]
\centering
\includegraphics[width=.43\textwidth]{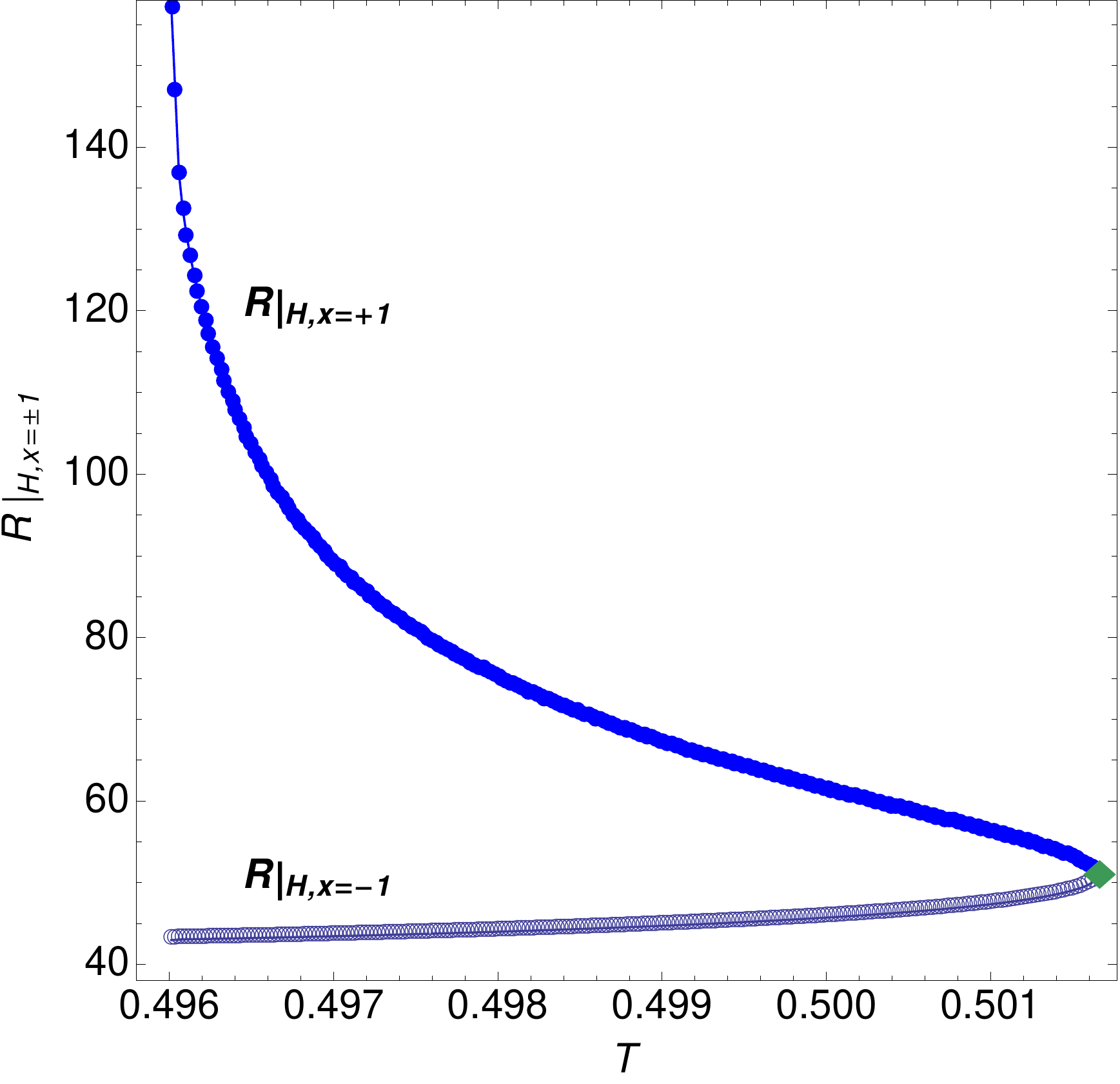}
\hspace{0.5cm}
\includegraphics[width=.45\textwidth]{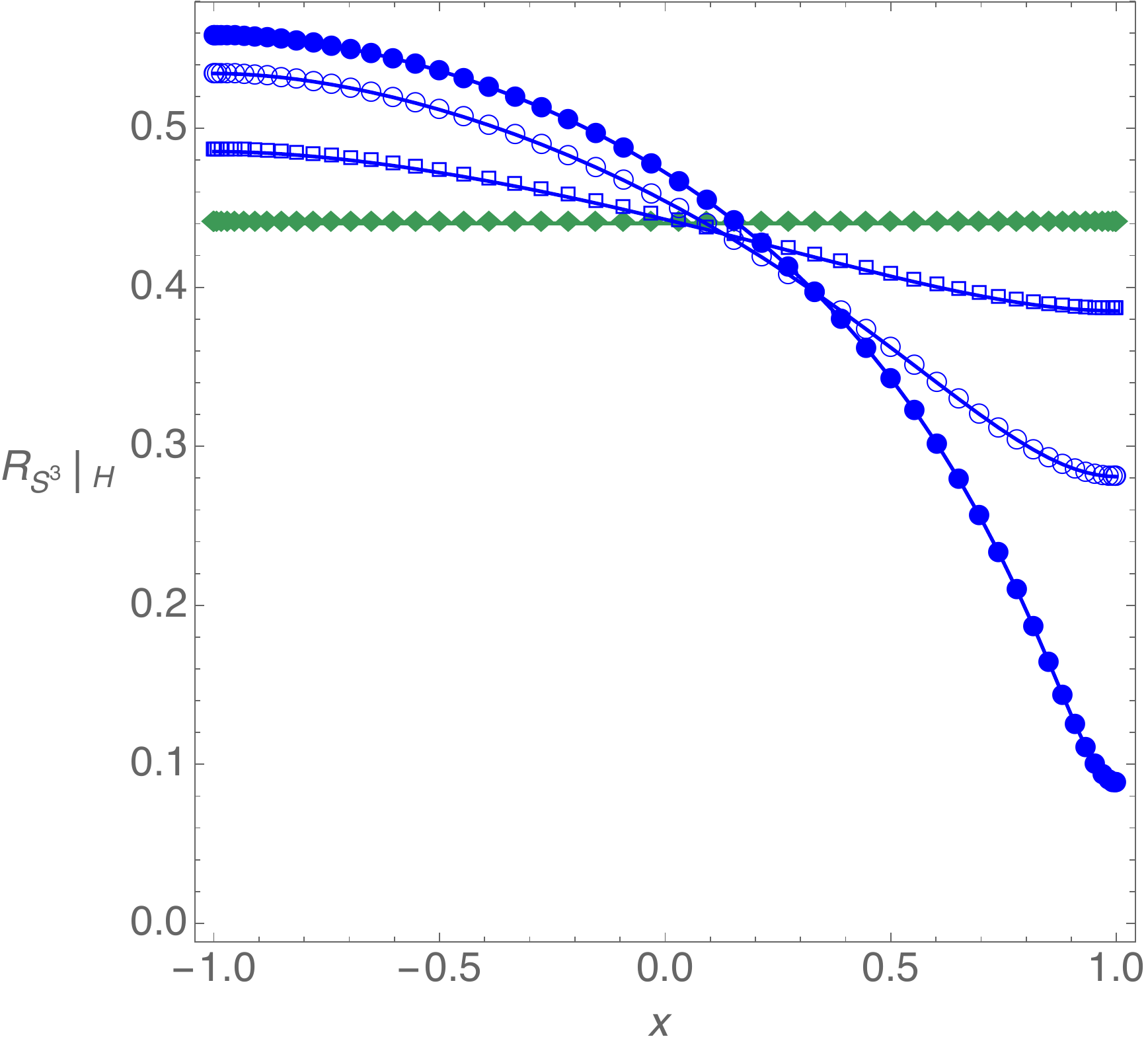}
\caption{ {\it Left Panel}:  Ricci scalar of the $\ell=1$ lumpy BHs evaluated at the horizon and at the north ($x=1$; upper curve) and south poles ($x=-1$; lower curve) of the S$^5$.  {\it Right Panel}:  Radius of the S$^3$ evaluated at the horizon as a function of the polar angle $x$ of the S$^5$ for the $\ell=1$ lumpy BH. 
The green diamonds are for the solution closer to the GL merger ($T=0.50167$), while the blue dots describe the lumpy solution with the lowest temperature ($T=0.49444$) we have reached.  In between we have two other curves with intermediate temperatures, namely $T=0.50120$ (empty squares)  and $T=0.49898$ (circles).  These solutions appear to be approaching a localised black hole.}\label{Fig:ricciH}
\end{figure}

Now we attempt to analyse the approach of the lumpy solutions towards the conical mergers.  Let us first discuss the $\ell=1$ case.  In the left panel of Fig. \ref{Fig:ricciH}, we plot the Ricci scalar of the induced horizon geometry on each of the poles of the S$^5$ as a function of temperature.  We see that the Ricci scalar is getting large at one pole and small at the other pole.  In the right panel of Fig. \ref{Fig:ricciH}, we plot the radius of the S$^3$ at the horizon as a function of the polar variable $x$ of the S$^5$ for four different temperatures.  As the temperature decreases and we move away from the GL zero mode we find that this radius is getting small at the South pole ($x=1$), consistent with the conjectured conical merger. As we mentioned earlier, we suspect topologically S$^8$ BHs on the other side of this conical merger.

\begin{figure}[ht]
\centering
\includegraphics[width=.5\textwidth]{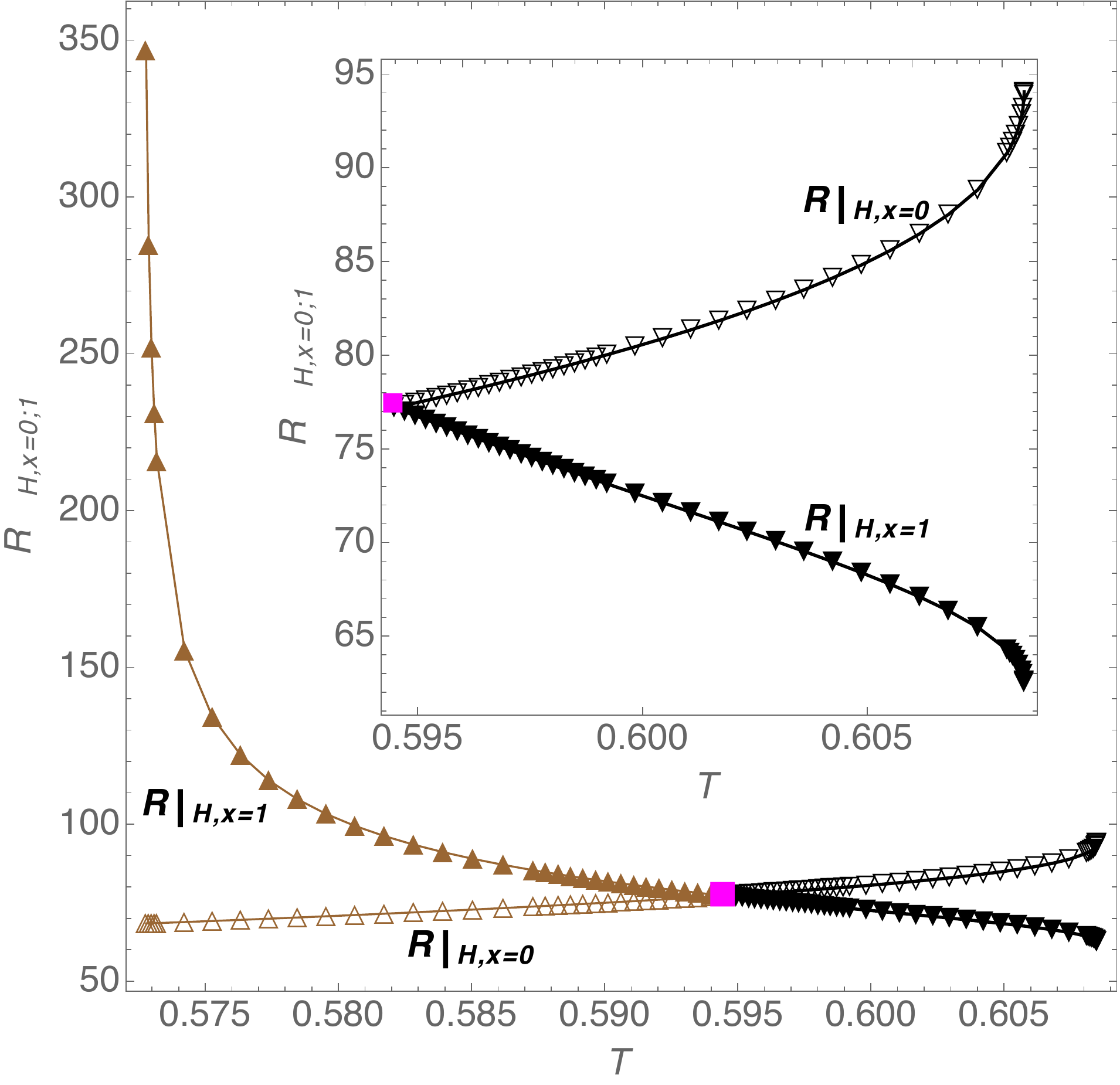}
\caption{Ricci scalar of the induced horizon geometry for the $\ell=2$ lumpy double BH branch (black inverted triangles)  and $\ell=2$  lumpy black belt branch (brown triangles). We show this quantity evaluated both at the pole ($x=1$; filled triangles) and at the equator ($x=0$; empty triangles) of the S$^5$.}\label{Fig:ricciHL2}
\end{figure}

\begin{figure}[ht]
\centering
\includegraphics[width=.45\textwidth]{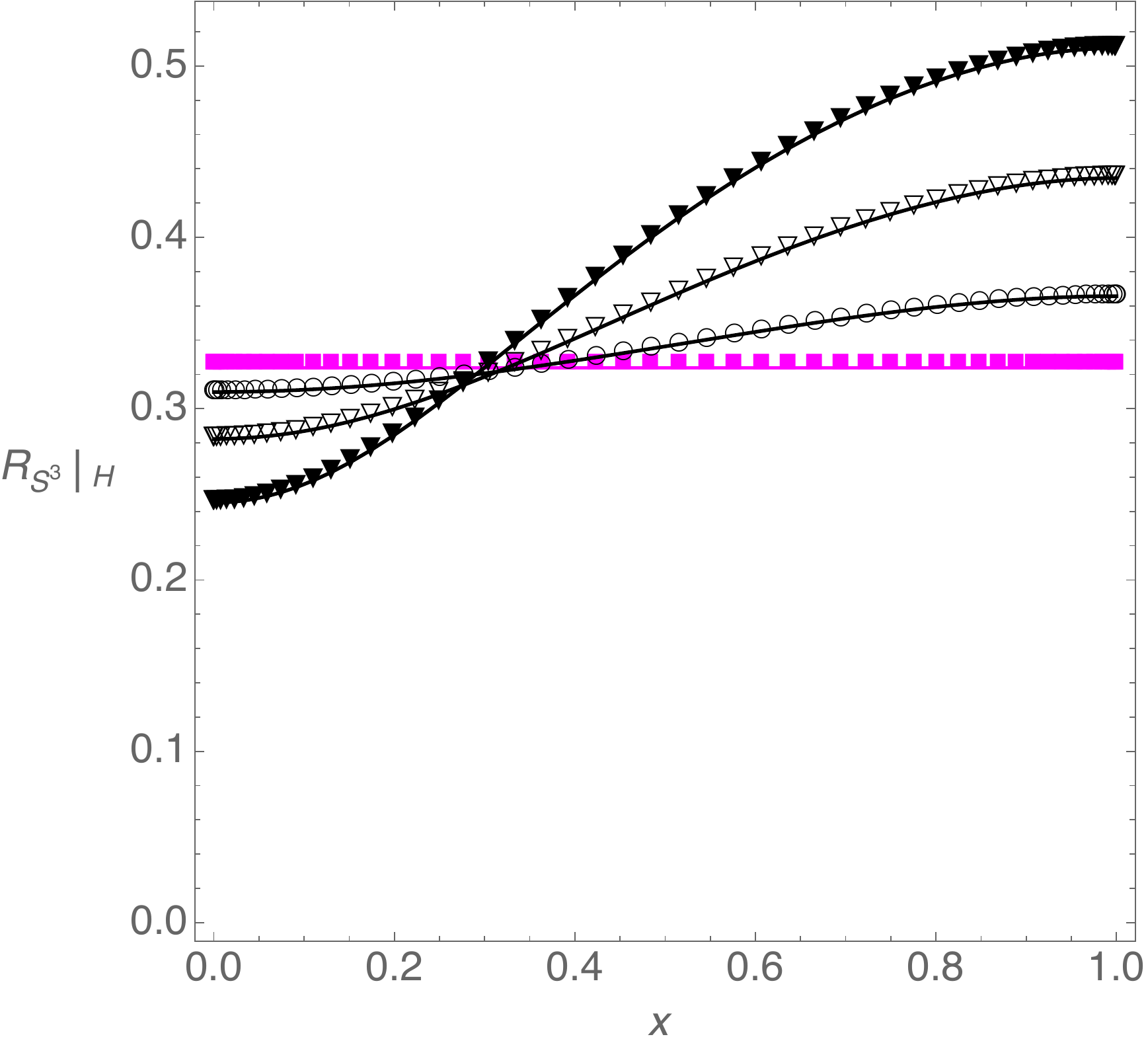}
\hspace{0.5cm}
\includegraphics[width=.45\textwidth]{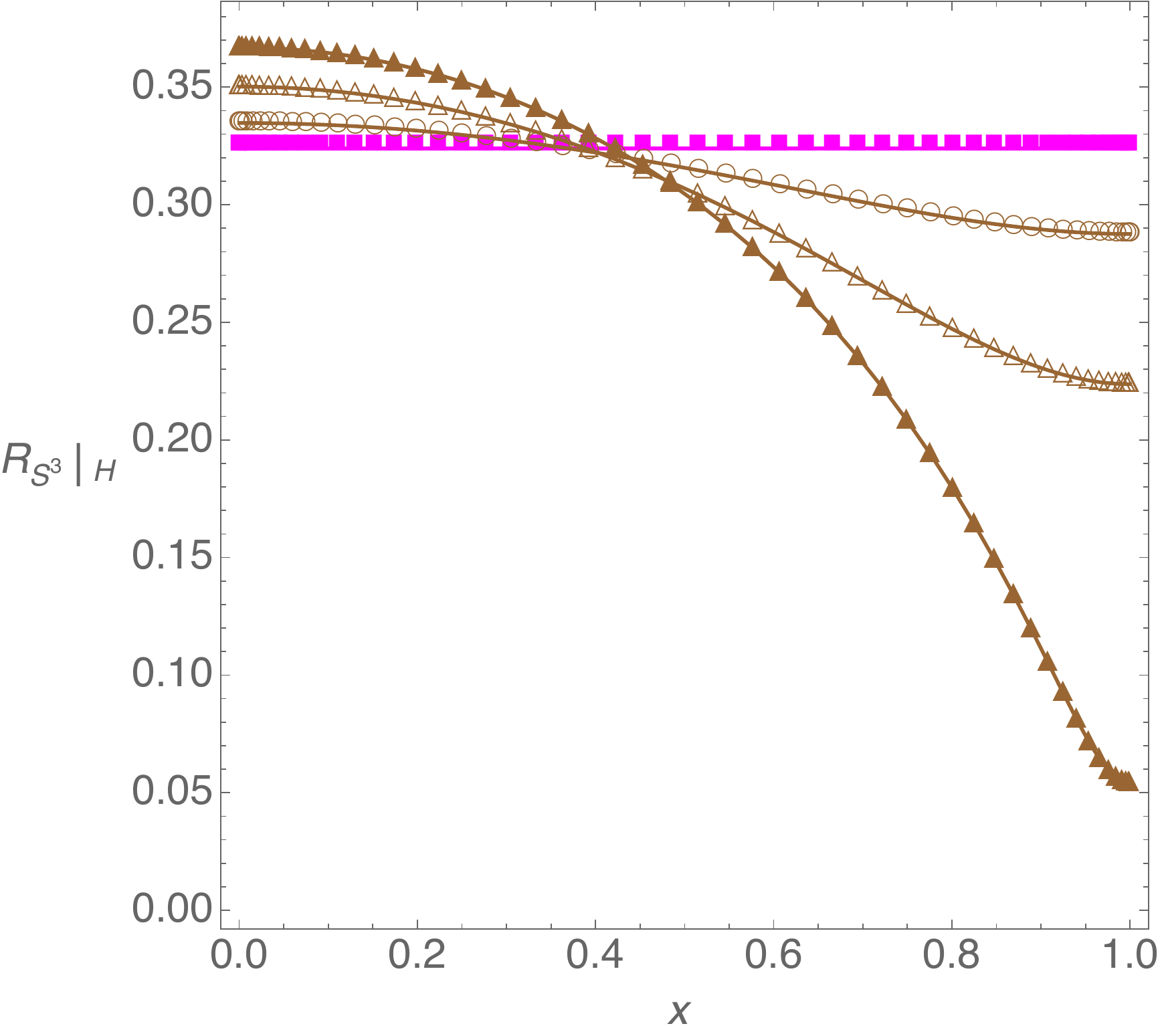}
\caption{Radius of the S$^3$ evaluated at the horizon of the $\ell=2$ lumpy solutions as a function of the polar variable $x$ of the S$^5$. {\it Left panel}: $\ell=2$ lumpy double BH branch with the magenta squares being the solution closer to the GL merger (with $T=0.59448$), and the black filled triangles being the solution with the highest temperature ($T=0.60849$). In between we also present the solutions with $T=0.59922$ (circles) and $T=0.60552$ (empty inverted triangles). {\it Right panel}: $\ell=2$ black belt branch with the magenta squares being the solution closer to the GL merger (with $T=0.59448$), and the brown filled triangles being the solution with the lowest temperature ($T=0.57278$). In between we also present the solutions with $T=0.59015$ (circles) and $T=0.58283$ (empty triangles).}\label{Fig:radiusL2}
\end{figure}

Now we proceed with the $\ell=2$ case.  In Fig. \ref{Fig:ricciHL2}, we plot the Ricci scalar of the induced horizon geometry at one of the poles of the S$^5$ and at the equator, both as a function of temperature.  The curvature of the black belt branch is getting large at the poles. In Fig. \ref{Fig:radiusL2}, we plot the radius of the S$^3$ on the horizon as a function of the polar angle $x$ of the S$^5$ for four different temperatures.  In the right panel, we see that close to the conjectured conical merger, the S$^3$ radius of the black belt branch gets very small on the poles ($x=1$) of the S$^5$.  In the left panel, we see that the S$^3$ radius of the double BH branch is decreasing on the equator ($x=0$) of the S$^5$ as we approach the conical merger, though we are still somewhat far from this conjectured merger.  This also explains why the induced Ricci scalar is not yet appreciably large at the equator for this family (see Fig. \ref{Fig:ricciHL2}).  

\section{Discussion and Prospects \label{sec:discussion}}
Let us now summarise our findings.  As predicted by \cite{Banks:1998dd,Peet:1998cr}, the authors of \cite{Hubeny:2002xn} found that small AdS$_5$-$\mathrm{Schw}\times$S$^5$ BHs suffer from a Gregory-Laflamme instability. This instability contains zero modes where new stationary solutions are expected to exist.  We constructed the ``lumpy" solutions corresponding to the $\ell=1$ and $\ell=2$ modes that break the $SO(6)$ symmetry of the S$^5$ down to $SO(5)$.  Due to the symmetries of the linear perturbations, the $\ell=1$ mode contains one branch while the $\ell=2$ mode contains two branches.  We called the $\ell=2$ branches the ``double black hole" branch and the ``black belt" branch in anticipation of the ensuing (conjectured) conical mergers.  In the microcanonical ensemble, only parts of the double BH branch of the $\ell=2$ mode is preferred over AdS$_5$-Schw$\times$S$^5$ BHs, but we do not have any evidence that it actually dominates the ensemble.  None of these phases are preferred in the canonical ensemble, which is likely dominated by large AdS$_5$-Schw$\times$S$^5$ BHs.  

We have good numerical evidence that these solutions approach conical mergers.  We also have evidence that the $\ell=1$ lumpy BHs transitions to a topologically S$^8$ BH that sits on one of the poles of the S$^5$ (see Fig. \ref{Fig:phases}.b).  We also have evidence that the $\ell=2$ double BH branch transitions to two S$^8$ BHs, localised on each of the poles of the S$^5$ (Fig. \ref{Fig:phases}.c), and the black belt branch transitions to an s$^4\times$S$^4$ black hole with the smaller s$^4$ wrapping around the larger S$^4$ equator of the S$^5$ (Fig. \ref{Fig:phases}.d).

\begin{figure}[ht]
\centering
\includegraphics[width=.495\textwidth]{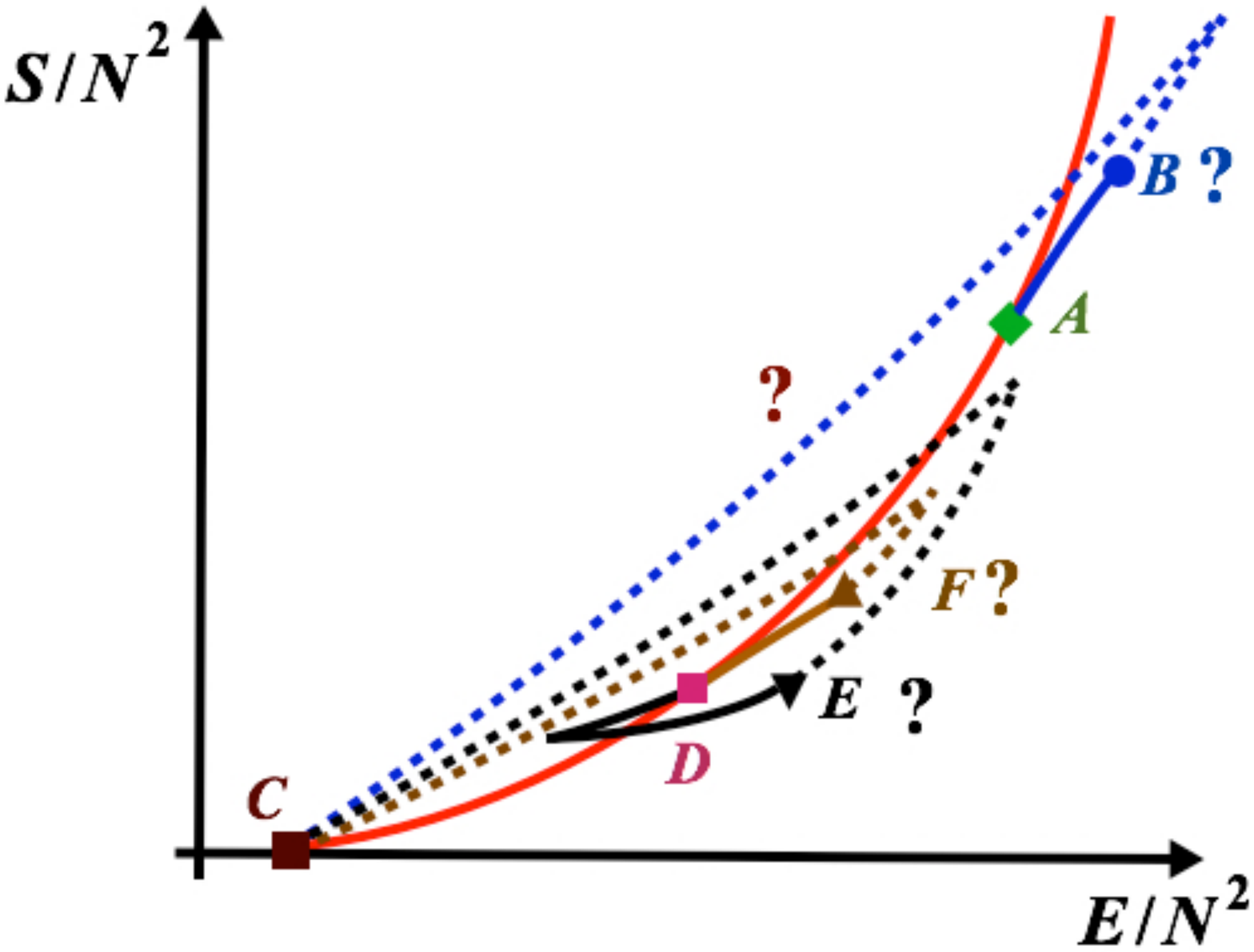}
\includegraphics[width=.495\textwidth]{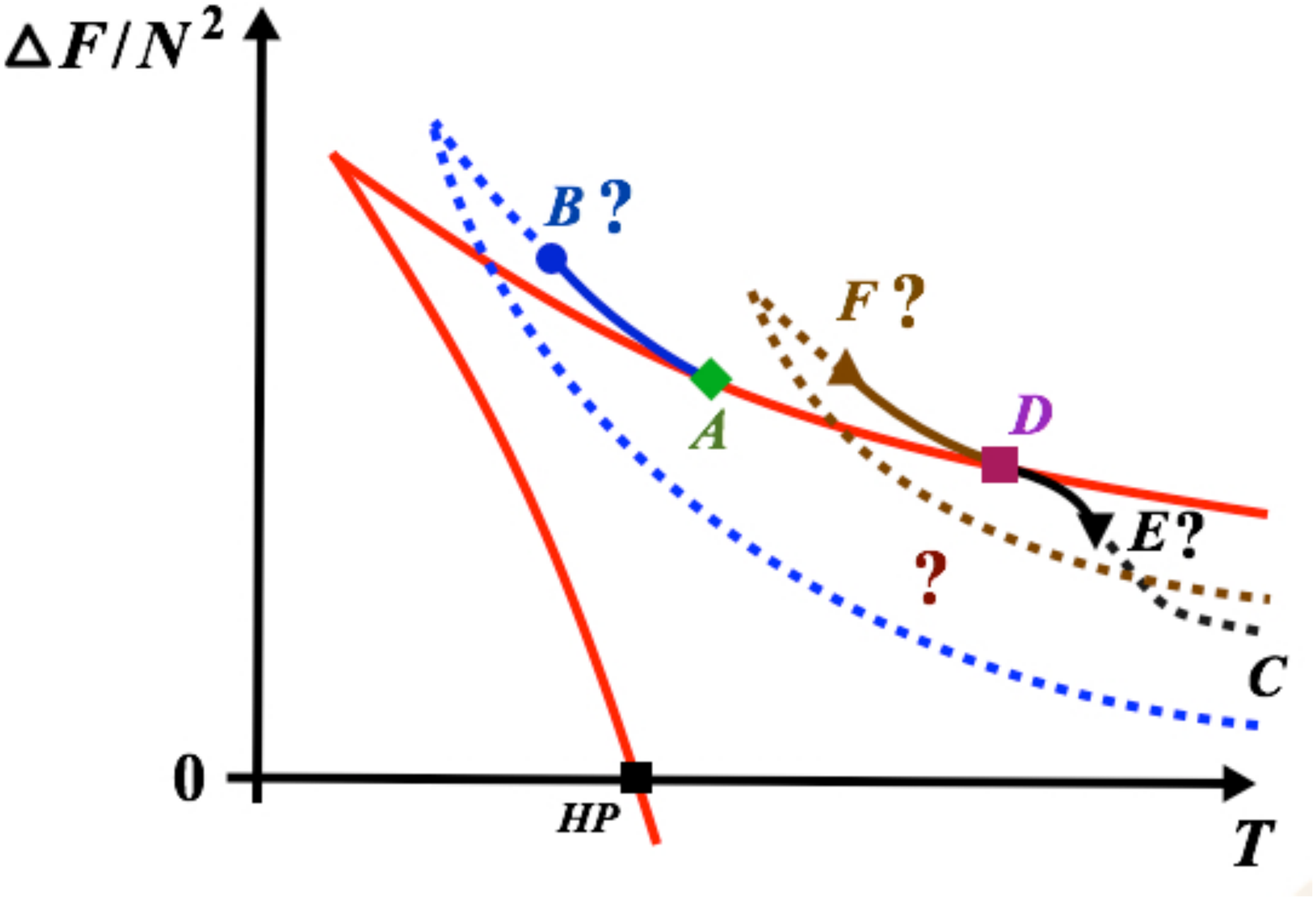}
\caption{Conjectured phases diagram of asymptotically AdS$_5\times$S$^5$ static BHs in the microcanonical ({\it Left Panel}) and canonical ({\it Right Panel}) ensembles ($\Delta F \equiv {\cal F}-{\cal F}_{AdS_5}$; $L\equiv 1$). The red  line is the AdS$_5$-Schw$\times$S$^5$ BH, the blue line $AB$ describes the $\ell=1$ {\it lumpy} AdS$_5\times$S$^5$ BH family with horizon topology $S^3\times$S$^5$. The green diamond $A$ is the $\ell=1$ zero mode.  The blue dashed line $BC$ represents the conjectured {\it localised} AdS$_5\times$S$^5$ BHs with horizon topology S$^8$. We expect there to be some point $B$ which is the conical merger between these solutions.  The turning points must be cusps to be consistent with the first law.  The black square is the Hawking-Page (HP) critical point. The magenta square $D$ is the $\ell=2$ zero mode with the double BH branch along the brown curve and the black belt branch along the black curve.  There are conjectured points $E$ and $F$ that mark conjectured conical mergers.}\label{Fig:phasediag2}
\end{figure}

Let us now speculate on the complete phase diagram which we conjecture to be something resembling Fig. \ref{Fig:phasediag2}. Consider the microcanonical ensemble (\emph{Left Panel}) and solutions with $\ell=1$. Small localised BHs should look muck like $d=10$ small asymptotically flat Schwarzschild BHs whose entropy scales as $S\sim E^{8/7}$. We can compare this  with that of small AdS$_5$-$\mathrm{Schw}\times$S$^5$ BH, whose entropy scales as $S\sim E^{3/2}$. We therefore conclude that the entropy of a small localised BH is larger than the entropy of a small AdS$_5$-Schw$\times$S$^5$ BH. We further expect these S$^8$ BHs to merge with the lumpy BHs with horizon topology $S^3\times$S$^5$ at some conical merger point (point $B$ in Fig. \ref{Fig:phasediag2}). A simple curve that satisfies these two properties is the dashed blue curve $CB$ sketched in the {\it Left Panel} of Fig. \ref{Fig:phasediag2} (the turning point of this curve and similar curves in the diagram must be a cusp to be consistent with the first law of thermodynamics). If this conjectured  curve turns out to be correct, there will be a first order phase transition where the entropy of the localised BHs and the AdS$_5$-Schw$\times$S$^5$ BHs are equal at the same energy.  Such a phase transition can be interpreted in the dual field theory as spontaneous symmetry breaking, though here this would be a first-order transition rather than second order. 

Consider now the $\ell=2$ solutions. Both the double BH and black belt should be unstable to the formation of a single localised BH. Indeed, the double BH is an unstable equilibrium configuration and a small perturbation should make the two BHs merge into a single localised BH. Similarly, if we slightly perturb the black belt along  the polar S$^5$ direction, it is reasonable to expect that it will also collapse into a localised BH at the pole. Another possible instability mechanism is the fragmentation of the black belt into one or an array of BHs along the equator of the S$^5$, and these should again be unstable and merge into a single localised BH. For these reasons, we expect that the double BH and black belt should be less preferred phases than the $\ell=1$ localised BH, both in the microcanonical and canonical ensembles. The simpler scenario with these properties is described by the dashed black (CE) and brown (CF) curves in Fig. \ref{Fig:phasediag2}.  


Consider now the canonical ensemble ({\it Right Panel} of Fig. \ref{Fig:phasediag2}). Again, small localised S$^8$ BHs resemble 10-dimensional asymptotically flat Schwarzschild BHs whose free energy scales as ${\cal F}\sim T^{-7}$.  Small AdS$_5$-$\mathrm{Schw}\times$S$^5$ BHs have a free energy that scales as  ${\cal F}\sim T^{-2}$.  Thus, for high temperatures, the localised S$^8$ BHs are likely to have a lower free energy than small AdS$_5$-$\mathrm{Schw}\times$S$^5$ BHs.  Though, \emph{large} AdS$_5$-$\mathrm{Schw}\times$S$^5$ BHs have a free energy that scales as $\mathcal F\sim T^4$, and they are perturbatively stable so they are likely to dominate the canonical ensemble.  

We should mention that, in the sketched phase diagrams of Fig. \ref{Fig:phasediag2}, we are probably oversymplifying the structure of the solutions near the conical mergers. Indeed, it might well be the case that the lumpy and localised branches will spiral towards the conical merger, leading to an infinite discrete non-uniqueness similar to the one found in \cite{Bhattacharyya:2010yg,Dias:2011tj,Gentle:2011kv,Dias:2014cia,Emparan:2014pra}. In the present case, we do not approach the conical mergers sufficiently enough to address this question.

Of course, to fully complete these phase diagrams, the localised solutions need to be constructed.  We leave this to future work that is currently in progress \cite{DSW:IIb2}.  We note that it is not necessary to resort to numerics to contribute to our understanding of this phase diagram.  In particular, small localised BHs and black belts should be well described by black branes, and are hence amenable to a matched asymptotic expansion or a blackfold approximation (similar to the analysis done in \cite{Harmark:2002tr,Harmark:2003yz,Gorbonos:2004uc,Dias:2007hg,Emparan:2007wm,Horowitz:2011cq,Suzuki:2012av} for localised BHs on a S$^1$ \cite{DSW:IIb2}).\footnote{We note that although asymptotically flat BHs only have a blackfold description when one of the spheres is odd, the s$^4\times$S$^4$ belts are not supported by angular momentum, but by the geometry of the S$^5$.}

We also note that we have studied but two modes in the entire spectrum of spherical harmonics on S$^5$, and we have only focused on those preserving an $SO(5)$ symmetry.  The full phase diagram is thus incredibly rich.  Though, since the localised BHs connected to the $\ell=1$ modes would possess a full $SO(4)\times SO(5)$ symmetry, they are likely the most symmetric of the single localised S$^8$ BHs, and are thus likely to be the entropically dominant phase for small energies in the microcanonical ensemble. For $\ell>1$, we can have multi-BH configurations localised on the S$^5$ even with different sizes. 

We have also only focused on global $AdS_5$ which corresponds to a field theory background on $\mathbb R\times \mathrm S^3$.  Other backgrounds such as $\mathbb M^{1,2}\times S^1$ \cite{Witten:1998zw} or BH backgrounds \cite{Marolf:2013ioa} yield gravity solutions with physics near the AdS scale.  It would be interesting to understand how breaking the symmetries of the S$^5$ will influence these geometries.

Our choice of boundary conditions ensures that the CFT dual to our gravity solution is $\mathcal{N}=4$ SYM \cite{Witten:1998qj}. However, there are other choices that are consistent with finite energy and the absence of ghosts. Most notably, we could have chosen boundary conditions that correspond on the CFT side to adding a relevant double trace deformation to $\mathcal{N}=4$ SYM \cite{Gubser:2002vv}. The effect of these boundary conditions on the phase diagram remains unclear.

There are also a number of other known AdS/CFT dualities such as those arising from $\mathrm{AdS}_4\times \mathrm S^7$ and $\mathrm{AdS}_7\times \mathrm S^4$.  Where the asymptotics are global $AdS_q\times \mathrm S^p$, we expect similar behaviour to what we have found in $\mathrm{AdS}_5\times \mathrm S^5$.  

\section*{Acknowledgments}
It is a pleasure to thank Veronika Hubeny, Mukund Rangamani, Harvey Reall, Kostas Skenderis and Marika Taylor for helpful discussions.  The authors thankfully acknowledge the computer resources, technical expertise, and assistance provided by CENTRA/IST. Some of the computations were performed at the cluster `Baltasar-Sete-S\'ois' and supported by the DyBHo-256667 ERC Starting Grant. O.D. acknowledges the kind hospitality of the Yukawa Institute for Theoretical Physics, where part of this work has been done during the workshop ``Holographic vistas on Gravity and StringsÓ, YITP-T-14-1, and the discussions held during the CERN workshop ``TH Institute on Numerical Holography". O.D. is supported by the STFC Ernest Rutherford grants ST/K005391/1 and ST/M004147/1. B.W. is supported by European Research Council grant no. ERC-2011-StG 279363-HiDGR. The research leading to these results has received funding from the European Research Council under the European Community's Seventh Framework Programme (FP7/2007-2013) / ERC grant agreement no. [247252].

\begin{appendix}

\section{Kaluza-Klein holography\label{KKholography}}
In the main text, we have omitted any details concerning Kaluza-Klein (KK) holography \cite{Skenderis:2006uy} which we will supply here in this appendix.  There are three tasks that required the use of this KK formalism:  computing the energy, computing the VEVs of the scalar fields, and determining the appropriate asymptotic boundary conditions that correspond to turning off sources on the field theory. 

The gauge invariant formalism of KK holography was developed by Skenderis and Taylor in \cite{Skenderis:2006uy}. Previous studies useful for this endeavour are \cite{Kim:1985ez,Gunaydin:1984fk,Lee:1998bxa,Lee:1999pj,Arutyunov:1999en}, and KK holography is further discussed and applied in \cite{Skenderis:2006di,Skenderis:2007yb}.   We will review this formalism in some detail, following \cite{Skenderis:2006uy}.  We will also need to extend the results of \cite{Skenderis:2006uy} to our system where odd harmonics are excited.  Moreover, we use a different harmonic representation for the S$^5$. We will try to be self-contained but refer the reader to \cite{Skenderis:2006uy} and \cite{Kim:1985ez,Lee:1998bxa} for a more thorough exposition.

The aim of KK holography  is to first dimensionally reduce solutions with AdS$_p \times X^q$ asymptotics to solutions on  AdS$_p$, then apply holographic renormalisation to compute field theory quantities on the boundary of AdS$_p$  \cite{Skenderis:2006uy}.  In our case (asymptotically AdS$_5 \times$S$^5$ solutions), the dimensional reduction requires expanding any solution as a sum of harmonics of the S$^5$.   These harmonic modes are interpreted as fields in the reduced AdS$_5$ theory.  The behaviour of these fields on the boundary of AdS$_5$ give VEVs of operators on the dual conformal field theory.  

The dimensional reduction obtains the effective $d=5$ fields $\Psi$ from some $d=10$ fields $\psi$.  In general, the map between $\Psi$ and $\psi$ is highly nonlinear.  However, if we are only interested in computing VEVs in the dual field theory, we only need the field $\Psi$ up to some order in a Fefferman-Graham expansion off the AdS$_5$ boundary.  That is, for any particular VEV, we can write $\Psi$ as some polynomial of $\psi$ and its derivatives, truncating at a particular order  \cite{Skenderis:2006uy}.  

For example, the quadratic expansion for a field $\Psi^k$ takes the form
\be
\Psi^k = \psi^k+\sum_{lm}\left(J_{klm}\psi^l\psi^m+ L_{klm}D_\mu\psi^l D^\mu\psi^m \right)+{\mathcal O}([\psi^k]^3),
\ee
for some constants $J_{klm}$ and $L_{klm}$.  If $\Psi^k$ is dual to an operator of dimension $k$, then we would require expanding $\Psi^k$ off the boundary in a Feffermann-Graham expansion to $\mathcal O(z^k)$.  The quadratic terms with $l+m=k$ also contribute to such an expansion.  Higher order terms contribute as well, but for a given $k$, we can truncate this expansion  \cite{Skenderis:2006uy}.  In our case, it suffices to stop at quadratic order.  

At this point, it would be useful to give a brief overview of the lengthy calculation to follow.  We begin in section \ref{sec:KKexpansion} by writing our lumpy BH solutions as deformations of global AdS$_5 \times$S$^5$, and expressing those deformations as a sum of S$^5$ harmonics $\sum\tilde\psi^\ell Y_\ell$.  We then carry out a Feffermann-Graham expansion of the coefficients of these harmonics $\tilde\psi$ to the order needed to extract VEVs.

To avoid gauge issues, we now need to rewrite these coefficients in a gauge invariant way, keeping up to quadratic terms in the number of fields (recall we only need up to second order to extract the VEVs of interest).  In section \eqref{sec:KKO1}, we write down the gauge invariant quantities at linear order in the number of fields (and call these collective quantities $\hat\psi$), and also show that most of these gauge-invariant fields obey an effective Klein-Gordon equation for a massive scalar.  We proceed with quadratic order in section \ref{sec:KKO2}, and call the resulting quantities $\psi$.  The second-order fields $\psi$ obey an inhomogeneous Klein-Gordan equation for a massive scalar.  The source term depends on the square of linear-order fields $\hat\psi$ and their derivatives.  
 
We perform the KK reduction in section \ref{sec:KK5d}. There, we will obtain the effective 5-dimensional field $\Psi$ in terms of the 10-dimensional $\hat\psi$ and $\psi$.  In section \ref{sec:KKrenormalization} we will obtain the effective 5-dimensional action, namely \eqref{5dAction}, which describes KK scalars subject to a certain potential and living in the 5-dimensional background $G$ with a negative (5-dimensional) cosmological constant. Section \ref{sec:KKrenormalization} ends with the Einstein equation \eqref{5dEin}-\eqref{5dEinT} that the reduced graviton $G$ obeys. This equation has a non-trivial energy-momentum tensor.

Section \ref{sec:KKrenormalization} also takes the 5-dimensional gravitational ($G$) and scalar fields ($\Psi$) and applies the standard holographic renormalisation procedure \cite{deHaro:2000xn}. We first introduce the Fefferman-Graham coordinate $Z=Z(z)$ for the 5-dimensional solution, and then do the standard Fefferman-Graham expansion off the 5-dimensional AdS boundary $Z=0$. We can then construct the associated holographic stress tensor \eqref{holoT} and VEVs \eqref{vevS2S3} of the most relevant KK scalars. In particular, from the holographic stress tensor we can then read off the expression for the energy, which we use in the main text.  In the process, we also explain our physical motivation for our choice of asymptotic boundary conditions.  

Summarising our notation, $\tilde{\psi}$ describes the coefficients of the 10-dimensional harmonic expansion around global AdS$_5 \times$S$^5$; $\hat{\psi}$ represents gauge invariant quantities at linear order in the number of fields; $\psi$ describes gauge invariant quantities at quadratic order; and finally $\Psi$ describes the reduced 5-dimensional KK field (i.e. capital letters denote 5-dimensional fields and lower-case letters always refer to 10-dimensional fields).

\subsection{Lumpy AdS$_5 \times$S$^5$ BHs expanded in spherical harmonics of the S$^5$ \label{sec:KKexpansion}}
Our lumpy BH solutions are asymptotically AdS$_5 \times$S$^5$.  Before proceeding with a dimensional reduction to AdS$_5$, we first need to expand these solutions in terms of harmonics on the S$^5$.  There are scalar, vector, and tensor harmonics, which are defined by their transformations on the S$^5$.  The details of the harmonic expansion differ for each type, but only the scalar harmonics are consistent with our preserved symmetries (a $SO(5)$ subgroup of $SO(6)$), so we will only focus on these harmonics.

If the S$^5$ is written as
\begin{equation}
d\Omega_5^2=\frac{4\, dX^2}{2-X^2}+\left(1-X^2\right)^2 d\Omega_4^2\;,
\end{equation}
the regular (axisymmetric) scalar spherical harmonics are given by 
 \begin{equation}\label{Harmonic}
Y_\ell(X)=\frac{\sqrt{(\ell +2) (\ell +3)}}{2^{\frac{1}{2} (\ell +1)} \sqrt{3}} \, _2F_1\left(-\ell ,\ell
   +4;\frac{5}{2};\frac{1}{2} \left(1+X\sqrt{2-X^2} \right)\right),
\end{equation}
and satisfy 
 \begin{equation}\label{HarmonicEOM}
 \Box_{S^5} Y_\ell(X)=\Lambda_\ell  \,Y_\ell(X), \qquad \hbox{with} \quad \Lambda_\ell=-\ell(\ell+4), \quad \ell=0,1,2,\ldots
 \end{equation}
The quantum number $\ell$ is a measure of the number of nodes along the polar direction $X$ that was quantised by requiring regularity at the poles $X=\pm 1$ of the S$^5$; we set the azimuthal quantum number $m=0$ because these modes would further break the $SO(5)$ symmetry. We have chosen a normalisation in \eqref{Harmonic} so that
 \begin{equation}\label{HarmonicNorm}
\int_{S^5} Y_{\ell_1} Y_{\ell_2} = z(\ell_1) \delta^{\ell_1 \ell_2}, \qquad \hbox{with} \quad 
z(\ell) = \frac{\Omega_5}{2^{\ell-1} (\ell+1) (\ell+2)}\,, \quad \Omega_5=\pi^3.
\end{equation}

Now let us expand our lumpy BH solutions in terms of these harmonics.  First, we write the fields of the solution as a deformation of AdS$_5\times S^5s$:
\begin{eqnarray}
&& g_{MN} = g^o_{MN} + h_{MN}, \\
&& F_{MNPQR} =  F^o_{MNPQR} + f_{MNPQR}\;,\nonumber
\end{eqnarray}
where $\{ g^o, F^o_{(5)}\}$ is global AdS$_5\times$S$^5$.  Here, $h$ and $f$ need not be small.  The field fluctuations abound global AdS$_5 \times$S$^5$ thus admit the harmonic expansion:
\begin{eqnarray} \label{coef:H}
h_{\mu \nu}(z,X) &=& \sum_{\ell} \tilde{h}^{\ell}_{\mu \nu}(z) Y_{\ell}(X)\,, \qquad
h_{\mu a} (z,X)=  \sum_{\ell} \tilde{B}^{\ell}_{\mu}(z) D_a Y_{\ell}(X)\,, \nonumber \\
h_{(ab)}(z,X) &=& \sum_{\ell} \tilde{\phi}^{\ell}(z) D_{(a} D_{b)} Y_{\ell}(X) )\,, \qquad
h_{a}^a(z,X) = \sum_{\ell} \tilde{\pi}^{\ell}(z) Y_{\ell}(X)\,, 
\end{eqnarray}
and 
\begin{eqnarray}  \label{coef:F}
f_{\mu \nu \rho\sigma\tau}(z,X) &=& \sum_{\ell}
5 D_{[\mu}  \tilde{b}^{\ell}_{\nu \rho\sigma\tau]}(z) Y_{\ell}(X)\,, \qquad
f_{a \mu \nu \rho \sigma}(z,X) = \sum_{\ell} \tilde{b}^{\ell}_{\mu \nu \rho \sigma}(z) D_a Y_{\ell}(X)\,,  \nonumber \\ 
f_{ab \mu \nu \rho}(z,X) &=& 0\,,\qquad   
f_{a b c \mu \nu}(z,X) = 0\,,    \\ 
f_{abcd\mu}(z,X)  &=& \sum_{\ell} 
D_\mu \tilde{b}^{\ell}(z) \,\epsilon_{abcd}{}^e D_e Y_{\ell}(X)\,, \qquad
f_{a b c d e}(z,X) = \sum \tilde{b}^{\ell}(z)\, \Lambda_{\ell} \,\epsilon_{abcde} Y_{\ell}(X),  \nonumber
\end{eqnarray}
where we use the symmetric traceless notation $A_{(ab)} = \frac{1}{2} (A_{ab}+A_{ba}) - \frac{1}{5} g_{ab} A_a^a$. It follows from the field equations that there is an algebraic relation between the coefficients $\tilde{b}^{\ell}_{\nu \rho\sigma\tau}$ and $\tilde{b}^{\ell}$. Therefore we do not discuss the fluctuations $\tilde{b}^{\ell}_{\nu \rho\sigma\tau}$ any further. Note that the case  $\ell=1$ is special because $D_{(a} D_{b)} Y_{\ell}=0$ so  $\tilde{\phi}^{\ell=1}$ is not defined.  The case $\ell=0$ is also special since $\Lambda_\ell=0$,   $D_{a} Y_{\ell}=0$ and  $D_{(a} D_{b)} Y_{\ell}=0$; therefore, $\tilde{\phi}^{\ell=0}$, $\tilde{B}_{\mu}^{\ell=0}$ and $\tilde{b}^{\ell=0}$ are not defined. The expansion of all other fields start at $\ell=0$.

In anticipation of reading off field theory quantities from the boundary of the reduced AdS$_5$, let us also perform a Feffermann-Graham expansion off the boundary.  The ``background" fields $\{ g^o, F^o_{(5)}\}$ that give AdS$_5\times$S$^5$ are expanded as 
\begin{eqnarray}\label{AdS5xS5:FG}
&& ds_o^2 = \frac{L^2}{z^2}\left[dz^2 -\left( 1+\frac{1}{2}\,z^2+ \frac{1}{16}\,z^4 +{\cal{O}}(z^5)\right)\! dt^2+ \!\left( 1-\frac{1}{2}\,z^2+ \frac{1}{16}\,z^4 +{\cal{O}}(z^5)\right) \! d\Omega_3^2 \right] \!+L^2 d\Omega_5^2 \nonumber\\ 
&& F^{o}_{\mu \nu \rho\sigma\tau} = \epsilon_{\mu \nu \rho\sigma\tau}, \qquad
F^{o}_{abcde} = \epsilon_{abcde}\,,
\end{eqnarray}
where we have stopped at ${\cal O}(z^4)$ which contains terms necessary to compute the holographic stress tensor and the VEVs of scalar operators. Note that the factors of $z^2/2$ are present because the background is global AdS$_5$, not planar AdS$_5$, and the factors of $z^4/16$ come from the conformal anomaly of AdS$_5$.  We have chosen a conformal frame where the boundary geometry of the AdS$_5$ is $\mathbb R\times S^3$.  

Now let us proceed with the first few harmonic coefficients in \eqref{coef:H} and \eqref{coef:F}. Performing a Taylor expansion of the lumpy BHs up to $O(z^4)$, or $(L^2/z^2)O(z^4)$ for fields on the AdS$_5$ base space, the equations of motion yield (henceforward, we set $L= 1$): 
\begin{eqnarray}  \label{coef:lumpy}
&& \hspace{-0.5cm} \tilde{h}^{\ell=0}_{\mu \nu} =  C_N \, \frac{y_+^2}{3072}\,{\biggl[}y_+^2 {\biggl(} 5 \beta_2+45 \,C_0 \,\beta_2^2+12 (16 \,\delta_{0}+\delta_4-192){\biggr)} -2304{\biggr]} \, z^2 \, \eta_{\mu\nu}, \nonumber \\
&& \tilde{h}^{\ell=1}_{\mu \nu} =  0, \qquad  \tilde{h}^{\ell=2}_{\mu \nu} =   \frac{y_+^2}{128} \sqrt{\frac{3}{10}} \beta_2 {\biggl(} 80+ \left(60 \,C_2+23 \,y_+^2\, \beta_2\right)z^2 {\biggr)}  \eta_{\mu\nu}, \nonumber \\
&& \tilde{h}^{\ell=3}_{\mu \nu} = - \frac{y_+^3}{12} \sqrt{\frac{5}{2}}\, \gamma_3 \, z \, \eta_{\mu\nu}, \qquad  \tilde{h}^{\ell=4}_{\mu \nu} =   \frac{\sqrt{7} \, y_+^4}{3072}\,{\biggl(} 278\, \beta_2^2+25 \beta_2-100 \delta_4  {\biggr)}  \,z^2\, \eta_{\mu\nu}\,; \nonumber \\
&& \hspace{-0.5cm}  \tilde{B}_{\mu}^{\ell=1}=\tilde{B}_{\mu}^{\ell=2}=\tilde{B}_{\mu}^{\ell=3}=\tilde{B}_{\mu}^{\ell=4}=0\,; \nonumber \\
&& \hspace{-0.5cm}  \tilde{\phi}^{\ell=2}=\frac{y_+^2}{512} \sqrt{\frac{15}{2}} \beta_2  \left[ 32 z^2 +\left(y_+^2\,\beta_2 +20\right) z^4 \right], \nonumber \\
&&   \tilde{\phi}^{\ell=3}=   \frac{y_+^3}{72} \sqrt{\frac{5}{2}}\, \gamma_3  \, z^3,  \qquad \tilde{\phi}^{\ell=4}=- \frac{5 \sqrt{7}}{36864} y_+^4  \left(34 \beta_2^2+5 \beta_2-20 \delta_4\right) z^4\,;
 \nonumber \\
&& \hspace{-0.5cm} \tilde{\pi}^{\ell=0}=  \frac{5}{256}  \, y_+^4 \,\beta_2^2\,  z^4,\qquad \tilde{\pi}^{\ell=1}=0,  \nonumber \\
&& \tilde{\pi}^{\ell=2}= -  \frac{y_+^2}{128} \sqrt{\frac{15}{2}}\, \beta_2 {\biggl(} 64 z^2+ \left( 17 y_+^2\, \beta_2+20\right) z^4 {\biggr)}, \nonumber \\
&& \tilde{\pi}^{\ell=3}=  
\frac{y_+^3}{3} \sqrt{\frac{5}{2}} \, \gamma_3\, z^3,\qquad  \tilde{\pi}^{\ell=4}= -  \frac{5}{768} \sqrt{7}\,  y_+^4  \left(34 \beta_2^2+5 \beta_2-20 \delta_4 \right) z^4\,; \nonumber \\
&& \hspace{-0.5cm} \tilde{b}^{\ell=1}=0,\qquad \tilde{b}^{\ell=2}=\sqrt{\frac{3}{10}}\frac{y_+^2}{1024}\,\beta_2\,{\biggl(}160 \,z^2 + \left(31 \,y_+^2 \,\beta_2 +60\right) z^4 {\biggr)},  \nonumber \\
&& \tilde{b}^{\ell=3}=-\frac{y_+^3}{72} \sqrt{\frac{5}{2}}\,\gamma_3 \, z^3, \qquad 
\tilde{b}^{\ell=4}=\frac{\sqrt{7}\, y_+^4 }{24576}   \left(116 \beta_2^2+25 \beta_2-100 \delta_4 \right)z^4,
\end{eqnarray}
where  $\eta_{\mu\nu}=\hbox{diag}\{-1,0,\eta_{ij} \}$ with $\eta_{ij}$ being the line element of a unit radius S$^3$ on the AdS$_5$ base space. To shorten the presentation, we have introduced the auxiliary constants $\{ C_N,C_0,C_2 \}$ such that $\{ C_N,C_0,C_2 \}= \{ 1,1,1 \}$ for $\mu=\nu=t $, while $\{ C_N,C_0,C_2 \}= \{ -1/3,-1/3,1/9 \}$ for $\mu=\nu=x^i)$ in $\tilde{h}_{\mu\nu}^{\ell=0}$ and $\tilde{h}_{\mu\nu}^{\ell=2}$. These harmonic coefficients depend on the horizon radius $y_+$ and on four undetermined constants $\{ \beta_2, \gamma_3, \delta_0, \delta_4\}$ that appear in the Taylor expansion off-the boundary $z=0$. 
In this expansion, we have already imposed appropriate asymptotic boundary conditions (BCs) that eliminate several extra undetermined constants that would appear in the expansion. 
We will defer our discussion of these BCs to a later section \ref{sec:KKrenormalization}; see in particular the BCs \eqref{BCs:bdry} and Fig. \ref{Fig:confdim} therein.

For reference, the leading terms of the transformation that bring the lumpy BH from the $\{x,y\}$ coordinates of our ansatz \eqref{lumpyansatz} into the FG coordinates $\{ z,X\}$ are
\begin{eqnarray}
y &=& 1-\frac{y_+}{2}\,z-\frac{y_+^2}{8} \,z^2-\frac{1}{192} y_+\left( 24+y_+^2 \left[ 12+\beta_2 \left(1-6X^2(2-X^2)\right) \right] \right)z^3+\cdots\,,\nonumber \\
x &=& X-\frac{3}{32}\,y_+^2 \,\beta_2 \,X(2-3X^2+X^4) z^2\nonumber \\
&& +\frac{1}{96} \sqrt{2-X^2} \left(1-X^2\right)y_+^3  \left[ 3 \gamma_1+2 \gamma_3 \left(X^4-2 X^2-1\right)\right]z^3+\cdots  
\end{eqnarray}

\subsection{Gauge invariance and field equations at linear order\label{sec:KKO1}}
Not all of the coefficients in \eqref{coef:H}, \eqref{coef:F} are independent.  Under a gauge transformation $x\rightarrow x+\xi$, the fluctuations transform as $h\rightarrow\delta h$ and $f\rightarrow\delta f$, where to linear order in the number of fields, these are given by
\begin{eqnarray} \label{gaugeTransf}
&& \delta h_{MN} = (D_M \xi_N +  D_N \xi_M) + (D_M \xi^P h_{PN} + D_N \xi^P h_{MP}
+\xi^P D_P h_{MN}); \\
&& \delta f_{MNPQR} = 5 D_{[M} \xi^S F^o_{NPQR]S} +  (5 D_{[M} \xi^S f_{NPQR]S}
+\xi^S D_S f_{MNPQR}). \nonumber
\end{eqnarray}
The gauge parameter $\xi^M(z,X)$ itself has a scalar harmonic expansion,
\begin{equation}
\xi_\mu(z,X)=\sum_{\ell_1} \xi^{\ell_1}_\mu (z) Y_{\ell_1}(X)\,, \qquad \qquad
\xi_a(z,X) =\sum_{\ell_1}  \xi^{\ell_1}(z) D_a Y_{\ell_1}(X). 
\end{equation}

To avoid further gauge issues, we would like to use gauge-invariant quantities, which in this section we do to linear order in the number of fields. At linear order in the fluctuation, only the leading terms in the gauge transformation \eqref{gaugeTransf} contribute and the coefficients in the harmonic expansion \eqref{coef:H} and \eqref{coef:F} transform as (as justified above we do not need to discuss $\tilde{b}_{\mu \nu \rho \sigma}^{\ell}$)
\begin{eqnarray} \label{gauge:O1}
&& \delta\tilde{h}_{\mu \nu}^{\ell} = D_\mu \xi_\nu^{\ell} +  D_\nu \xi_\mu^{\ell}, \quad\hbox{for} \quad \ell\geq 0\,; \qquad
 \delta\tilde{B}_{\mu}^{\ell} = D_\mu \xi^{\ell} + \xi_\mu^{\ell}, \quad\hbox{for} \quad \ell\geq 1\,; \\
&& \delta\tilde{\phi}^{\ell} = 2 \xi^{\ell}, \quad\hbox{for} \quad \ell\geq 2\,; \qquad
 \delta\tilde{\pi}^{\ell} = 2 \Lambda_{\ell} \xi^{\ell}, \quad\hbox{for} \quad \ell\geq 0\,; \qquad
 \delta \tilde{b}^{\ell} = \xi^{\ell}, \quad\hbox{for} \quad \ell\geq 1. \nonumber 
\end{eqnarray}
As we have mentioned earlier, some field coefficients are not defined for $\ell=0$ and/or $\ell=1$.

It follows that for $\ell \geq 2$ one can define three gauge invariant quantities:
\begin{eqnarray} \label{inv:O1}
\bullet \:\:  \hbox{For} \quad \ell \geq 2: \quad \hat{\pi}^{\ell} &=& \tilde{\pi}^{\ell} - \Lambda_{\ell} \tilde{\phi}^{\ell} \,,\nonumber \\
\hat{h}^{\ell}_{\mu \nu} &=& \tilde{h}_{\mu \nu}^{\ell} 
- D_{\mu} \hat{B}^{\ell}_{\nu}
-D_\nu \hat{B}^{\ell}_{\mu}\,, \nonumber \\
 \hat{b}^{\ell} &=&  \tilde{b}^{\ell} - \frac{1}{2} \tilde{\phi}^{\ell}\,;  \\
 \hbox{with auxiliary field}&& \hat{B}_{\mu}^{\ell} = \tilde{B}^{\ell}_{\mu} 
- \frac{1}{2} D_\mu \tilde{\phi}^{\ell} \qquad \Rightarrow \qquad
\delta\hat{B}_{\mu}^{\ell} = \xi_\mu^{\ell}.\nonumber
\end{eqnarray}
For $\ell=1$, since $\tilde{\phi}^{\ell}$ is not defined, the system is effectively described by one less gauge invariant quantity. One can define the  gauge invariant quantities:
\begin{eqnarray} \label{inv:O1:l1}
\hspace{0.6cm}\bullet \:\: \hbox{For} \quad \ell=1: \quad \hat{\pi}^{1} &=& 0\,, \nonumber \\
\hat{h}^{1}_{\mu \nu} &=& \tilde{h}_{\mu \nu}^{1}
- D_{\mu} \hat{B}^{1}_{\nu}
-D_\nu \hat{B}^{1}_{\mu}\,, \nonumber \\
 \hat{b}^{1} &=&  \tilde{b}^{1} - \frac{1}{2 \,\Lambda_{1}}\, \tilde{\pi}^{1}\,;  \\
 \hbox{with auxiliary field}&& \hat{B}_{\mu}^{1} = \tilde{B}^{1}_{\mu} 
- \frac{1}{2 \,\Lambda_{1}}\, D_\mu \tilde{\pi}^{1} \qquad \Rightarrow \qquad
\delta\hat{B}_{\mu}^{1} = \xi_\mu^{1}.\nonumber
\end{eqnarray}
The case $\ell=0$ is  special since $\tilde{\phi}^{\ell}$, $\tilde{B}_{\mu}^{\ell}$ and $\tilde{b}^{\ell}$ are not defined. One also has  $\delta\tilde{\pi}^{\ell=0}=0$ since $\Lambda_{\ell=0}=0$.
 It follows that $ \tilde{\pi}^{\ell=0}$ is itself already gauge invariant and $ \hat{b}^{\ell=0} $ is not defined. This leaves the gauge invariant quantities:
\begin{eqnarray} \label{inv:O1:l0}
\hspace{-5.6cm}\bullet \:\:  \hbox{For} \quad \ell=0: \quad \hat{\pi}^{0} &=& \tilde{\pi}^{0}\,, \nonumber \\
\hspace{-5.6cm} \hat{h}^{0}_{\mu \nu} &=& \tilde{h}_{\mu \nu}^{0}
+\frac{1}{3} \tilde{\pi}^{0} g_{\mu\nu}^o\,, \nonumber \\
\hspace{-5.6cm} \hat{b}^{0} & & \hbox{not defined}. 
\end{eqnarray}
Note that $\tilde{h}^{\ell=0}=\tilde{h}^0$ is just a deformation of the background metric and the gauge invariant combination $\hat{h}^{0}_{\mu \nu} $ was chosen because it obeys the linearised field equations.

With these gauge invariant quantities at linear order, one can introduce the gauge invariant combinations
\begin{eqnarray} \label{comb:O1}
\hat{s}^{\ell} = \frac{1}{20 (\ell+2)} \left[ \hat{\pi}^{\ell} 
- 10 (\ell+4) \hat{b}^{\ell}\right ], \qquad (\ell\geq 1); \qquad 
\hat{t}^{\ell} = \frac{1}{20 (\ell+2)} \left( \hat{\pi}^{\ell} + 10 \,\ell \, \hat{b}^{\ell}\right ), \qquad (\ell\geq 0); \nonumber \\
\end{eqnarray}
with inverse relations
$\hat{b}^{\ell}=-\hat{s}^{\ell}+\hat{t}^{\ell}$ and $\hat{\pi}^{\ell}=10k\hat{s}^{\ell}+10(\ell+4)\hat{t}^{\ell}$.
To leading order in the number of fields, these satisfy the equations of motion
\begin{equation}\label{EOM:shat}
\Box \, \hat{s}^{\ell} = \ell (\ell-4) \, \hat{s}^{\ell},  \quad \hbox{for}\quad \ell \geq 1\,; \qquad\qquad
\Box \, \hat{t}^{\ell} = (\ell+4) (\ell+8) \, \hat{t}^{\ell},   \quad \hbox{for}\quad \ell \geq 0, 
\end{equation}
where $\Box$ is the D'Alembertian in AdS$_5$. That is, the scalar field $\hat{s}^\ell$ has mass $m^2_{\psi_\ell} =\ell(\ell-4)$ (in AdS$_5$ radius units $L= 1$) with the conformal dimension of the dual operators being $\{\Delta_+,\Delta_-\}=\{\ell, 4-\ell \}$, while the scalar field $\hat{t}^\ell$ has mass $m^2_{\psi_\ell} =\ell(\ell+4)(\ell+8)$ which corresponds to the conformal dimensions $\{\Delta_+,\Delta_- \}=\{\ell+8,-\ell-4 \}$. 

The massive KK gravitons also couple to the scalar harmonics (actually, the instability of the AdS$_5$-Schw BH occurs in the KK graviton sector with $\ell\geq 1$) and are described by the  transverse and traceless
fields
\begin{equation}\label{KK:O1}
\hat{\phi}_{(\mu \nu)}^{\ell} = \hat{h}^{\ell}_{(\mu \nu)} - \frac{1}{(\ell+1)(\ell+3)} 
D_{(\mu} D_{\nu)} \left( \frac{2}{5} \hat{\pi}^{\ell} - 12 \hat{b}^{\ell} \right), 
\quad \ell > 0.
\end{equation}
which obey the equation
\begin{equation}\label{KKeom:O1}
\Box \, \hat{\phi}_{(\mu \nu)}^{\ell} = \left[\ell(\ell+4)-2 \right ] \,\hat{\phi}_{(\mu \nu)}^{\ell}, \quad \ell > 0.
\end{equation}
Thus, the KK gravitons (spin 2) have mass $m^2_{\psi_\ell} = \ell(\ell+4)$ and the conformal dimensions of the dual operators are $\{\Delta_+,\Delta_- \}=\{\ell+4,-\ell \}$.
For $\ell=0$, the combination $\hat{h}^{\ell}_{\mu \nu}$ in \eqref{inv:O1:l0} obeys the 5-dimensional linearized Einstein equations (the shift by $\tilde{\pi}^{0}$ can be traced back to the Weyl transformation required to write the 5-dimensional action in the Einstein frame).

\subsection{Gauge invariance and field equations at quadratic order\label{sec:KKO2}}

At quadratic order in the fluctuation, all the terms in the gauge transformation \eqref{gaugeTransf} contribute. 
For our purposes, it is not necessary to discuss the KK gravitons with $\ell>0$, so for compactness we omit these fields from our discussion. The interested reader can find the analysis of these modes in the original papers.

To discuss the gauge transformations, it is necessary to first project the fields into the spherical harmonic basis. This projection introduces the following integrals of the spherical harmonics ($z(\ell)$ was introduced in the normalisation \eqref{HarmonicNorm}):
\begin{eqnarray}
&&\int_{S^5} D_{(a}  D_{b)} Y_{\ell_1} D_{(a} D_{b)}Y_{\ell_2} = 
z(\ell) q(\ell) \delta_{\ell_1 \ell_2}\,, \qquad \hbox{with} \quad q(\ell) \equiv \frac{4}{5} \,\ell  (\ell -1) (\ell +4) (\ell +5), \nonumber \\
&& a_{123} \equiv \int_{S^5} Y_{\ell_1} Y_{\ell_2} Y_{\ell_3}\,, \qquad \qquad
b_{123}\equiv\int_{S^5} Y_{\ell_1} D_a Y_{\ell_2} D^a Y_{\ell_3} \,, 
\nonumber \\
&&c_{123} \equiv\int_{S^5}  D^{(a}  D^{b)} Y_{\ell_1} D_a Y_{\ell_2} D_b Y_{\ell_3} \,,\qquad\qquad
d_{123} \equiv\int_{S^5}  Y_{\ell_1} D^{(a} D^{b)} Y_{\ell_2} D_a D_b Y_{\ell_3} \,,
\nonumber \\
&&e_{123}\equiv \int_{S^5} D^{(a}  D^{b)} Y_{\ell_1} \left(2 D_a D^c Y_{\ell_2} D_{(c} D_{b)} Y_{\ell_3}
+ D^c Y_{\ell_2} D_c D_{(a} D_{b)} Y_{\ell_3}\right).
\end{eqnarray}

Up to second order, the coefficients in the harmonic expansion \eqref{coef:H} and \eqref{coef:F} transform as (recall that $\Omega_5=\pi^3$)
\begin{eqnarray} \label{gauge:O2}
&&\hspace{-1.5cm} \bullet \:\: \hbox{For} \:\:\:\ell_1\geq 2:  \nonumber\\
\delta \tilde{\pi}^{\ell_1} &=&2 \Lambda_{\ell_1} \xi^{\ell_1} + \frac{\Omega_5}{z(\ell_1)} \sum_{\ell_2,\ell_3\geq 1}
\left[ 2 \tilde{\phi}^{\ell_2} \xi^{\ell_3}  d_{123}
+\left(\frac{2}{5} \Lambda_{\ell_2} \xi^{\ell_2} \tilde{\pi}^{\ell_3} 
+ \xi^{\mu}_{\ell_2} D_\mu \tilde{\pi}^{\ell_3}\right) a_{123}
\right. \nonumber \\ 
&& \left. 
+ \left( \xi^{\ell_2} \tilde{\pi}^{\ell_3} 
+ 2 \xi^{\mu}_{\ell_2} \tilde{B}_{\mu}^{\ell_3} \right)
b_{123} \right]\!, 
           \nonumber \\
\delta \tilde{\phi}^{\ell_1} &=& 2 \xi^{\ell_1}+\frac{1}{ q(\ell_1)} \frac{\Omega_5}{z(\ell_1)} \sum_{\ell_2,\ell_3\geq 1}
\left(\xi^{\ell_2} \tilde{\phi}^{\ell_3} e_{123}
+ \xi^{\mu}_{\ell_2} D_\mu \tilde{\phi}^{\ell_3} d_{213} 
+ \frac{2}{5} \xi^{\ell_2} \tilde{\pi}^{\ell_3} d_{312}
+ 2 \xi^{\mu}_{\ell_2} \tilde{B}_{\mu}^{\ell_3} c_{123}
\right),\nonumber \\
\delta \tilde{b}^{\ell_1} &=& \xi^{\ell_1}+\frac{1}{\Lambda_{\ell_1}}\frac{\Omega_5}{z(\ell_1)} \sum_{\ell_2,\ell_3\geq 1} 
\left(\xi^{\mu 2} D_\mu  \tilde{b}^{\ell_3} + \Lambda_{\ell_2}   \tilde{b}^{\ell_2} \xi^{\ell_3} \right)
(b_{123} + \Lambda_{\ell_3} a_{123});
\end{eqnarray}
\begin{eqnarray} \label{gauge:O2:l1}
&&\hspace{-10.8cm}  \bullet \:\: \hbox{For} \:\:\:\ell_1=1:  \nonumber\\
&& \hspace{-9.2cm}  \delta\tilde{\pi}^{1} = 2 \Lambda_{1} \xi^{1}\,,\nonumber \\
&& \hspace{-9.2cm} \delta\tilde{\phi}^{1} \quad \hbox{not defined}, \nonumber \\
&& \hspace{-9cm}  \delta \tilde{b}^{1} = \xi^{1};
\end{eqnarray}
\begin{eqnarray} \label{gauge:O2:l0}
&&\hspace{-2cm}  \bullet \:\:  \hbox{For} \:\:\:\ell=0:  \nonumber\\
&& \hspace{-0.5cm} \delta \tilde{\pi}^0 = \sum_{\ell \geq 1} \frac{z(\ell)}{\Omega_5} \left(2 \xi^{\ell} \tilde{\phi}^{\ell} q(\ell)
+\frac{2}{5} \Lambda_{\ell} \xi^{\ell} \tilde{\pi}^\ell 
+ \xi^{\mu}_\ell  D_\mu \pi^\ell 
-(\xi^\ell \tilde{\pi}^\ell + 2 \xi^{\mu}_\ell B_{\mu}^\ell) \Lambda_\ell \right),  \nonumber \\
&& \hspace{-0.5cm}\delta\tilde{\phi}^{0} \quad \hbox{not defined}, \nonumber \\
&& \hspace{-0.5cm}\delta \tilde{b}^{0} \quad \hbox{not defined},
\end{eqnarray}
For notational convenience, in these expressions and all the expressions of this section, whenever we have $\tilde{\phi}^{\ell}$ with $\ell=1$ we mean $\tilde{\pi}^{\ell=1}/\Lambda_{\ell=1}$. This means, e.g. that   $\sum_{\ell_2,\ell_3\geq 1}  \tilde{\phi}^{\ell_2} \xi^{\ell_3}  d_{123}$ in \eqref{gauge:O2} is a short-hand notation for $\sum_{\ell_2,\ell_3\geq 1}  \tilde{\phi}^{\ell_2} \xi^{\ell_3}  d_{123} \equiv \sum_{\ell_2=2,\ell_3\geq 1}
 \frac{\tilde{\pi}^{\ell_2}}{\Lambda_{\ell_2}} \xi^{\ell_3}  d_{123} + \sum_{\ell_2\geq 2,\ell_3\geq 1}
 \tilde{\phi}^{\ell_2} \xi^{\ell_3}  d_{123}$. Similarly,  one uses the short-hand notation $\sum_{\ell \geq 1} \frac{z(\ell)}{\Omega_5} \xi^{\ell} \tilde{\phi}^{\ell} q(\ell)\equiv \frac{z(1)}{\Omega_5} \xi^{1} \frac{\tilde{\pi}^{1}}{\Lambda_1} q(1)+\sum_{\ell \geq 2} \frac{z(\ell)}{\Omega_5} \xi^{\ell} \tilde{\phi}^{\ell} q(\ell)$ in \eqref{gauge:O2:l0}.

Using these transformations one can check that the gauge invariant quantities to quadratic order are:\footnote{To get \eqref{inv:O2}, symmetrise \eqref{gauge:O2} and promote $\{ \xi^{\ell} \to \tilde{\phi}^{\ell}/2 , \xi_\mu^{\ell}\to \hat{B}_{\mu}^{\ell} \}$ as follows from \eqref{gauge:O1} and \eqref{inv:O1}.}
\begin{eqnarray}\label{inv:O2}
&&\hspace{-1.5cm} \bullet \:\:  \hbox{For} \:\:\:\ell_1\geq 2:  \nonumber\\
 \pi^{\ell_1} &=& \hat{\pi}^{\ell_1} - \frac{\Omega_5}{2 z(\ell_1)}\sum_{\ell_2,\ell_3 \geq 1}\left[ 
\left( \frac{2}{5} \Lambda_{\ell_2} a_{123} + b_{123}
- \frac{2 \Lambda_{\ell_1} }{5 q(\ell_1)} d_{312} \right) \tilde{\phi}^{\ell_2} \hat{\pi}^{\ell_3} 
+ \left(d_{123} -\frac{\Lambda_{\ell_1}}{q(\ell_1)} e_{123} 
\right . \right.   \nonumber\\
&& \left . \left.
+\Lambda_{\ell_3}  \left(\frac{1}{5} \Lambda_{\ell_2} a_{123} + \frac{1}{2} b_{123}
- \frac{\Lambda_{\ell_1}}{5 q(\ell_1)} d_{213} \right) \right ) \tilde{\phi}^{\ell_2} \tilde{\phi}^{\ell_3} 
\right .     \nonumber\\
&& \left . 
+ 2 \hat{B}_{\mu}^{\ell_2}
\left (D^{\mu} \hat{\pi}^{\ell_3}a_{123} + \hat{B}^{\mu}_{\ell_3} (b_{123} - 
\frac{2 \Lambda_{\ell_1}}{q(\ell_1)} c_{123}) \right) \right], 
\nonumber \\
 b^{\ell_1} &=& \hat{b}^{\ell_1} +  \frac{\Omega_5}{z(\ell_1)} \sum_{\ell_2,\ell_3 \geq 1} \left[
\frac{\Lambda_{\ell_3}}{2 \Lambda_{\ell_1}} 
\tilde{\phi}^{\ell_2} \hat{b}^{\ell_3} b_{312}
+ \frac{1}{10 q(\ell_1)} d_{312} \tilde{\phi}^{\ell_2} \hat{\pi}^{\ell_3}
+  \left (\frac{\Lambda_{\ell_3}}{8 \Lambda_{\ell_1}} b_{312}+ 
\frac{\Lambda_{\ell_3}}{20 q(\ell_1)} d_{213} \right. \right. 
\nonumber \\
&&  \left.  \left.  
+ \frac{1}{8 q(\ell_1)} 
e_{123} \right )\tilde{\phi}^{\ell_2} \tilde{\phi}^{\ell_3} + \hat{B}_{\mu}^{\ell_2} 
\left ( \frac{1}{2 q(\ell_1)} \hat{B}^{\mu}_{\ell_3} c_{123}
+ \frac{1}{\Lambda_{\ell_1}} D^{\mu} \hat{b}^{\ell_3}  b_{213} \right) 
\right ]. 
\end{eqnarray}
\begin{eqnarray}\label{inv:O2:l1}
&&\hspace{-11.4cm} \bullet \:\:  \hbox{For} \:\:\:\ell_1= 1:  \nonumber\\
&&\hspace{-9.3cm} \pi^{\ell_1} = \hat{\pi}^{\ell_1}=0, \nonumber\\
&&\hspace{-9.3cm} b^{\ell_1} = \hat{b}^{\ell_1}.
\end{eqnarray}
\begin{eqnarray}\label{inv:O2:l2}
&&\hspace{-2.4cm} \bullet \:\: \hbox{For} \:\:\:\ell_1= 0:  \nonumber\\
\pi^0 &=& \tilde{\pi}^0 + \sum_{\ell \geq 1}\frac{z(\ell)}{\Omega_5} \left(\frac{3}{10} \Lambda_\ell  \tilde{\phi}^\ell \hat{\pi}^\ell
-\frac{1}{4} \Lambda_\ell(\Lambda_\ell+8) \tilde{\phi}^\ell \tilde{\phi}^\ell 
- \hat{B}^{\mu}_\ell D_{\mu} \hat{\pi}^\ell 
+ \Lambda_\ell \hat{B}^{\mu}_\ell \hat{B}_{\mu}^{\ell}
\right),\nonumber\\
b^{0} & & \hbox{not defined},
\end{eqnarray}
where the linear order quantities $\hat{\pi}^\ell $ and $\hat{B}_{\mu}^{\ell}$ are defined in \eqref{inv:O1}-\eqref{inv:O1:l1}.

Although we do not discuss the details of massive KK gravitons (i.e. $h_{\mu\nu}^{\ell}$ with $\ell\geq 1$; see \cite{Skenderis:2006uy}), the properties of the massless KK graviton $h_{\mu\nu}^{0}$ will be fundamental for our later analysis. 
Under a gauge transformation this field transforms as
\begin{eqnarray} \label{gauge:O2:l0graviton}
&& \delta \tilde{h}_{\mu \nu}^{0} = \left[ D_\mu \xi_\nu^{0} +  D_\nu \xi_\mu^{0} \right] + {\biggl[} D_\mu \xi^{\alpha}_{0} \tilde{h}_{\alpha \nu}^{0}+ D_\nu \xi^{\alpha}_{0} \tilde{h}_{\alpha \mu}^{0}+\xi^{\alpha}_{0} D_\alpha \tilde{h}_{\mu \nu}^{0}  \\ 
&& \hspace{1.3cm} + \sum_{\ell>1} \frac{z(\ell)}{\Omega_5}\left( D_\mu \xi^{\alpha}_\ell \tilde{h}_{\alpha \nu}^{\ell}+ D_\nu \xi^{\alpha}_\ell \tilde{h}_{\alpha \mu}^{\ell}+ \xi^{\alpha}_\ell D_\alpha \tilde{h}_{\mu \nu}^{\ell} - \Lambda_\ell (\xi^{\ell} \tilde{h}_{\mu \nu}^{\ell} + 2 D_{(\mu} \xi^{\ell} \tilde{B}_{\nu)}^\ell) \right) {\biggr]}, \nonumber 
\end{eqnarray}
and thus the corresponding gauge invariant massless graviton is given by ($g_{\mu \nu}^o$ is the AdS$_5$ metric)
\begin{eqnarray}\label{inv:O2:l2:grav}
h_{\mu \nu}^0 &=& \tilde{h}_{\mu \nu}^0 +\frac{1}{3} \pi^0 g_{\mu \nu}^o
-\sum_{\ell \geq 1} \frac{z(\ell)}{\Omega_5}  \left(-\frac{1}{2}\Lambda_\ell (\phi^\ell \hat{h}_{\mu \nu}^\ell
+ \frac{1}{2} D_\mu \phi^\ell D_\nu \phi^\ell)  \right. \\
&&\left. + D_{\mu} \hat{B}^{\sigma}_\ell \hat{h}_{\nu \sigma}^\ell + D_{\nu}
\hat{B}^{\sigma}_\ell \hat{h}^{\ell}_{\mu \sigma}  +
\hat{B}^{\sigma}_\ell D_\sigma \hat{h}_{\mu \nu}^\ell 
+ D_\mu \hat{B}^{\sigma}_\ell D_\nu \hat{B}_{\sigma}^\ell + 
\hat{B}^{\sigma}_\ell \hat{B}_{\sigma}^\ell g_{\mu \nu}^o
-\hat{B}_{\mu}^\ell \hat{B}_{\nu}^\ell \right), \nonumber 
\end{eqnarray}


With these gauge invariant quantities at quadratic order one can introduce the gauge invariant combinations
\begin{eqnarray} \label{comb:O2}
s^{\ell} &=& \frac{1}{20 (\ell+2)} \left[ \pi^{\ell} 
- 10 (\ell+4) b^{\ell}\right ], \qquad \hbox{for} \quad \ell\geq 1,   \nonumber \\
t^{\ell} &=& \frac{1}{20 (\ell+2)} \left( \pi^{\ell} + 10 \,\ell \, b^{\ell}\right ), \qquad  \hbox{for} \quad \ell\geq 0,  
\end{eqnarray}
which obey the equations of motion
\begin{eqnarray}\label{EOM:stO2}
&& \left( \Box -m^2_{s^\ell} \right) s^{\ell} = \frac{1}{2 (\ell+2)} \left[(\ell+4)(\ell+5) Q_1^\ell + Q_2^\ell
+ (\ell+4) (D_{\mu}Q_3^{\mu\:\ell} + Q_4^\ell) \right], \nonumber \\
&& \left( \Box -m^2_{t^\ell} \right) t^{\ell} = \frac{1}{2 (\ell+2)} \left [ \ell(\ell-1) Q_1^\ell + Q_2^\ell
- \ell (D_{\mu}Q_3^{\mu\:\ell} + Q_4^\ell) \right ], 
\end{eqnarray}
where $\Box$ is again the D'Alembertian in AdS$_5$. The quantities $\{Q_1^\ell, Q_2^\ell, Q_3^{\mu\:\ell}, Q_4^\ell \}$ in the RHS of these two equations are the {\it same} as those defined in equations (3.27)-(3.34) of \cite{Lee:1998bxa}. On the other hand, the masses of these scalar fields are the same as that of the linear fields $\hat{s}^\ell$ and $\hat{t}^\ell$
\begin{equation}\label{masses}
m^2_{s^\ell}=\ell(\ell-4)\,,\qquad m^2_{t^\ell}=(\ell+4)(\ell+8)\,.
\end{equation}

Letting $\psi_\ell=\{s^\ell,t^\ell\}$ and $m^2_{\psi_\ell}=\{m^2_{s^\ell},m^2_{t^\ell}\}$, and using the equations of motion for $\hat{s}^\ell$ \eqref{EOM:shat} and its derivatives one finds that the equations of motion \eqref{EOM:stO2} can be put in the form
\begin{equation}\label{EOM:O2v1}
\left( \Box -m^2_{\ell_1} \right) \psi_{\ell_1} = \sum_{\ell_2,\ell_3 \geq 1} \left[
D_{\ell_1\ell_2 \ell_3} \hat{s}^{\ell_2}\hat{s}^{\ell_3} 
+ E_{\ell_1\ell_2 \ell_3} \nabla_\mu \hat{s}^{\ell_2} \nabla^\mu \hat{s}^{\ell_3} 
+ F_{\ell_1\ell_2 \ell_3} \nabla^{(\mu}\nabla^{\nu)} \hat{s}^{\ell_2}
\nabla_{(\mu}\nabla_{\nu)}\hat{s}^{\ell_3} \right],
\end{equation}
for some coefficients $D_{\ell_1\ell_2 \ell_3}, E_{\ell_1\ell_2 \ell_3}, F_{\ell_1\ell_2 \ell_3}$. 

\subsection{5-dimensional KK description of the lumpy BHs \label{sec:KK5d}}

Now we are in position to obtain the KK map.  One can remove the derivative terms on the RHS of \eqref{EOM:O2v1} by a field redefinition $\psi_\ell \to \Psi_\ell$. The inverse of this relation gives the the reduced $d=5$ field $\Psi_\ell=\{{\cal S}^\ell,{\cal T}^\ell\}$ in terms of the $d=10$ field $\psi_\ell=\{ s^\ell,t^\ell \}$  and is given by
\begin{equation}\label{redef10to5v1}
\Psi_{\ell_1} = w(\psi_{\ell_1})\left[ \psi_{\ell_1}  - \sum_{\ell_2, \ell_3} \left(
J_{\ell_1\ell_2 \ell_3}\hat{s}^{\ell_2}\hat{s}^{\ell_3} + L_{\ell_1\ell_2 \ell_3}\nabla^\mu \hat{s}^{\ell_2} \nabla_\mu \hat{s}^{\ell_3} \right) \right],
\end{equation}
with the normalisation factor $w(\psi_\ell)$ and  coefficients $L,J$ given by
\begin{eqnarray}
&& w(s^\ell) = \sqrt{ \frac{8 \ell (\ell-1) (\ell+2)}{(\ell+1)}\frac{z(\ell)}{\Omega_5}}, \qquad
w(t^\ell) = \sqrt{ \frac{8 (\ell+2)(\ell+4)(\ell+5)}{(\ell+3)}\frac{z(\ell)}{\Omega_5}}; \nonumber \\
&& L_{\ell_1\ell_2 \ell_3} = {1 \over 2} F_{\ell_1\ell_2 \ell_3},\qquad 
J_{\ell_1\ell_2 \ell_3} ={1 \over 2}  E_{\ell_1\ell_2 \ell_3} 
+ {1 \over 4} F_{\ell_1\ell_2 \ell_3} \left(m_{\psi_{\ell_1}}^2 -m_{s^{\ell_2}}^2 - m_{s^{\ell_3}}^2 + 8\right).
\end{eqnarray}
The field equation for the reduced field is then
\begin{equation}\label{EOM:5dPsiv1}
\left( \Box -m^2_{\psi_{\ell_1}} \right) \Psi_{\ell_1}  = 
\sum_{\ell_2,\ell_3 \geq 1} \lambda_{\ell_1\ell_2 \ell_3} \hat{s}^{\ell_2} \hat{s}^{\ell_3},
\end{equation}
where
\begin{equation}\label{EOM:5dPsi2}
\lambda_{\ell_1\ell_2 \ell_3} = 
D_{\ell_1\ell_2 \ell_3} - \left( m_{s^{\ell_2}}^2 + m_{s^{\ell_3}}^2 - m_{\psi_{\ell_1}}^2 \right) J_{\ell_1\ell_2 \ell_3}
- {2 \over 5} L_{\ell_1\ell_2 \ell_3} m_{s^{\ell_2}}^2 m_{s^{\ell_3}}^2.
\end{equation}

In practice, for our lumpy AdS$_5\times$S$^5$ BH and up to the relevant order $O(z^4)$, only the sources associated to $\hat{s}^{2}\equiv \hat{s}^{\ell=2}$ contribute in the RHS of equations \eqref{EOM:O2v1}, \eqref{redef10to5v1} and \eqref{EOM:5dPsiv1}. Accordingly, up to this order we can rewrite them simply as
\begin{eqnarray}
\left( \Box -m^2_{\psi_\ell} \right) \psi_\ell  &=& D_{\psi 22} (\hat{s}^2)^2 
+ E_{\psi22} D_{\mu} \hat{s}^2 D^{\mu} \hat{s}^2
+ F_{\psi22} D_{(\mu} D_{\nu)} \hat{s}^2 D^{(\mu} D^{\nu)} \hat{s}^2,\label{EOM:10dpsilumpy}\\
\Psi_\ell &=& w(\psi_\ell) \left(\psi_\ell - J_{\psi22} (\hat{s}^2)^2 
- L_{\psi22} D_\mu \hat{s}^2 D^\mu \hat{s}^2 \right), \label{EOM:10dpsiTo5dPsilumpy} \\
\left( \Box -m^2_{\psi_\ell} \right) \Psi_{\ell}  &=& 
\lambda_{\Psi 22} (\hat{s}^2)^2 , \label{EOM:5dPsilumpy}
\end{eqnarray}
where the coefficients $\{D_{\psi 22}, E_{\psi 22}, F_{\psi 22}\}$, $\{J_{\psi22},L_{\psi22}\}$ and $\lambda_{\Psi 22}$ for the several $\psi_\ell$  are given in Table \ref{Table:DEFJL}. \footnote{Note that these coefficients depend on the harmonic representation we use for the S$^5$.} Explicitly, the 5-dimensional scalar fields of the lumpy AdS$_5\times$S$^5$ BH are
\begin{eqnarray}\label{5dscalars}
{\cal S}^1  &=& 0,\qquad {\cal S}^2=-\frac{1}{8} \sqrt{\frac{5}{3}} \,\beta_2 \,y_+^2 \left(z^2+ \frac{1}{6}  \left( 3+ \beta_2\, y_+^2 \right)z^4 \right),  \nonumber \\
{\cal S}^3  &=& \frac{1}{32} \sqrt{\frac{5}{6}} \, y_+^3 \,\gamma_3\, z^3\,,\qquad  {\cal S}^4= \sqrt{\frac{7}{3}}\,\frac{y_+^4}{76800} {\biggl(} 500 \delta_4+\beta_2 (194 \beta_2-125) {\biggr)} z^4\,;\nonumber \\
{\cal T}^0&=& 0  \,, \qquad {\cal T}^1 = 0,\qquad {\cal T}^2=0\,, \qquad
{\cal T}^3  = 0, \qquad {\cal T}^4= \frac{27}{640 \sqrt{10}} \,y_+^4\, \beta_2^2 \, z^4\,. 
\end{eqnarray}

\begin{table}
\begin{center}
\begin{tabular}{|c|||c|c|c|c|||c|c|c|c|c|}
\hline
\   & ${\cal S}^1$ & ${\cal S}^2$& ${\cal S}^3$ & ${\cal S}^4$ & ${\cal T}^0$ & ${\cal T}^1$ & ${\cal T}^2$ & ${\cal T}^3$ & ${\cal T}^4$  \\
\hline \hline
$D_{\psi 22}$ & $0$ & $-16 \sqrt{\frac{2}{15}}$  & $0$ & $-\frac{428 \sqrt{7}}{125}$ & $\frac{229}{75}$ & $0$ & $\frac{304}{25}\sqrt{\frac{6}{5}}$ & $0$ & $\frac{1084 \sqrt{7}}{125}$ \\
$E_{\psi 22}$ & $0$ & $\frac{2}{5}\sqrt{\frac{6}{5}}$  & $0$ & $\frac{3 \sqrt{7}}{5}$ & $-\frac{11}{20}$ & $0$ & $-\frac{6}{5}\sqrt{\frac{6}{5}}$ & $0$ & $-\frac{\sqrt{7}}{5}$\\
$F_{\psi 22}$ & $0$ & $\frac{1}{3}\sqrt{\frac{2}{15}}$  & $0$ & $\frac{8 \sqrt{7}}{225}$ & $\frac{1}{60}$ & $0$ & $\frac{1}{15}\sqrt{\frac{2}{15}}$ & $0$ & $\frac{3 \sqrt{7}}{50}$ \\
\hline
$J_{\psi 22}$ & $0$ & $\frac{8}{5}\sqrt{\frac{2}{15}}$ & $0$ & $\frac{199 \sqrt{7}}{450}$ & $-\frac{3}{40}$ & $0$ & $-\frac{8}{15}\sqrt{\frac{2}{15}}$ & $0$ & $\frac{79\sqrt{7}}{50}$ \\
$L_{\psi 22}$ & $0$ & $\frac{1}{3 \sqrt{30}}$ & $0$ & $\frac{4 \sqrt{7}}{225}$ & $\frac{1}{120}$ & $0$ & $\frac{1}{15 \sqrt{30}}$ & $0$ & $\frac{3 \sqrt{7}}{100}$ \\
\hline
$\lambda_{\Psi 22}$ & $0$ & $-\frac{128}{9 \sqrt{15}}$ & $0$ & $0$ & $0$ & $0$ & $0$ & $0$ & $\frac{5184}{5}\sqrt{\frac{2}{5}}$ \\
\hline
\end{tabular}
\caption{\label{Table:DEFJL} Coefficients $\{D_{\psi 22}, E_{\psi 22}, F_{\psi 22}\}$, $\{J_{\psi22},L_{\psi22}\}$ and $\lambda_{\Psi 22}$ for the several $\psi_\ell=\{ s^\ell,t^\ell \}$ or $\Psi_\ell=\{{\cal S}^\ell,{\cal T}^\ell\}$ that appear in \eqref{EOM:10dpsilumpy}-\eqref{EOM:5dPsilumpy}. (Note that these coefficients depend on the harmonic representation we use for the S$^5$).}
\end{center}
\end{table}

A similar treatment for the massless KK graviton field, including a field redefinition, allows us to write the $d=5$ graviton $G_{\mu\nu}$ in terms of the $d=10$ fields. Again up to the relevant order $(1/z^2)O(z^4)$, in the lumpy AdS$_5\times$S$^5$ BH only contributions sourced by $\hat{s}^{\ell=2}\equiv\hat{s}^{2}$ contribute. The reduced metric then reads (recall that $g_{\mu \nu}^o$ is the AdS$_5$ metric)
\begin{equation} \label{5dGab}
G_{\mu \nu} = h^{0}_{\mu \nu} 
- \frac{1}{12} \left[
\frac{2}{9} D_{\mu} D^{\sigma} \hat{s}^2 D_{\nu} D_{\sigma} \hat{s}^2 
- \frac{10}{3} \hat{s}^2 D_{\mu} D_{\nu} \hat{s}^2 +
 \left(\frac{10}{9}(D \hat{s}^2)^2 - \frac{32}{9} (\hat{s}^2)^2\right) g^{o}_{\mu
  \nu} \right ].
\end{equation}
This yields the following non-vanishing components
\begin{eqnarray}  \label{5dGab:explicit}
G_{zz} &=& \frac{1}{z^2}-\frac{65}{4608}\,y_+^4\,\beta_2^2  \, z^2, \\
 G_{ij} &=& \eta_{ij} \left[ \frac{1}{z^2} +\frac{K_0}{2}\right. \nonumber \\
&& \hspace{0.7cm} \left. 
+ \frac{ y_+^4 \left[85 \beta_2^2 \, K_2+30 \beta_2\, K_1+72 K_1 (16 \delta_{0}+\delta_4-192)\right]-13824\, K_1 \, y_+^2+1152}{18432} \,z^2 \right], \nonumber 
\end{eqnarray}
where  $\eta_{ij}=\hbox{diag}\{-1,\eta_{\hat{i}\hat{j}} \}$ with $\eta_{\hat{i}\hat{j}}$ being the line element of a unit radius S$^3$ and, to shorten the presentation, we introduced the auxiliary constants $\{ K_0, K_1,K_2 \}$ such that $\{ K_0, K_1,K_2  \}= \{ 1,1,1 \}$ for $i=j=t $, while $\{ K_0, K_1,K_2  \}= \{ -1,-1/3,31/17 \}$ for components $i=j$ on the S$^3$.

The field equations \eqref{EOM:5dPsilumpy} can be obtained from a 5-dimensional action, namely
\begin{equation} \label{5dAction}
S_{5d} = 
\frac{N^2}{2 \pi^2} \int d^5 x \sqrt{-G}\, \left[ \frac{1}{4} R-3+ \sum_{\Psi_\ell} \left(
\frac{1}{2} G^{\mu \nu} \partial_\mu \Psi_\ell \,\partial_\nu \Psi_\ell + V(\Psi_\ell)  \right) \right],
\end{equation}
with $G_{\mu\nu}$ given by \eqref{5dGab}  (recall that Newton's constant $G_5$ is given by \eqref{newton}). The first two contributions in this action are the Einstein and cosmological terms (recall $L=1$) that admit AdS$_5$ as a solution. Up to order $O(z^4)$, the potentials in this action are 
\begin{eqnarray} \label{5dPotential}
&&V(\Psi_\ell)=\frac{1}{2} m^2_{{\cal S}^2} ({\cal S}^2)^2-\frac{16}{3 \sqrt{15}} ({\cal S}^2)^3,\qquad \hbox{for} \quad \Psi_\ell={\cal S}^2\,,  \nonumber \\
&&V(\Psi_\ell)=\frac{1}{2} m^2_{{\cal T}^4} ({\cal T}^4)^2-48({\cal T}^4)^2=O(z^5),\qquad \hbox{for} \quad \Psi_\ell={\cal T}^4\,,  \nonumber \\
&&V(\Psi_\ell)=\frac{1}{2} m^2_{\psi_\ell} \left(\Psi_\ell \right)^2=O(z^5),\qquad \hbox{otherwise}.
\end{eqnarray}
Variation of the 5-dimensional action w.r.t. the scalar fields indeed yields the massive Klein-Gordon equations \eqref{EOM:5dPsilumpy}; note that $\Box=\Box_{AdS_5}=\Box_G$ up to order $O(z^4)$.
 
The Einstein equation that follows from the  5-dimensional action \eqref{5dAction} is
\begin{equation} \label{5dEin}
R_{\mu \nu}[G] = 2 \left(-2 G_{\mu\nu}+T_{\mu \nu} - \frac{1}{3} G_{\mu \nu} T^\sigma_\sigma \right),
\end{equation}
where $R_{\mu \nu}$ is the Ricci tensor of $G_{\mu \nu}$ and the energy-momentum tensor reads
\begin{eqnarray}\label{5dEinT}
T_{\mu \nu}&=& \sum_{\Psi_\ell} \left[ \partial_\mu \Psi_\ell \,\partial_\nu \Psi_\ell - G_{\mu \nu} \left( \frac{1}{2} 
(\partial \Psi_\ell)^2 + V(\Psi_\ell) \right) \right] \nonumber \\
&=& \partial_\mu {\cal S}^2 \,\partial_\nu {\cal S}^2 - G_{\mu \nu} \left( \frac{1}{2} 
(\partial {\cal S}^2)^2 + V({\cal S}^2) \right) +{\cal O}(z^5).
\end{eqnarray}
As indicated by the second equality up to the relevant ${\cal O}(z^4)$ only the scalar field ${\cal S}^{\ell=2}$ contributes to the stress tensor.

\subsection{Holographic renormalisation and Stress tensor. \label{sec:KKrenormalization}}

We can now apply the standard holographic renormalisation procedure to the 5-dimensional solution \cite{deHaro:2000xn}. 
Introduce the Fefferman-Graham coordinate $Z=z-\frac{65}{36864}\,y_+^4\,\beta_2^2  \, z^5+O(z^6)$ for the 5-dimensional metric $G_{\mu\nu}$, and denote the boundary coordinates collectively by $X$. Furthermore, collectively denote the scalar fields of the system by $\Phi=\{{\cal S}^\ell,{\cal T}^\ell,\phi_{KK}^\ell\}$ (where $\phi_{KK}^\ell$ are the massive KK gravitons described at linear order by \eqref{KK:O1}-\eqref{KKeom:O1}) and recall that the conformal dimensions $\Delta$ (and $\Delta_-=4-\Delta$) of the operators dual to these fields that are given in Table \ref{table:confDim}. 
\begin{table}
\begin{center}
\begin{tabular}{| l| | l | l | l |}
\hline
\  & $\:\:\: {\cal S}^\ell$ & $\:\:\:  {\cal T}^\ell$ & $\:\:\:\phi_{KK}^\ell$  \\
\hline \hline
$\Delta_+=\Delta$ & $\ell$ & $ \ell+8$ & $\ell+4$ \\
\hline
$\Delta_-=4-\Delta$ & $4-\ell$ & $-\ell-4$ & $-\ell$ \\
\hline
\end{tabular}
\caption{\label{table:confDim}Conformal dimensions $\Delta_\pm$ of the scalar fields ${\cal S}^\ell, {\cal T}^\ell$ and of the KK gravitons $\phi_{KK}^\ell$. }
\end{center}
\end{table}

\begin{figure}[ht]
\centering
\includegraphics[width=.6\textwidth]{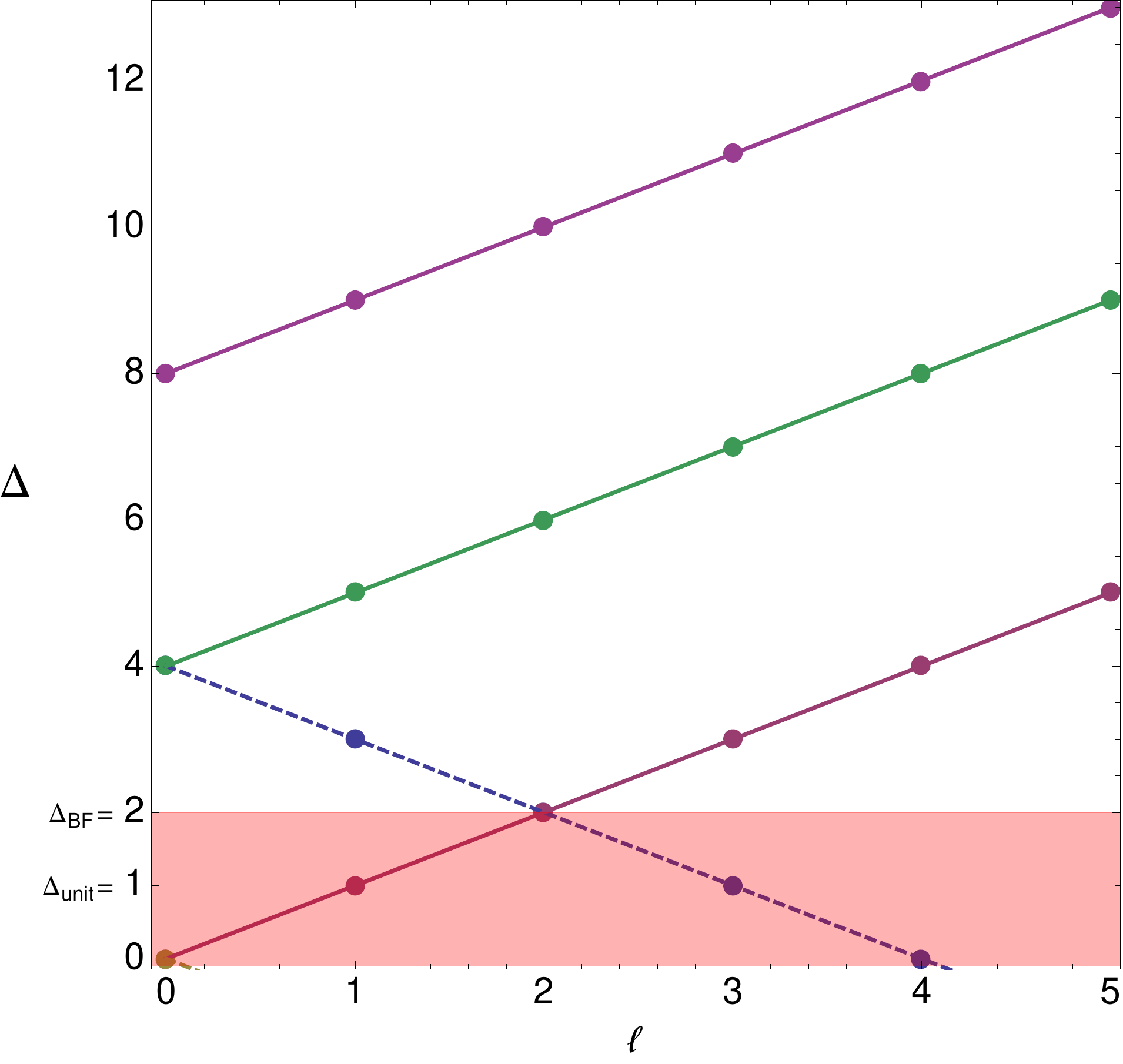}
\caption{Conformal dimensions of the several operators that are present in the lumpy system. From top to bottom the points connected by a continuous line are the conformal dimensions $\Delta_+$ of the dual operator of the scalars ${\cal T}^\ell$, KK gravitons $\phi^{KK}_\ell$, and scalars ${\cal S}^\ell$. The dots connected by a dashed line represent the conformal dimension $\Delta_-=4-\Delta_+$; in the region displayed, the plot shows only the dashed line associated with the scalars ${\cal S}^\ell$.}\label{Fig:confdim}
\end{figure}

The expansion around the holographic boundary $Z=0$ for the 5-dimensional metric $G_{\mu\nu}$, and scalar fields $\Phi=\{{\cal S}^\ell,{\cal T}^\ell,\phi_{KK}^\ell\}$ is then
\begin{eqnarray} \label{FGexpansion}
&& \hspace{-2cm} ds_5^2 =
\frac{dZ^2}{Z^2} 
+ \frac{1}{Z^2}\left[G^{(0)}_{ij}(X) + Z^2 G^{(2)}_{ij}(X) 
+ Z^{4} \left(G^{(4)}_{ij}(X) + \log Z^2 H^{(4)}_{ij}(X)\right)+\cdots \right] dX^i dX^j; 
\nonumber \\
\Phi^2(X,Z) &=& 
Z^2 \left(\log Z^2 \Phi^2_{(0)}(X) + \tilde{\Phi}_{(0)}^2(X) + \cdots \right), \qquad \hbox{for} \quad \Delta=\Delta_{BF}=2; 
\nonumber \\
\Phi^\Delta(X,Z) &=& Z^{(4-\Delta)} \Phi^\Delta_{(0)}(X) + \cdots +Z^\Delta \Phi_{(2\Delta-4)}^\Delta(X) + \cdots,
\qquad \hbox{for} \quad  \Delta > 2\;,
\end{eqnarray}
where the $\Delta=2$ case saturates the 5-dimensional Breitenlohner-Freedman (BF) bound, i.e.  $\Delta_+=\Delta_- = \Delta_{BF}$. In Fig. \ref{Fig:confdim}, we plot the conformal dimensions of the several scalar fields $\Phi=\{{\cal S}^\ell,{\cal T}^\ell,\phi_{KK}^\ell\}$ as a function of the harmonic quantum number $\ell$. The BF bound is saturated only for the field ${\cal S}^{\ell=2}$. 

In the above off-boundary expansion, the non-normalisable modes $G_{(0)ij}, \Phi^2_{(0)}, \Phi^\Delta_{(0)}$ are source terms for the boundary QFT stress tensor and dual operators of dimension $\Delta=2$ and $\Delta$, respectively. On the other hand, the normalizable modes, namely  $G_{ij}^{(4)}, \tilde{\Phi}^2_{(0)}, \Phi^\Delta_{(2\Delta-4)}$ are determined by solving the field equations \eqref{5dEin} and \eqref{EOM:5dPsilumpy} of \eqref{5dAction} subject to regular (ingoing) boundary conditions at the BH horizon. All other coefficients $G_{ij}^{(k)}, \Phi^\Delta_{(k)}$ of \eqref{FGexpansion}, typically represented by dots, are expressed as a function of the (non-)normalisable modes and their derivatives. 

At this point we can discuss the boundary conditions (BCs) that we impose in the  holographic boundary $Z=0$.
We do not want to deform the boundary background so we fix $G^{(0)}_{ij}$ to be the static $R_t\times S^3$ metric as a Dirichlet boundary condition.  Moreover, we do not want to deform the boundary CFT, i.e. we require vanishing sources in the scalar fields, so we impose vanishing Dirichlet boundary conditions on the fields $\Phi$. Altogether we thus have the asymptotic boundary conditions:
\begin{eqnarray} \label{BCs:bdry}
G^{(0)}_{ij}(X) &=& G^{(0)}_{ij}{\bigl |}_{R_t\times S^3}\nonumber \\
\Phi^2_{(0)}(X) &=& 0, \qquad \hbox{for} \quad \Delta=\Delta_{BF}=2; 
\nonumber \\
 \Phi^\Delta_{(0)}(X) &=& 0, \qquad \hbox{for} \quad \Delta > 2. 
\end{eqnarray}
These BCs can be discussed in more detail with the aid of Fig. \ref{Fig:confdim}.  {\it \`A priori}, we could impose BCs that would allow operators with $\Delta\geq \Delta_{unit}$, where $\Delta_{unit}=1$ is the unitarity bound. However, the dual CFT of our system is ${\cal N}=4$ SU(N) SYM. In this special case, the requirement that the norm of the supercurrent of the theory is positive definite requires that we exclude conformal dimensions in the range $\Delta_{unit} \leq \Delta<\Delta_{BF}$ (ultimately this means that we have a dual $SU(N)$ gauge theory and not $U(N)$; for further details see e.g. section 2.6 of \cite{Witten:1998qj}). Therefore, we impose BCs such that only dual operators with conformal dimension equal or higher than the BF bound, $\Delta\geq \Delta_{BF}=2$, are present. This means that the BCs we impose are such that sources of operators in (and below) the red area of Fig. \ref{Fig:confdim} vanish. 

In addition, some modes above this region (the source terms  with $\{\Delta_-,\ell\}=\{2,2\}$, and $\{\Delta_-,\ell\}=\{4,0\}$) are eliminated by the BCs $\Phi^2_{(0)}=0$ and $\Phi^4_{(0)}=0$ on the scalars ${\cal S}^\ell$. There is however a special mode that we cannot exclude with our BCs namely, the mode with $\{\Delta_-,\ell\}=\{3,1\}$. This is a pure gauge mode and therefore it does not appear in any physical quantity.\footnote{Consequently, it is associated to a parameter that does not appear in the harmonic coefficients \eqref{coef:lumpy} up to the order in $z$ that we display because  \eqref{coef:lumpy} shows only the order needed to compute the relevant physical charges and vevs.}   We cannot remove this mode since we are using the deTurck method. The gauge is fixed after solving the equations, and cannot be imposed from the equations of motion alone (see discussion associated to \eqref{EOM:deTurck}). 

We can now discuss the undetermined constants in the asymptotic expansion \eqref{coef:lumpy} that are permitted by the BCs \eqref{BCs:bdry}. The coefficient $\tilde{\Phi}_{(0)}^2$ in \eqref{FGexpansion}, with  $\{\Delta_+,\ell\}=\{2,2\}$, is proportional to the parameter $\beta_2$ appearing in \eqref{coef:lumpy}. In \eqref{FGexpansion}, the normalisable modes $ \Phi_{(2\Delta-4)}^\Delta$ of the scalars ${\cal S}^\ell$  associated to $\{\Delta_+,\ell\}=\{3,3\}$ and $\{\Delta_+,\ell\}=\{4,4\}$ are, respectively, proportional to the parameters $\gamma_3$ and $\delta_4$ present in  \eqref{coef:lumpy}. Finally, the fourth constant $\delta_0$ that appears in the boundary expansion \eqref{coef:lumpy} describes a KK graviton with $\{\Delta_+,\ell\}=\{4,0\}$. Normalisable modes with $\Delta_+>4$ $-$ see Fig. \ref{Fig:confdim} $-$ appear only at an order in $z$ higher than the one displayed in \eqref{coef:lumpy}. They are also present in our solution but we do not discuss them further because they do not contribute to the mass of the lumpy BHs. 

The normalisable modes are related to the holographic 1-point functions that give the VEVs $\langle T_{ij} \rangle$, $\langle {\cal{O}}^2 \rangle$ and  $\langle {\cal{O}}^\Delta \rangle$ of the dual operators, via the standard holographic renormalisation procedure. In particular, the vev of the operators ${\cal O}_{{\cal S}^2}$ and ${\cal O}_{{\cal S}^3}$  are ($\tilde{\Phi}_{(0)}^2$ and $\Phi_{(2)}^3$ are read directly from \eqref{5dscalars}) 
\begin{eqnarray}\label{vevS2S3} 
&& \left \langle {\cal O}_{{\cal S}^2} \right\rangle = \frac{N^2}{\pi^2}\, \tilde{{\cal S}}_{(0)}^2=-\frac{N^2}{\pi^2}\,\frac{1}{8} \sqrt{\frac{5}{3}} \,y_+^2 \,\beta_2  ,\nonumber\\ 
&& \left \langle {\cal O}_{{\cal S}^3} \right\rangle = \frac{N^2}{\pi^2}\, \Phi_{(2)}^3= \frac{N^2}{\pi^2}\,  \frac{1}{8} \sqrt{\frac{5}{6}}  \, y_+^3 \,\gamma_3,
\end{eqnarray}
and the holographic stress tensor is
\begin{eqnarray} \label{holoTij}
\langle T_{ij} \rangle &=&  \frac{N^2}{2 \pi^2}\left[
G_{ij}^{(4)} +\frac{1}{3} \tilde{{\cal S}}_{(0)}^2 G_{ij}^{(0)}+
\frac{1}{8}\left({\rm Tr} \,G_{(2)}^2 -({\rm Tr} \,G_{(2)})^2\right) G_{ij}^{(0)}\right. \\
&&\hspace{1cm} -\left.\frac{1}{2} (G_{(2)}^2)_{ij} + \frac{1}{4} G_{ij}^{(2)} {\rm Tr} \,G_{(2)}
+\frac{3}{2} H_{ij}^{(4)} 
+\left(\frac{2}{3} {\cal S}^2_{(0)} - \tilde{{\cal S}}_{(0)}^2\right)
{\cal S}^2_{(0)} G_{ij}^{(0)}\right],\nonumber
\end{eqnarray}
which, for the lumpy BHs, explicitly reads
\begin{equation} \label{holoT}
\langle T_{ij} \rangle =  \frac{N^2}{2 \pi ^2} \left[ \frac{3}{16}+\frac{3}{4}\,y_+^2-\frac{y_+^4}{3072} {\biggl(}30 \beta_2^2+5 \beta_2+12 (16 \delta_{0}+\delta_4-192){\biggr)} \right]\, {\rm diag} \left\{ 1, \frac{1}{3}\,\eta_{\hat{i} \hat{j}} \right\},
\end{equation}
with $\eta_{\hat{i}\hat{j}}$ being the line element of a unit radius S$^3$. This holographic stress tensor is conserved, $ {}^{(0)}\nabla_i \langle T^{ij} \rangle =0$, and it is traceless, $\langle T^{\:i}_i \rangle=0$.\footnote{The lumpy BH asymptotes to global AdS$_5$ which is conformal to the Einstein Static universe $R_t\times S^3$, and thus conformal to flat space. Therefore the gravitational conformal anomaly vanishes. Moreover a possible contribution, both to the conservation equation and trace, of the form $\Phi^\Delta_{(0)}{\cal O}_\Phi$ is not present because we impose Dirichlet boundary conditions in the scalar field, $\Phi^\Delta_{(0)}\equiv 0$.}

An important holographic quantity that we want to extract from \eqref{holoT} is the energy of the solution. This is done by pulling-back $\langle T_{ij} \rangle$ to a 3-dimensional spatial hypersurface  $\Sigma_t$, with unit normal $n$ and induced metric $\sigma^{ij}=G_{(0)}^{ij}+n^{i}n^{j}$, and contracting it with the Killing vector $\xi=\partial_t$ that generates time translations. The integral of this quantity gives the desired energy 
\begin{eqnarray} \label{energy0}
E&=&-\int_{\Sigma_t}\sqrt{\sigma} \langle T_i^j \rangle \xi^i n_j\nonumber\\
&=&\frac{N^2}{3072}{\biggl [}576+2304 \,y_+^2(1+y_+^2) -y_+^4 {\biggl (}5\, \beta_2+30 \,\beta_2^2+12\left(16 \,\delta_{0}+\delta_4\right){\biggr)}{\biggr ]}.
\end{eqnarray}
This is the main result of this Appendix. In the main text we will use \eqref{energy0} (rewritten in \eqref{energy} with factors of $L$  restored) to determine the energy of the lumpy BHs.  The energy of the AdS$_5$-$\mathrm{Schw}$ BH, $E_{SAdS_5}/N^2=(3/4) y_+^2 \left(y_+^2+1\right) + 3/16$, is recovered when we set the lumpy parameters to zero, $\beta_2=\delta_{0}=\delta_4=0$.

\section{Numerical details and validity\label{sec:numerics}}

In this appendix we discuss the validity of our numerical results, while giving further details of the numerical construction of the AdS$_5\times$S$^5$ lumpy BHs.

\begin{figure}[ht]
\centering
\includegraphics[width=.48\textwidth]{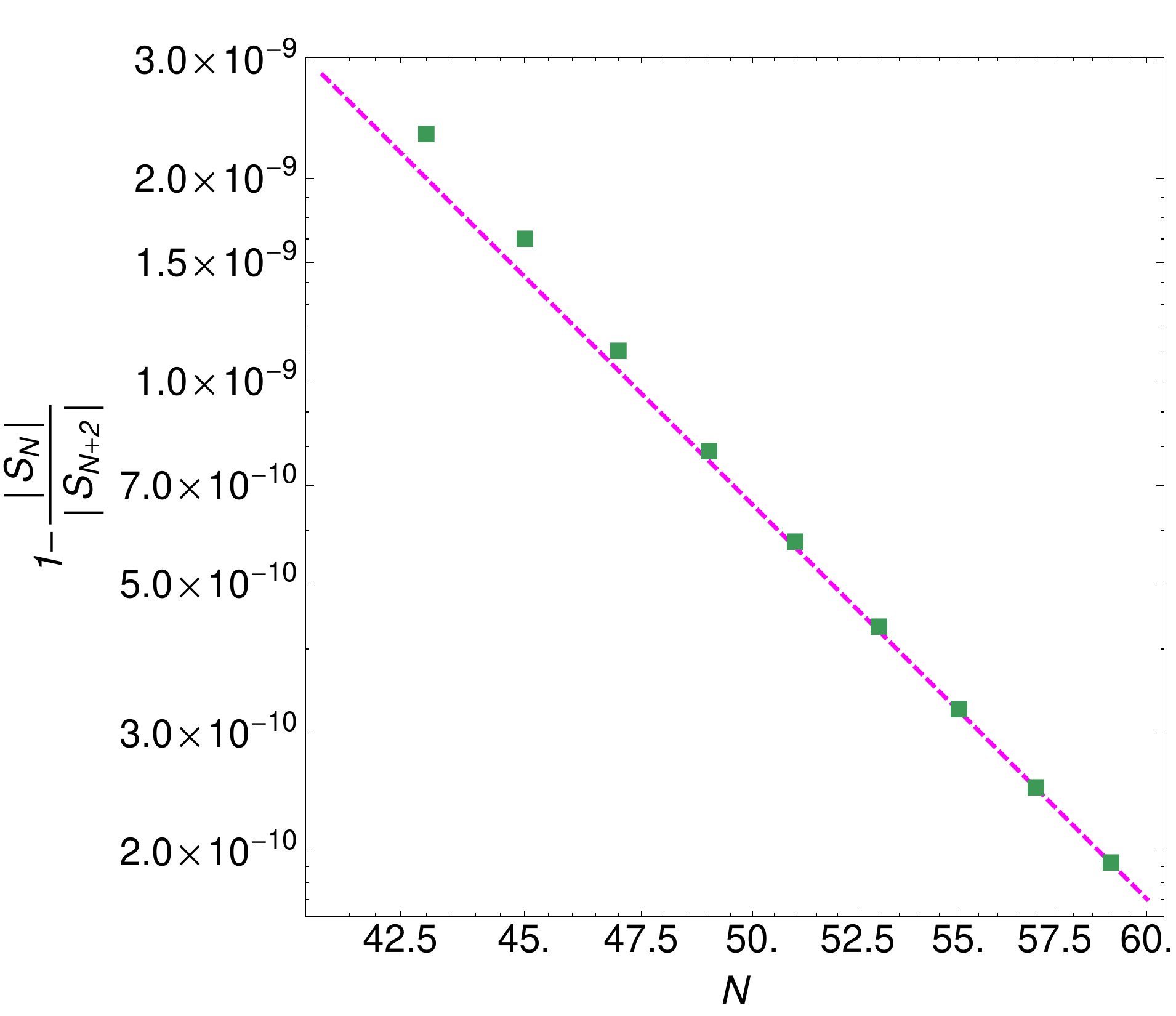}\qquad
\includegraphics[width=.45\textwidth]{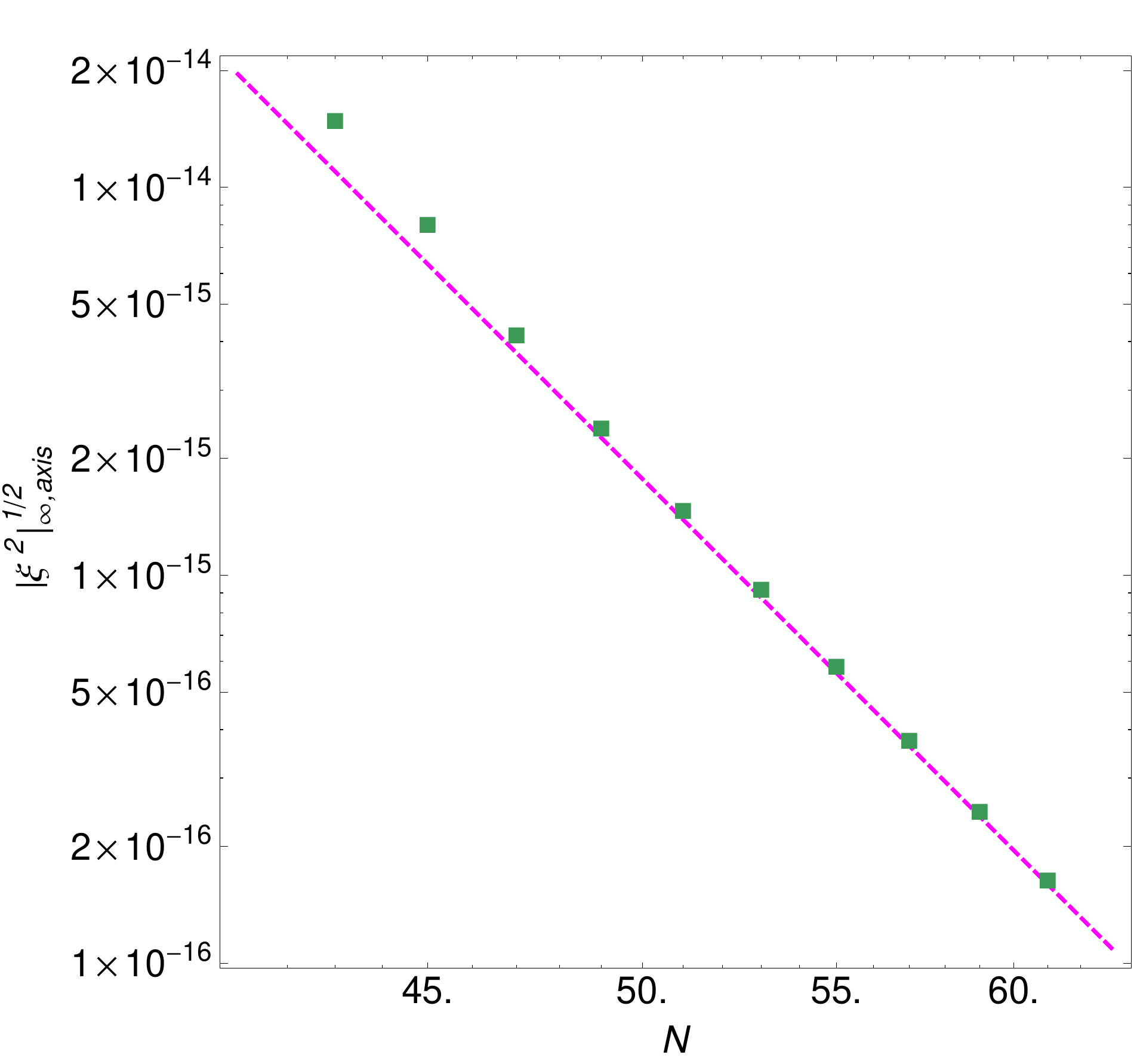}
\caption{{\bf Left panel:} Convergence test for the entropy of the lumpy BHs. We plot $1-|S(N)|/|S(N+2)|$, as a function of the number of grid points $N$.  {\bf Right panel:} Convergence test of the square root of the norm of the DeTurck vector at the asymptotic boundary and at the rotation axis, $\sqrt{|\xi^2|}_{\infty,axis}$, as a function of the grid points $N$. Both plots are for lumpy BHs with $T=0.50065$ (i.e. $y_+=0.44225$).}\label{fig:converge}
\label{fig:convex}
\end{figure}

We start by testing the numerical convergence.  We use pseudospectral collocation methods, and thus we expect exponential convergence with increasing number of grid points. We demonstrate this convergence with the panel  of Fig.~\ref{fig:converge}.  In this figure, as a typical example of our results, we consider a lumpy BH at constant temperature $T=0.50065$, and show how its entropy changes as the number $N$ of grid points is varied.

Next, we test numerical convergence of the norm of the deTurck vector $\xi^2$, defined below \eqref{EOM:deTurck}. The Einstein-de Turck method solves Einstein equations in the gauge $\xi^M=0$. Therefore, the norm of the deTurck vector is a measure of how well this gauge condition is satisfied, and verifies that we have a proper solution to the Einstein equations and not a DeTurck soliton with $\xi^M\neq0$.  On the right panel of Fig.~\ref{fig:converge}, we take a lumpy BH at constant temperature and plot the square root of the norm of the DeTurck vector (evaluated at the asymptotic boundary, $y=1$, and at the rotation axis $x=1$) as a function of the grid points. Again we confirm the presence of exponential convergence. We find that $\sqrt{|\xi^2|}<10^{-12}$ everywhere.  

The first law and the energy read from KK holography provide a final important test of our numerics. We can independently compute the energy, entropy and temperature using  KK holography. Therefore, we can test whether the first law, $dE=T dS$, is satisfied. We find this to be the case as mentioned below \eqref{energy} and in the discussions  associated with Figs. \ref{Fig:microcanonicalL1} and \ref{Fig:canonicalL1}.  

\end{appendix}


\bibliography{refs}{}
\bibliographystyle{JHEP}

\end{document}